\documentclass[aps,physrev,amssymb,amsmath,superscriptaddress,preprintnumbers,reprint, floatfix]{revtex4-2}

\usepackage[caption=false]{subfig}

\usepackage{cases}

\usepackage{graphicx}
\usepackage{dcolumn}
\usepackage[normalem]{ulem}
\usepackage{float}
\usepackage{color}
\usepackage{xcolor}

\usepackage[utf8]{inputenc}  
\usepackage[T1]{fontenc}     
\usepackage{newtxtext,newtxmath}  
\usepackage{enumitem}        

\usepackage{hyperref}
\hypersetup{
  colorlinks=true,
  linkcolor=red,
  filecolor=blue, 
  urlcolor=blue,
  citecolor=blue,
}

\usepackage{textcase}

\newcommand{\ld}{\ell_\mathrm{d}}
\newcommand{\lp}{\ell_\mathrm{p}}
\newcommand{\tp}{\tau_\mathrm{p}}
\newcommand{\sref}[1]{Sec.~\ref{#1}}

\begin{document}

\preprint{\href{https://doi.org/10.1103/PhysRevE.111.024128}{PHYSICAL REVIEW E {\bf 111}, 024128 (2025)}}

\title{Hydrodynamics of a hard-core active lattice gas}

\author{Ritwik Mukherjee}
\email{ritwik.mukherjee@icts.res.in}
\affiliation{International Centre for Theoretical Sciences, Tata Institute of Fundamental Research, Bangalore 560089, India}

\author{Soumyabrata Saha}
\email{soumyabrata.saha@tifr.res.in}
\affiliation{Department of Theoretical Physics, Tata Institute of Fundamental Research, Mumbai 400005, India}

\author{Tridib Sadhu}
\email{tridib@theory.tifr.res.in}
\affiliation{Department of Theoretical Physics, Tata Institute of Fundamental Research, Mumbai 400005, India}

\author{Abhishek Dhar}
\email{abhishek.dhar@icts.res.in}
\affiliation{International Centre for Theoretical Sciences, Tata Institute of Fundamental Research, Bangalore 560089, India}

\author{Sanjib Sabhapandit}
\email{sanjib@rri.res.in}
\affiliation{Raman Research Institute, Bangalore 560080, India}

\begin{abstract}
We present a fluctuating hydrodynamic description of an active lattice gas model with excluded volume interactions that exhibits motility-induced phase separation under appropriate conditions. For quasi-one dimension and higher, stability analysis of the noiseless hydrodynamics gives quantitative bounds on the phase boundary of the motility-induced phase separation in terms of spinodal and binodal. Inclusion of the multiplicative noise in the fluctuating hydrodynamics describes the exponentially decaying two-point correlations in the stationary-state homogeneous phase. Our hydrodynamic description and theoretical predictions based on it are in excellent agreement with our Monte Carlo simulations and pseudospectral iteration of the hydrodynamics equations. Our construction of hydrodynamics for this model is not suitable in strictly one-dimension with single-file constraints, and we argue that this breakdown is associated with micro-phase separation. 
\end{abstract}

\maketitle

\section{Introduction}
Active matter commonly refers to nonequilibrium systems with energy injections at the microscopic scale that independently drives individual constituents and often leads to large scale self-organized structures~\cite{2022_Bowick_Symmetry,2017_Ramaswamy_Active,2010_Ramaswamy_The,2013_Marchetti_Hydrodynamics}. Although they are inspired by living matter, active systems have been realized in synthetic objects such as Janus particles~\cite{1995_Vicsek_Novel}, vibrated rods~\cite{2012_Deseigne_Vibrated}, driven granular systems~\cite{2010_Deseigne_Collective}, motor driven robots~\cite{2021_Wang_Emergent,2022_Wood_Biohybrid}, and even in quantum matter~\cite{2023_Khasseh_Active}. The prominence of the field comes from the novel emergent collective behaviors that often comes as a surprise from an equilibrium viewpoint. Beside their potential applicability~\cite{2018_Frangipane_Dynamic,2021_Fodor_Active}, these emergent many-body structures pose an intriguing challenge for nonequilibrium statistical mechanics. In recent decades, major efforts have been devoted to developing a theoretical understanding based on effective field theories~\cite{2022_Cates_Active} and large scale computer simulations~\cite{2020_Shaebani_Computational,2021_Klamser_Kinetic,2023_Sabass_Computational}. However, even for simple characteristic many-body phases of active matter, namely the Motility-Induced Phase Separation (MIPS)~\cite{2015_Cates_Motility,2023_Byrne_An}, there is no clear bottom-up theoretical understanding, analogous to equilibrium statistical mechanics for liquid-gas coexistence, that could relate the large scale phases to the underlying microscopic dynamics.

A promising bottom-up approach for many-body dynamics is in terms of fluctuating hydrodynamics, that has been phenomenally successful in capturing nonequilibrium fluctuations in diffusive transport models~\cite{2015_Bertini_Macroscopic}, Hamiltonian dynamical systems~\cite{2018_Doyon_Exact,2014_Spohn_Nonlinear}, and even in quantum integrable models~\cite{2016_Bertini_Transport,2016_Castro_Emergent,2018_Nardis_Hydrodynamic}. Hydrodynamics description has also proven tremendously effective in active matter~\cite{2002_Simha_Hydrodynamic,2005_Toner_Hydrodynamics,2018_Julicher_Hydrodynamic,2019_Partridge_Critical,Langford2023}, although a rigorous derivation relating to microscopic dynamics is available only for a handful of cases. One such example is an active lattice gas~\cite{2018_Houssene_Exact,2024_Erignoux}, whose hydrodynamics was proven in mathematical terms and its fluctuation extension incorporating a stochasticity has been recently obtained~\cite{2021_Agranov_Exact,2023_Agranov_Macroscopic} using intriguing mapping to a well-known ABC exclusion model. However, the model involves an exchange dynamics between particles that seems unphysical, presumably incorporated to sustain MIPS. Most well-known off-lattice active dynamics~\cite{2015_Cates_Motility,2018_Digregorio_Full,2018_Klamser_Thermodynamic,2018_Caballero_From} are for impassable particles with finite volume hard core which provides the much-needed activity-induced caging effect to sustain MIPS. In fact, volume exclusion is expected in natural examples of living or synthetic active matter.

In this article, we propose a lattice model, that is inspired by~\cite{2018_Houssene_Exact}, but without the exchange dynamics. The dynamics of our model is closer to realistic off-lattice dynamics of active matter that are known to exhibit MIPS~\cite{2015_Cates_Motility}. 
For our model in the quasi-one-dimension of a periodic two-lane ladder lattice, there is a stable MIPS state at certain range of high activity and density. For this model, we derive the fluctuating hydrodynamics using a field-theoretical approach that is generalizable and independent of the previous method in~\cite{2021_Agranov_Exact,2023_Agranov_Macroscopic}. Within the hydrodynamics we show how dynamical instabilities of the coupled equations of hydrodynamic fields quantitatively bound the MIPS boundaries. Moreover, we find that correlations from the fluctuating hydrodynamics have excellent agreement with their Monte Carlo results and capture the qualitative features seen~\cite{2021_Szamel_Long} in well-known off-lattice active matter models. Our construction of the hydrodynamics straightforwardly extends for the higher dimensional generalization of the model and in particular, they capture the existence of MIPS in two dimensions.

The rest of the article is organized as follows. In~\sref{s:quasi-1D} we precisely define the model in the quasi-one-dimensional (quasi-1D) ladder lattice and present its fluctuating hydrodynamic description in~\sref{s:hydro}. The spinodal and binodal bounds on the MIPS phase boundary obtained using stability analysis of the noiseless hydrodynamics is discussed in~\sref{s:spin-bin}. In~\sref{s:corr} we compute the spatial correlations of density and magnetization by incorporating the noise into the hydrodynamic description. Further generalizations to higher dimensions are presented in~\sref{s:dd}. The limitations of our hydrodynamic construction in strictly one-dimensional lattice are discussed in~\sref{s:1d} and conclude the article in~\sref{s:conclusion}. Additional details about the construction of the fluctuating hydrodynamics and numerical analysis are relegated to the Appendixes.

\begin{figure}[t]
\centering
\includegraphics[width=0.99\linewidth]{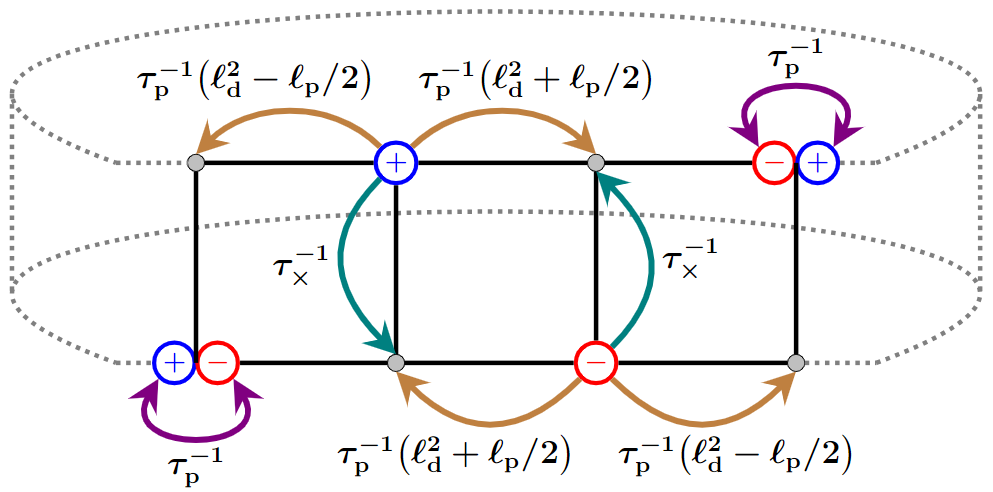}
\caption{\textbf{Quasi-1D model}: A schematic of the model defined in items (\ref{diff_drift})--(\ref{lane_cross}) on a two-lane ladder lattice with periodic boundary condition.}
\label{fig:quasi 1d model}
\end{figure}

\section{A quasi-1D model}
\label{s:quasi-1D}
We consider a two-species lattice gas with average density $\rho_0\in [0,1]$, on a periodic two-lane ladder-lattice, with each lane $j=\{1,2\}$ containing $L$ sites indexed as $i=\{1,\,2,\,\ldots,\,L\}$. To incorporate hard-core repulsion among particles, we enforce a simple exclusion condition that each site can contain at most one particle at a given instance. The two species denoted by $(+)$ and $(-)$, representing internal orientations of an active particle, follow a persistent dynamics on the lattice with intermittent switching of the orientations,
as defined below. (See Fig.~\ref{fig:quasi 1d model} for a schematic of the same.)
\begin{enumerate}

\item Biased diffusion: a $(+)$ particle at site $(i,j)$ hops to site $(i\pm 1,j)$ at rate $\tp^{-1}\left[(\ld/a)^2\pm (\lp/a)/2\right]$ while a $(-)$ particle at site $(i,j)$ hops to $(i\pm 1,j)$ site at rate $\tp^{-1}\left[(\ld/a)^2\mp (\lp/a)/2\right]$, provided the target site is empty. \label{diff_drift}

\item Tumbling: a $(+)$ particle at any site converts to a $(-)$ particle at rate $\tp^{-1}$ and vice versa. \label{tumble}

\item Lane crossing: a particle at a site $(i,1)$ on lane $1$ hops to the site $(i,2)$ on lane $2$ at rate $\tau_\times^{-1}$ and vice versa, provided the target site is empty. \label{lane_cross}

\end{enumerate}

The above dynamics mimics self-propelled particles with persistence time $\tp$, self-propulsion speed $\lp/\tp$, and thermal diffusivity $D=\ld^2/\tp$. The internal state $(\pm)$ represents the particle's preferred direction of motion, towards right and left, respectively. The ratio of the persistent length $\lp$ and the diffusive length $\ld$, known as the P\'eclet number $\mathrm{Pe}=\lp/\ld$, is a standard measure of activity. The parameter $a$ denotes lattice spacing, which we set to unity throughout this article.

Multilane lattice models are natural descriptions of protein transport along microtubule~\cite{2019_Ferro_Kinesin} and for breaking integrability in quantum transport~\cite{2024_Wienand_Emergence}. The quasi-one-dimensional geometry with the lane crossing effectively describes motion inside narrow channels, where particle size is comparable to the diameter of the channel, allowing occasional crossing~\cite{2012_Siems_Non,2020_Miron_Phase,2018_Wilke_Two,2021_Nandi_Dynein}.

A microscopic configuration of the system at a given time $\tau$ is specified in terms of the binary occupation variables $n^{\pm}_{i,j}(\tau)$ for each species $\sigma=(\pm)$ respectively, where $n^{\sigma}_{i,j}(\tau)=1$ or $0$ depending on whether the site $(i,j)$ is occupied by the species $\sigma$ or not. A more convenient choice to describe the dynamics is in terms of the total occupation variable, $n_{i,j}=n^{+}_{i,j}+ n^{-}_{i,j}$ and the polarization variable, $M_{i,j}=n^{+}_{i,j}-n^{-}_{i,j}$. Due to the exclusion condition, $n_{i,j}=\{0,1\}$ depending on the occupancy of the site, while $M_{i,j}=\{0,\pm 1\}$ depending on the species of the occupant. 

\section{A hydrodynamic description}
\label{s:hydro}

We now present a hydrodynamic description by coarse-graining these variables such that fast fluctuations are locally equilibrated over a hydrodynamic scale and slow modes smoothly evolve. For our dynamics the relevant length and time scales are $\ld$ and $\tp$. The hydrodynamics is defined in the rescaled coordinates $(x,t)\equiv(i/\ld,\tau/\tp)$, by taking $\ld, \lp $ and $\tp$ large, while keeping the diffusivity $D=\ld^2/\tp$ and the P\'eclet number $\mathrm{Pe}=\lp/\ld$ finite. For finite $\tau_\times\ll \tp$, the adjacent sites of the two lanes effectively equilibrate at the hydrodynamic scale, implying their evolution is described by the same coarse-grained variables $n_{i,j}(\tau)\simeq \rho(x,t)$ and $M_{i,j}(\tau)\simeq m(x,t)$, independent of the lane index $j$. 

\begin{figure}
\includegraphics[width=0.99\linewidth]{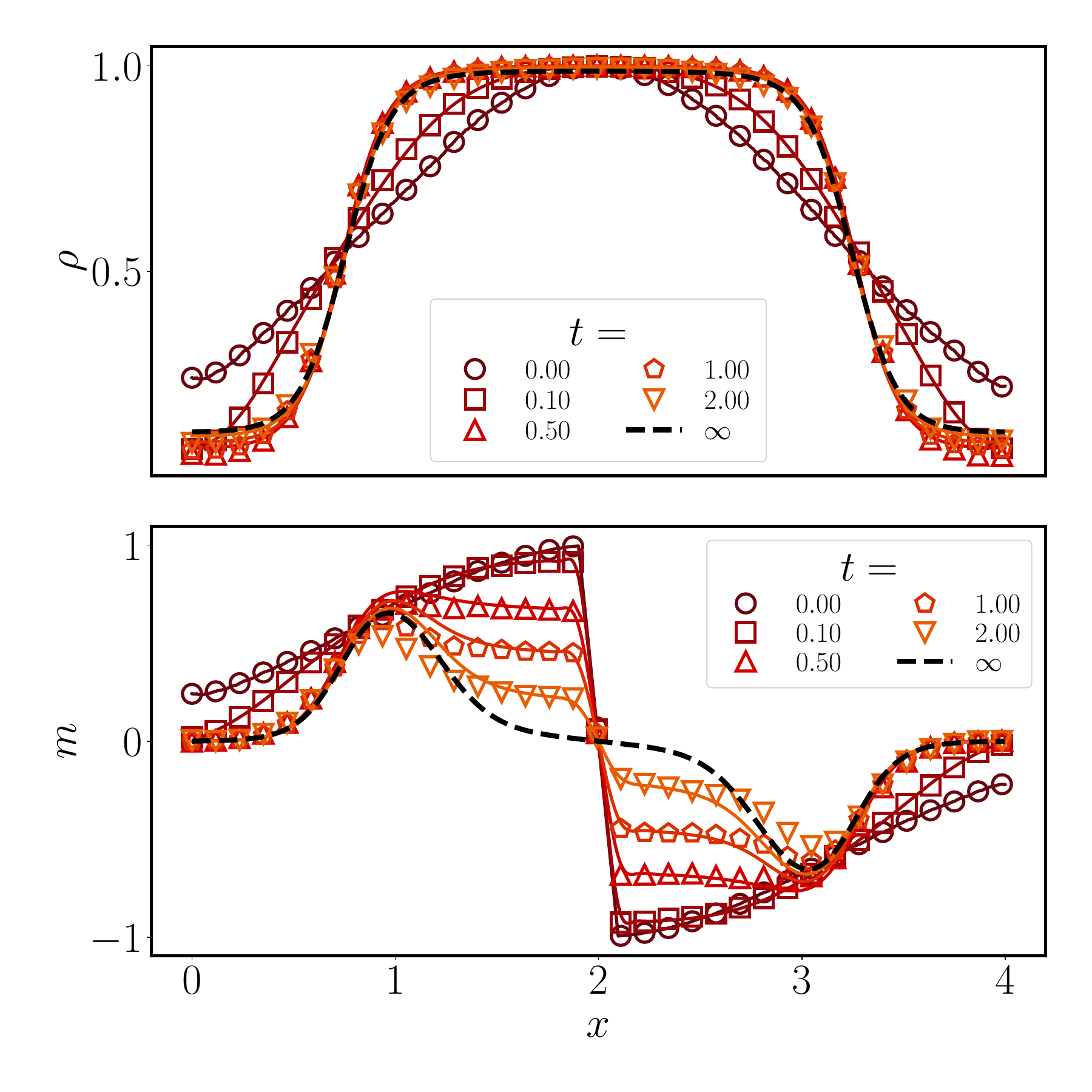}
\caption{\textbf{Test of hydrodynamics}: Time evolution of the hydrodynamic density $\rho(x,t)$ and polarization $m(x,t)$, starting from a Gaussian profile $\rho(x,0)$ centered around the middle of the system, with $(+)$ particles placed at the left half and $(-)$ particles at the right half. The solid lines represent noiseless hydrodynamic evolution~\eqref{fluctuate_hydro_quasi_1d}, while the points represent Monte Carlo simulations of the microscopic dynamics in (\ref{diff_drift})--(\ref{lane_cross}). The dashed lines indicate the steady-state profile from hydrodynamics. The plots are for bulk density $\rho_0=0.66$ and microscopic parameter values $\ld=1024$, $\tp= 2.5 \ld^2$, $\lp=10 \ld$, and $\tau_\times=0.1$, and system size $L=4\ld$, which corresponds to $\mathrm{Pe}=10$.}
\label{fig:timeEvolution}
\end{figure}

We show in Appendix~\ref{sect:action_fh_derive} that the evolution of $\rho(x,t)$ and $m(x,t)$ follows the coupled fluctuating hydrodynamics
\begin{subequations}
\label{fluctuate_hydro_quasi_1d}
\begin{align}
\partial_t\rho&=\partial_x^2\rho-\mathrm{Pe}\,\partial_x\big[m\,(1-\rho)\big]+\ld^{-1/2}\,\partial_x\eta_\rho,\\
\partial_tm&=(1-\rho)\,\partial_x^2m+m\,\partial_x^2\rho-\mathrm{Pe}\,\partial_x\big[\rho\,(1-\rho)\big]-2m\nonumber\\
&\qquad+\ld^{-1/2}\,\big(\partial_x\eta_m+2\,\eta_f\big),
\end{align}
\end{subequations}
where $\eta_\rho$, $\eta_m$ and $\eta_f$ are Gaussian noises each of zero mean with covariance $\left\langle\eta_{p}(x,t)\eta_q(x',t')\right\rangle=S_{p,q}\delta(x-x')\delta(t-t')$, with the mobility matrix $S_{\rho,\rho}=S_{m,m}=2\rho(1-\rho)$, $S_{\rho,m}=S_{m,\rho}=2m(1-\rho)$, $S_{f,f}=\rho$, and zero for the rest. The nonconservative noise $\eta_f$ is due to tumbling events in the dynamics. The same hydrodynamics extends for multilane generalizations with the number of lanes $\ll\ld$. The effective particle exchange rate along the quasi-one dimension is much reduced compared to the exchange rate in~\cite{2018_Houssene_Exact}, resulting in a different hydrodynamics.

Our detailed derivation for~\eqref{fluctuate_hydro_quasi_1d}, presented in Appendix~\ref{sect:action_fh_derive}, is based on an evaluation of the Martin-Siggia-Rose-Janssen-de Dominicis action~\cite{1973_Martin_Statistical,1976_Janssen_On,1978_Dominicis_Dynamics,1978_Dominicis_Field} of the stochastic differential equation~\eqref{fluctuate_hydro_quasi_1d} following a coarse-graining of the microscopic generator~\cite{2024_Saha_Large}. This method is generalizable for variations of the dynamics and dimensionality. It is fundamentally different from a derivation~\cite{2021_Agranov_Exact} for a related model~\cite{2018_Houssene_Exact}, which rests on mapping to stochastic diffusive models whose hydrodynamics is well tested. Such mapping relies on a very specific choice of the parameters, and not extendable for our model. Compared to the hydrodynamics in~\cite{2018_Houssene_Exact,2021_Agranov_Exact}, the diffusion matrix and the mobility matrix are significantly different.

\begin{figure}
\includegraphics[width=0.99\linewidth]{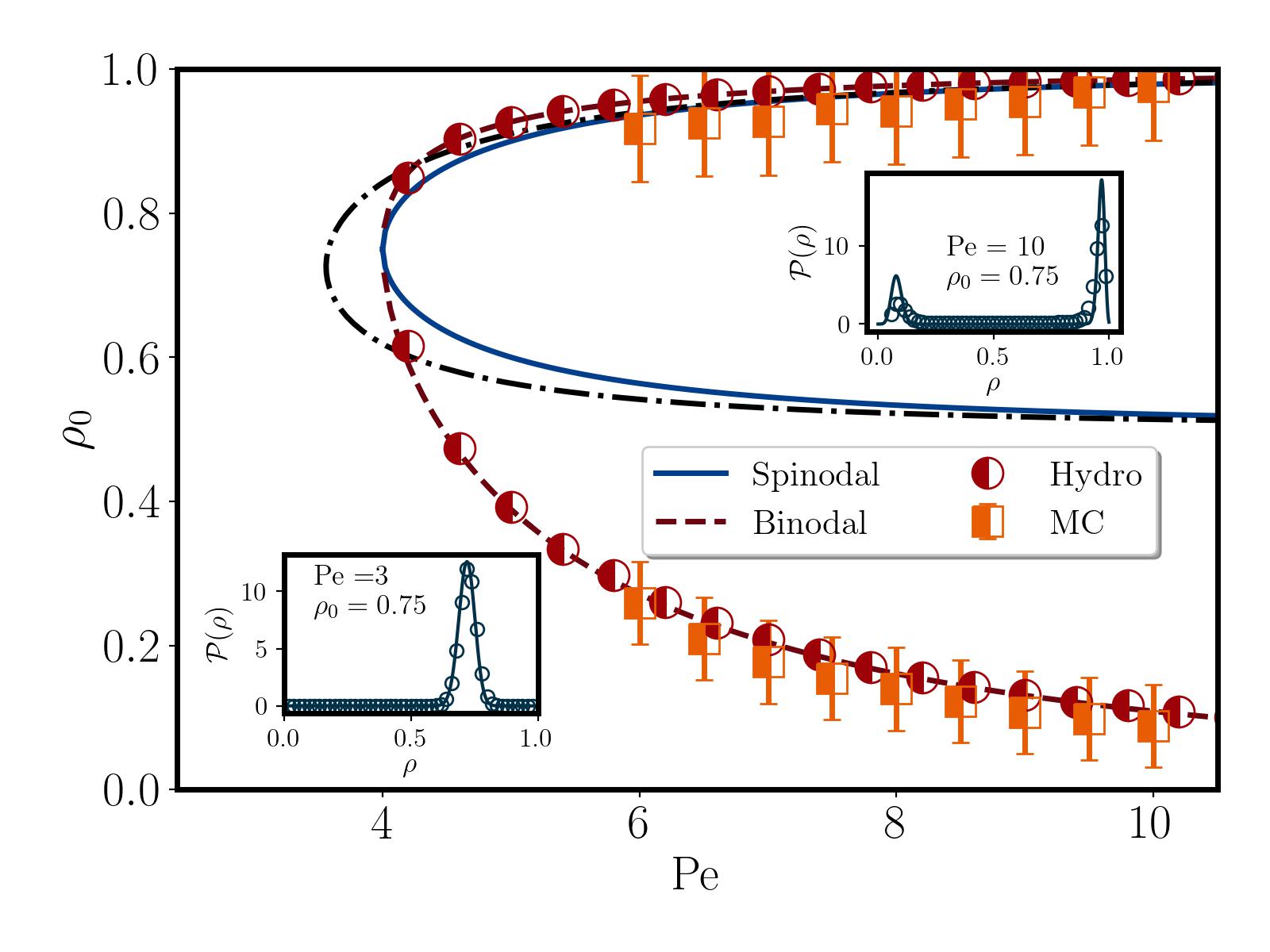}
\caption{\textbf{Phase diagram}: Spinodal and binodal, given by~\eqref{rho_lh} and~\eqref{binodal-condition} respectively, plotted with the solid and dashed lines. Along the dash-dotted line, the prefactor of the exponential in~\eqref{eq:correlation} diverges. The points represent the ``liquid'' and ``gas'' densities obtained from Monte Carlo simulations and numerical iteration of hydrodynamics. The error bars indicate the uncertainty due to the width of the density distribution around the peaks shown in the insets. The bimodal density distribution inside the spinodal region is an indication of MIPS, whereas the unimodal density distribution outside the binodal region is an indication of the homogeneous phase. In the inset, the points are from Monte Carlo simulations and the solid lines are from an analytical calculation in Appendix~\ref{numerical_evidence_loc_eq}. We use $L=4096$ in Monte Carlo simulations and hydrodynamics equation solved numerically using Fourier spectral method with $1024$ modes.}
\label{fig:phase_diag}
\end{figure}

In Fig.~\ref{fig:timeEvolution}, we compare the noiseless limit of hydrodynamics~\eqref{fluctuate_hydro_quasi_1d} with direct Monte Carlo simulation of the microscopic dynamics (\ref{diff_drift})--(\ref{lane_cross}). In the initial state, the left half of the system is occupied with $(+)$ particles and the right half with $(-)$ particles, each drawn from a Bernoulli distribution with a site-dependent average density $\langle n_i\rangle=\rho(i/\ld,0)$, shown in Fig.~\ref{fig:timeEvolution}. Hydrodynamic density and polarization profiles are obtained by coarse-graining over boxes containing $200$ sites. Only profiles of lane-one, averaged over $64$ ensembles, are shown in Fig.~\ref{fig:timeEvolution}. For iterating the hydrodynamics we numerically integrate~\eqref{fluctuate_hydro_quasi_1d} with zero noise using a Fourier-pseudospectral method~\cite{trefethen2000spectral}. 
An excellent agreement seen in the evolution of the profiles, asymptotically converging towards an inhomogeneous state, indicates a phase separation between high density and low-density regions.

\section{Spinodal and binodal analysis}
\label{s:spin-bin}

It is readily seen from~\eqref{fluctuate_hydro_quasi_1d} that uniform density $\rho(x)=\rho_0$ and $m(x)=0$ is a stationary solution of the noiseless hydrodynamics. An inhomogeneous phase-separated state, widely known as the MIPS for active matter systems, is an indication of the instability of this globally homogeneous solution. The spinodal curve on the $(\mathrm{Pe},\rho_0)$ plane describes the onset of instability against linear perturbations. A standard linear stability analysis of the noiseless hydrodynamics~\eqref{fluctuate_hydro_quasi_1d} (shown in Appendix~\ref{spinodal_calc}) predicts $\mathrm{Pe}^2(1-\rho_0)(2\rho_0-1)>2$ as the unstable regimes. Thus, for $0<\mathrm{Pe} <4$, a homogeneous state with any density $0<\rho_0<1$ is always stable under linear perturbations. On the other hand, for $\mathrm{Pe}>4$, the homogeneous state is unstable if 
\begin{equation}
\frac{3}{4}-\frac{1}{4}\sqrt{1-\left(\frac{4}{\mathrm{Pe}}\right)^2}< \rho_0<\frac{3}{4}+\frac{1}{4}\sqrt{1-\left(\frac{4}{\mathrm{Pe}}\right)^2}.
\label{rho_lh}
\end{equation}
The spinodal curve is shown in Fig.~\ref{fig:phase_diag}. Indeed, we see from Fig.~\ref{fig:timeEvolution} that if the parameters $(\mathrm{Pe},\rho_0)$ are chosen from within the spinodal region, the density profile $\rho(x,t)$ evolves to an inhomogeneous state. 
The spinodal region given by Eq.~\eqref{rho_lh} incidentally coincides with that for the model in~\cite{2018_Houssene_Exact}.

The binodal curve gives the two densities in a phase-separated state where a ``liquidlike'' high-density $\rho_l$ phase coexists with a ``gaslike'' low-density $\rho_g$ phase. Following~\cite{2018_Solon_Generalized_PRE,2018_Solon_Generalized_NJP}, we find that for our noiseless hydrodynamics~\eqref{fluctuate_hydro_quasi_1d} (details in Appendix~\ref{binodal_calc}), the two coexistence densities are determined by a pair of relations
\begin{equation}
g_0(\rho_g)=g_0(\rho_l) \quad\text{and}\quad
h_0(\rho_g)=h_0(\rho_l), 
\label{binodal-condition}
\end{equation} 
where $g_0(\rho)=\mathrm{Pe}\,\rho\,(1-\rho)-(2/\mathrm{Pe})\ln{(1-\rho)}$ and $h_0(\rho)=[2/(9\,\mathrm{Pe})]\,(1-\rho)^{-3}+(\mathrm{Pe}/6)\,(3-4\rho)\,(1-\rho)^{-2}$. We numerically solve the above pair of relations~\eqref{binodal-condition} and obtain the binodal curve, which we plot in Fig.~\ref{fig:phase_diag} along with the spinodal. Evidently, the volume fraction $v$ of the ``liquid'' region(s) is given by $\rho_lv+\rho_g(1-v)=\rho_0$, i.e., $v=(\rho_0-\rho_g)/(\rho_l-\rho_g)$. Therefore, for $\mathrm{Pe}>4$, for any global density $\rho_g<\rho_0<\rho_l$, it is possible to sustain a nonzero fraction $0<v<1$ of a liquid region within the analysis of the noiseless hydrodynamics. In particular, the parameters between the spinodal and the binodal curves correspond to a metastable region where, depending on the initial condition, the noiseless hydrodynamics~\eqref{fluctuate_hydro_quasi_1d} leads to either a homogeneous state or a phase-separated state (see Fig.~\ref{fig:metastability}). 

We also numerically iterate the noiseless hydrodynamics~\eqref{fluctuate_hydro_quasi_1d} for a long time so that it reaches a stationary state, and determine the steady-state ``liquid'' and ``gas'' densities, by choosing the minimum and maximum densities in the density profile, respectively. However, in Monte Carlo simulations, identifying the lowest and highest densities is less straightforward due to the fluctuating nature of the density profile. To address this, we extract density distributions and average them over steady-state profiles (depicted in the inset). The resulting bimodal density distribution indicates the presence of two distinct densities, characteristic of MIPS. 

\begin{figure}
\centering
\includegraphics[width=0.99\linewidth]{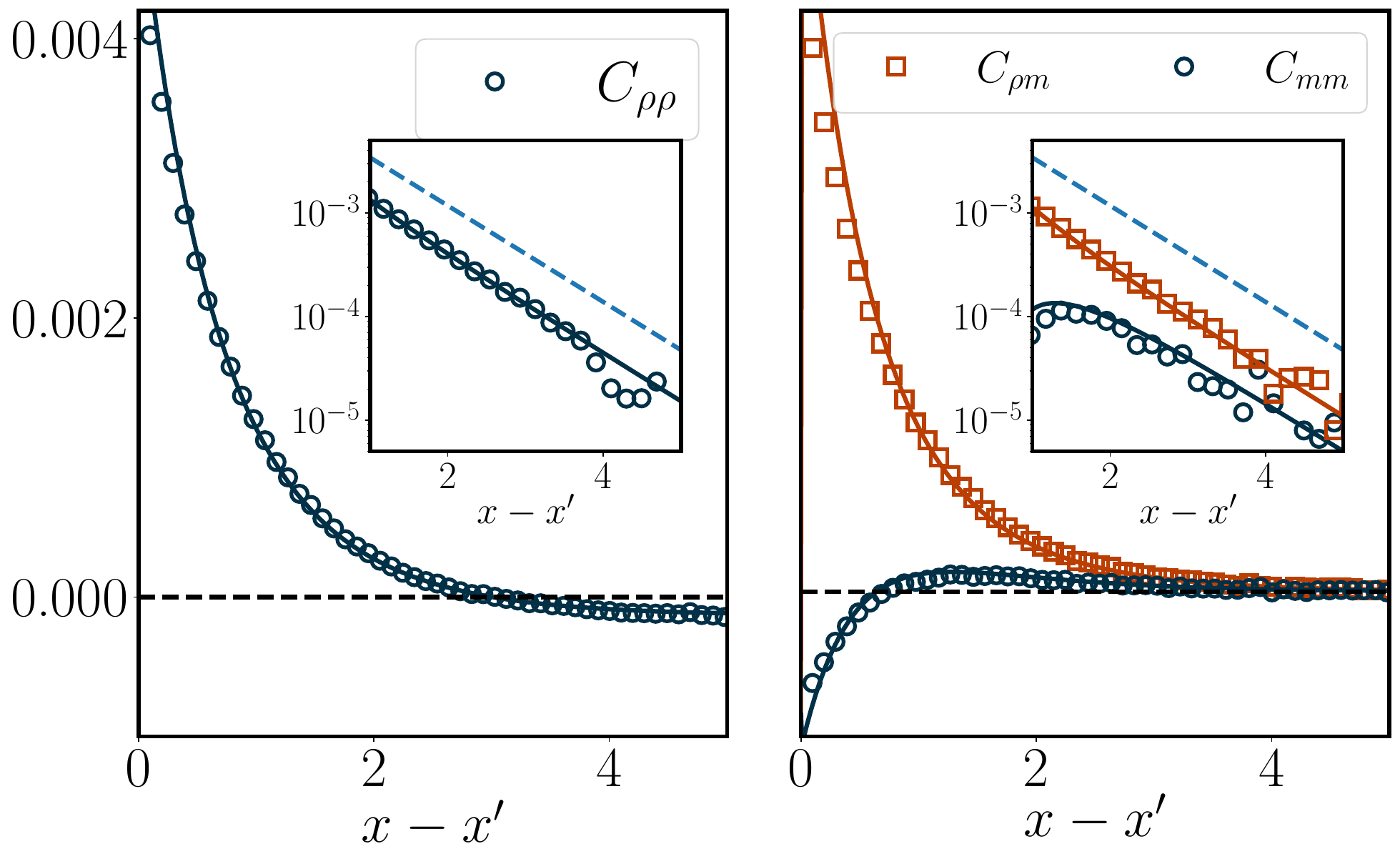}
\caption{\textbf{Correlations}: Solid lines plot the analytical results of the correlation functions whereas the points are from numerical simulations. The insets show the asymptotic decay of the correlation functions (with a constant shift in order to show up in the log-linear plot), 
with the dashed guiding lines plot~\eqref{eq:correlation} with a shift for better visibility. The parameters used in the simulations are $\ld=10.24$, $\tp=2\ld^2$, $\lp=2\ld$, $\tau_\times = 0.1$, $L=100\ld$ and $\rho_0=0.25$, and we average over $10^5$ statistically independent time steps to generate a clean curve for comparing with hydrodynamic predictions.}
\label{fig:correlations}
\end{figure}

\section{Two-point correlations}
\label{s:corr}

We now compute the correlation functions in the stationary-state homogeneous phase, using the fluctuating hydrodynamics of~\eqref{fluctuate_hydro_quasi_1d}. For example, 
the density-density correlation for far-separated points in the stationary-state homogeneous phase has the asymptote (see Appendix~\ref{two_pt_corr})
\begin{equation}
C_{\rho\rho}(x,x')\simeq\frac{A}{B \ld}\exp{\bigg(-\frac{|x-x'|}{\xi_1}\bigg)}, \label{eq:correlation}
\end{equation}
where $A=\mathrm{Pe}^2\rho_0^2\,(1-\rho_0)^2/\sqrt{1-{\rho_0}/{2}}$, $B= 2-\mathrm{Pe}^2 (1-\rho_0)(2-\rho_0)(2\rho_0-1)$, and the correlation length $\xi_1=\sqrt{1-{\rho_0}/{2}}$. The line along which the denominator $B$ vanishes, indicating breakdown of linearized hydrodynamics, is shown in Fig.~\ref{fig:phase_diag}. Its full form and the other correlation functions are given in Appendix~\ref{two_pt_corr}.
We compare our analytical correlation predictions with Monte Carlo simulations in Fig.~\ref{fig:correlations} and find excellent agreements. 
An indirect verification of the multiplicative noise terms in~\eqref{fluctuate_hydro_quasi_1d} comes from this comparison of correlations with Monte Carlo results.

\section{Generalization to $d$ dimensions} 
\label{s:dd}

There are several ways one can generalize our quasi-1D model to higher dimensions. The straightforward generalization takes the number of lanes to be $O(\ld)$ with $\tau_\times=\ld^2/\tp$. In this case, the noiseless hydrodynamic description along the \emph{active} direction remains the same as in the quasi-1D case (see Appendix~\ref{d_dimensions_general}). On the other hand, in an isotropic generalization, we define the $d$-dimensional version of the model on a periodic hypercubic lattice of $L^d$ sites and average density $\rho_0\in[0,1]$. The lattice sites indexed as $(i_1,i_2,\ldots,i_d)$, where $i_{k}=1,\,2,\,\ldots,\,L$ with $k=1,2,\ldots,d$. The system is composed of $2d$ distinct species denoted by $\sigma=1,2,\ldots,2d$ corresponding to the $2d$ possible directions of self-propulsion denoted by $\hat{r}_\sigma=\pm\hat{x}_1\,,\,\pm\hat{x}_2,\ldots,\pm\hat{x}_{d} $. In this model, a particle of $\sigma$ species hops at rate $\tp^{-1}(\ld^2+\lp/2)$ in the $\hat{r}_\sigma$ direction and at rate $\tp^{-1}(\ld^2-\lp/2)$ in the other $2d-1$ directions, provided that the target site is empty. In addition, a particle changes its species, i.e. tumbles, at rate $\tp^{-1}$.
The fluctuating hydrodynamics for this microscopic dynamics turns out to be (detailed derivation in Appendix~\ref{d_dimensions_general})
\begin{align}
&\partial_t\rho_\sigma=\,(1-\rho)\,\nabla^2\rho_\sigma+\rho_\sigma\,\nabla^2\rho-\mathrm{Pe}\,\vec{\nabla}[\rho_\sigma\,(1-\rho)]\cdot\hat{r}_\sigma\nonumber\\
&\quad-\sum_{\sigma'\neq\sigma}(\rho_\sigma-\rho_{\sigma'})+{\ld^{-d/2}}\,\left[\vec{\nabla}\cdot\vec{\eta}_\sigma+\sum_{\sigma'\neq\sigma}\eta_{\sigma\to\sigma'}\right]. \label{eq:hydrodynamics-d-dimensions}
\end{align}
Here, the conservative zero-mean Gaussian noise, $\vec{\eta}_\sigma=\bigl(\eta_\sigma^{(1)},\eta_\sigma^{(2)},\ldots,\eta_\sigma^{(d)}\bigr)$ arising from the biased-hopping dynamics, has the covariance $\bigl\langle\eta_\sigma^{(k)}(\vec{x},t)\,\eta_{\sigma'}^{(k')}(\vec{x}',t')\bigr\rangle=2\rho_\sigma(1-\rho)\,\delta_{\sigma,\sigma'}\,\delta_{k,k'}\,\delta(\vec{x}-\vec{x}')\,\delta(t-t')$ while the nonconservative zero-mean Gaussian noise, 
$\eta_{\sigma\to\sigma'}=-\eta_{\sigma'\to\sigma}$, arising from the tumbling dynamics, has the covariance $\bigl\langle\eta_{\sigma\to\sigma'}(\vec{x},t)\,\eta_{\psi\to\psi'}(\vec{x}',t')\bigr\rangle=(\rho_\sigma+\rho_{\sigma'})\,\delta_{\sigma,\psi}\delta_{\sigma',\psi'}\,\delta(\vec{x}-\vec{x}')\,\delta(t-t')$.

\begin{figure}[t!]
\centering
\includegraphics[width=0.99\linewidth]{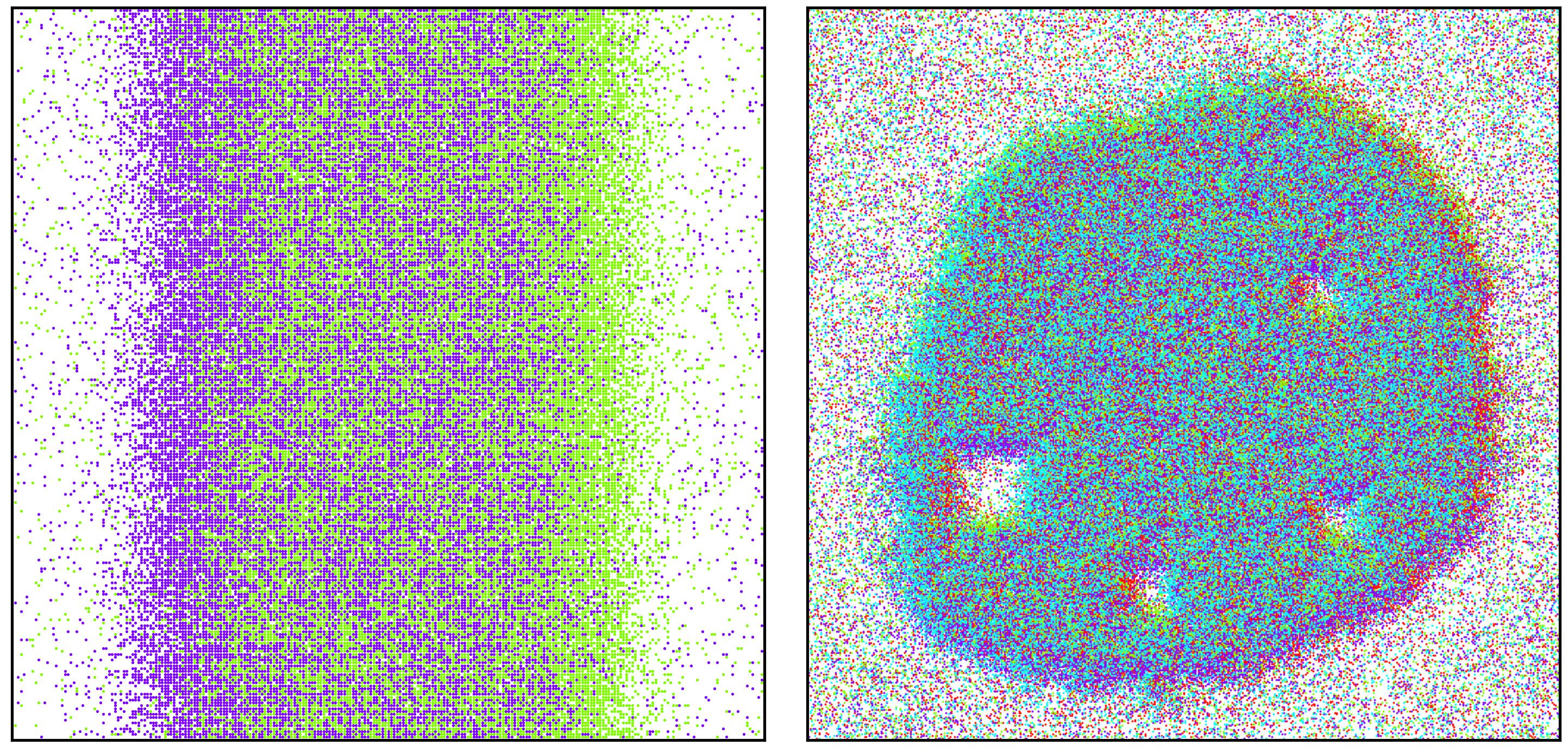}\caption{\textbf{MIPS}: Monte Carlo simulation for the two-dimensional models on a $512\times 512$ square lattice with $\rho_0=0.62$. Left figure corresponds to the two-species ($\rightarrow$ and $\leftarrow$) generalization and right to the four-species ($\rightarrow$, $\uparrow$, $\leftarrow$ and $\downarrow$) generalization. The colors indicate species (magenta$\equiv\rightarrow$, cyan$\equiv\uparrow$, green$\equiv\leftarrow$, red$\equiv\downarrow$). For the two-species model ${\ld=128}$ and for the four-species model ${\ld=32}$, with $\mathrm{Pe}=12$ and $\tp=\ld^2$ in both cases. The holes in the high-density phase were seen earlier in active systems~\cite{2018_Tjhung_Cluster,2023_Nejad_Spontaneous}. For the full time evolution see Appendix~\ref{supplement_vid} and~\cite{S_M}.}
\label{fig:2d mips}
\end{figure}

Since the analysis of the above hydrodynamics equation~\eqref{eq:hydrodynamics-d-dimensions} is similar to the quasi-1D case, we do not repeat it here. Instead, we perform Monte Carlo simulations for the two-dimensional microscopic model ($d=2$) with both generalizations. 
We observe MIPS in our simulations for some parameters, as shown in Fig.~\ref{fig:2d mips}, indicating that MIPS can emerge in two-dimensional systems, even without the exchange dynamics of~\cite{2018_Houssene_Exact}. Therefore, in dimensions higher than one, neither achieving a strong mixing effect nor observing the MIPS requires an exchange dynamics between the particles making our choice of dynamics naturally conducive for the desired results.

\section{A strictly 1D model}
\label{s:1d}

For the version of our model on a periodic one-lane lattice with the dynamics (\ref{diff_drift}) and (\ref{tumble}), exclusion interaction introduces single-file constraint, which preserves positional order of the particles. Our hydrodynamics construction, relying on an assumption of local equilibrium measure, predicts the same fluctuating hydrodynamics~\eqref{fluctuate_hydro_quasi_1d} as in the quasi-1D case, which predicts the same phase boundary as in Fig.~\ref{fig:phase_diag}. However, our Monte Carlo simulation of this 1D model does not support MIPS~\cite{2023_Mukherjee_Nonexistence} for the similar parameter ranges as in Fig.~\ref{fig:phase_diag}; neither do the correlations match with their hydrodynamics prediction (see Fig.~\ref{fig:full_correlation-1d}). Evidently, the predicted hydrodynamics does not work for this strictly 1D geometry, resonating similar findings~\cite{2021_Blondel_Stefan,2023_Touzo_Interacting}. This is further confirmed by a direct comparison of the noiseless evolution~\eqref{fluctuate_hydro_quasi_1d} of average fields with Monte Carlo simulation (see Fig.~\ref{fig:1d_nomips}).

\begin{figure}[t!]
\includegraphics[width=0.99\linewidth]{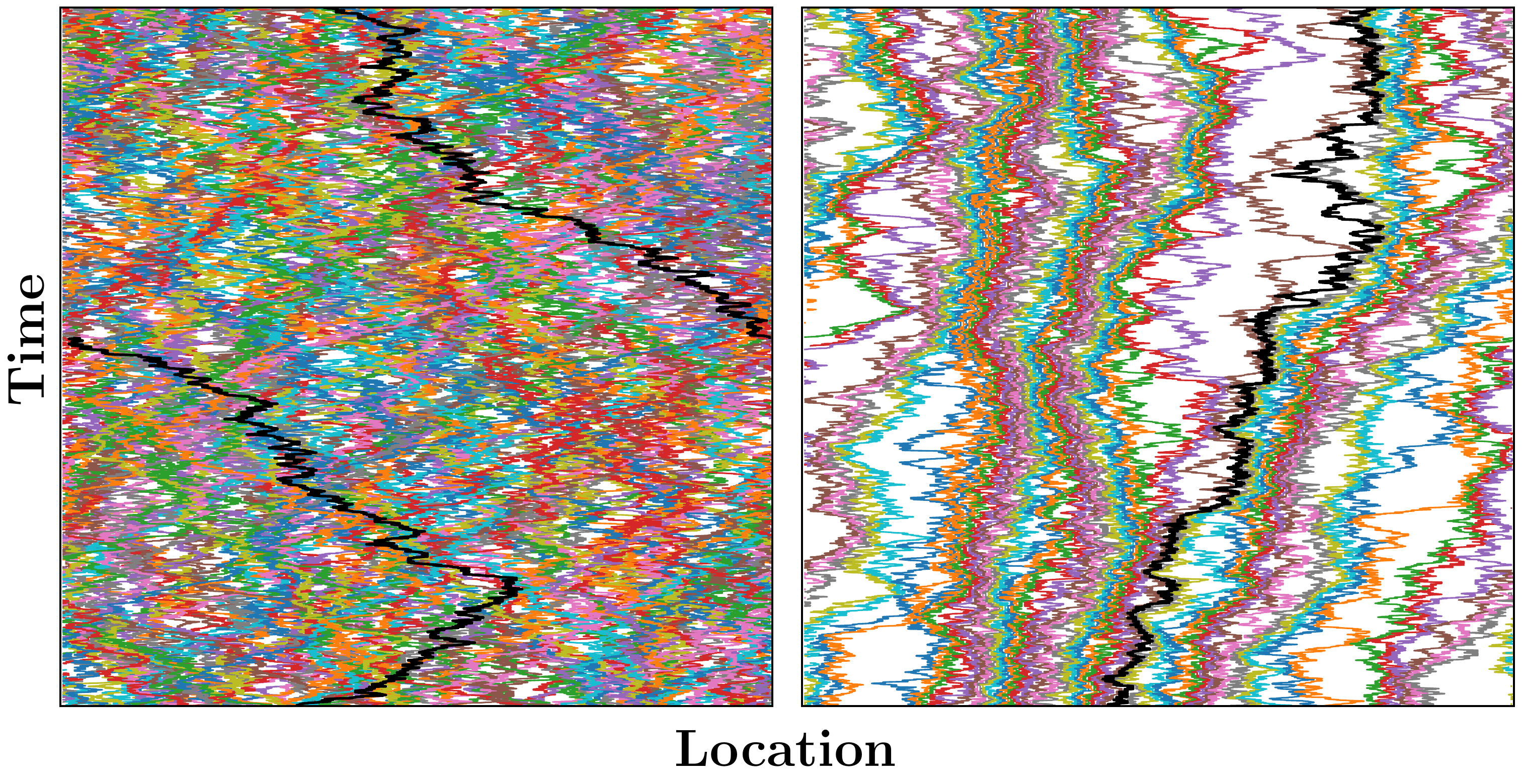}
\caption{\textbf{Kymographs}: Comparison of space-time trajectories of particles at long hydrodynamic times for two-lane (left) and one-lane (right) models. Black indicates a tracer. The figures correspond to parameter values ($\mathrm{Pe} = 6$, and $\rho_0=0.2$) outside the binodal curve in Fig.~\ref{fig:phase_diag}. Time evolution of the trajectories for different $\mathrm{Pe}$ are available in~\cite{S_M} [for more details see Appendix~\ref{supplement_vid}].}
\label{fig:kymograph}
\end{figure}

Hydrodynamics is by construction about effective dynamics of slow modes rising from local equilibrium of fast modes. Compared to higher dimensions, the single-file constraint in 1D preserves the local order of species, and therefore, local polarization at microscopic time scales $\ll \tp$. This fight between activity and local conservation, results in the formation of a large number of microclusters, that does not coarsen as seen in the kymographs of particles in Fig.~\ref{fig:kymograph}. Compared to the fluidlike evolution in the two-lane model in its homogeneous phase, the particle movements are severely restricted in the one-lane model at the same parameter values. This reduced effective diffusivity and drift of particles break our specific assumption about local equilibration (details in Appendix~\ref{numerical_evidence_loc_eq}), invalidating the hydrodynamics construction. 

\section{Conclusion}
\label{s:conclusion}

In the spirit of statistical mechanics, we have proposed a minimal model of interacting active particles where the simplest kind of steric interaction leads to macroscopic phase separation even in quasi-one dimension. It would be of interest to verify our prediction in experimental realizations of active transport inside narrow channels~\cite{2024_Das} or in optical lattices~\cite{2024_Wienand_Emergence}. The simplicity of our model allows us to theoretically analyze the macroscopic properties using a systematic bottom-up hydrodynamics theory. This hydrodynamic formulation potentially could serve as a minimal macroscopic description for MIPS, that is shared between different microscopic models, in the spirit of the Landau theory of phase transitions. There is a plethora of active dynamics~\cite{2018_Digregorio_Full,2022_Berthier_Disordered,2021_Klamser_Kinetic,2023_Mason_Exact} belonging to this class for which a detailed fluctuating hydrodynamics is not available. It would be interesting to see how the fluctuating hydrodynamics results for the present work compare with other models, including a full characterization of macroscopic fluctuations in terms of large deviations~\cite{2005_Bodineau_Distribution,2012_Bunin_Non,2016_Nyawo_Large,2018_Baek_Dynamical}. Our current work provides a crucial step in this promising direction.

\begin{acknowledgments}
We thank S Jose, J U Klamser, S Mukherjee, S Paul, K Ramola, S S Ray, and J C Sunil for stimulating discussions. AD and TS acknowledge financial support of the Department of Atomic Energy, Government of India, under Project Identification No. RTI 4001 and RTI 4002. AD acknowledges support from the J.C. Bose Fellowship (JCB/2022/000014) of the Science and Engineering Research Board of the Department of Science and Technology, Government of India. This research was supported in part by the International Centre for Theoretical Sciences (ICTS), participating in the program Statistical Physics of Complex Systems (code ICTS/SPCS2022/12) and 9th Indian Statistical Physics Community Meeting (code ICTS/ISPCM2024/4).
\end{acknowledgments}

\appendix

\begin{table*}[t]
\centering
\begin{tabular}{l r}
\hline
\hline
 & \\
$\quad\big(\mathcal{B}_{i,1}^+\,,\,\mathcal{B}_{i,1}^-\,,\,\mathcal{B}_{i,2}^+\,,\,\mathcal{B}_{i,2}^-\,,\,\mathcal{T}_{i,1}\,,\,\mathcal{T}_{i,2}\,,\,\mathcal{L}_i^+\,,\,\mathcal{L}_i^-\big)\quad$ & $\qquad\qquad\qquad\qquad\qquad$Rate$\qquad\qquad\qquad\qquad\qquad$\\[5.5pt]
\hline
 & \\
$(1\,,\,0\,,\,0\,,\,0\,,\,0\,,\,0\,,\,0\,,\,0)$ & $\tp^{-1}\,\big(\ld^2+\lp/2\big)\,n_{i,1}^+\,\big(1-n_{i+1,1}^+-n_{i+1,1}^-\big)$\\[11pt]
$(-1\,,\,0\,,\,0\,,\,0\,,\,0\,,\,0\,,\,0\,,\,0)$ & $\tp^{-1}\,\big(\ld^2-\lp/2\big)\,n_{i+1,1}^+\,\big(1-n_{i,1}^+-n_{i,1}^-\big)$\\[11pt]
$(0\,,\,1\,,\,0\,,\,0\,,\,0\,,\,0\,,\,0\,,\,0)$ & $\tp^{-1}\,\big(\ld^2-\lp/2\big)\,n_{i,1}^-\,\big(1-n_{i+1,1}^+-n_{i+1,1}^-\big)$\\[11pt]
$(0\,,\,-1\,,\,0\,,\,0\,,\,0\,,\,0\,,\,0\,,\,0)$ & $\tp^{-1}\,\big(\ld^2+\lp/2\big)\,n_{i+1,1}^-\,\big(1-n_{i,1}^+-n_{i,1}^-\big)$\\[11pt]
$(0\,,\,0\,,\,1\,,\,0\,,\,0\,,\,0\,,\,0\,,\,0)$ & $\tp^{-1}\,\big(\ld^2+\lp/2\big)\,n_{i,2}^+\,\big(1-n_{i+1,2}^+-n_{i+1,2}^-\big)$\\[11pt]
$(0\,,\,0\,,\,-1\,,\,0\,,\,0\,,\,0\,,\,0\,,\,0)$ & $\tp^{-1}\,\big(\ld^2-\lp/2\big)\,n_{i+1,2}^+\,\big(1-n_{i,2}^+-n_{i,2}^-\big)$\\[11pt]
$(0\,,\,0\,,\,0\,,\,1\,,\,0\,,\,0\,,\,0\,,\,0)$ & $\tp^{-1}\,\big(\ld^2-\lp/2\big)\,n_{i,2}^-\,\big(1-n_{i+1,2}^+-n_{i+1,2}^-\big)$\\[11pt]
$(0\,,\,0\,,\,0\,,\,-1\,,\,0\,,\,0\,,\,0\,,\,0)$ & $\tp^{-1}\,\big(\ld^2+\lp/2\big)\,n_{i+1,2}^-\,\big(1-n_{i,2}^+-n_{i,2}^-\big)$\\[11pt]
$(0\,,\,0\,,\,0\,,\,0\,,\,\pm 1\,,\,0\,,\,0\,,\,0)$ & $\tp^{-1}\,n_{i,1}^\pm$\\[11pt]
$(0\,,\,0\,,\,0\,,\,0\,,\,0\,,\,\pm 1\,,\,0\,,\,0)$ & $\tp^{-1}\,n_{i,2}^\pm$\\[11pt]
$(0\,,\,0\,,\,0\,,\,0\,,\,0\,,\,0\,,\,1\,,\,0)$ & $\tau_\times^{-1}\,n_{i,1}^+\,\big(1-n_{i,2}^+-n_{i,2}^-\big)$\\[11pt]
$(0\,,\,0\,,\,0\,,\,0\,,\,0\,,\,0\,,\,-1\,,\,0)$ & $\tau_\times^{-1}\,n_{i,2}^+\,\big(1-n_{i,1}^+-n_{i,1}^-\big)$\\[11pt]
$(0\,,\,0\,,\,0\,,\,0\,,\,0\,,\,0\,,\,0\,,\,1)$ & $\tau_\times^{-1}\,n_{i,1}^-\,\big(1-n_{i,2}^+-n_{i,2}^-\big)$\\[11pt]
$(0\,,\,0\,,\,0\,,\,0\,,\,0\,,\,0\,,\,0\,,\,-1)$ & $\tau_\times^{-1}\,n_{i,2}^-\,\big(1-n_{i,1}^+-n_{i,1}^-\big)$\\[5.5pt]
\hline
\hline
\end{tabular}
\caption{Possible values of the various dynamical events for the quantities in \eqref{eq:micro dynamics} and corresponding rates.}
\label{possible_current_and_rate}
\end{table*}

\section{{An action formulation for the quasi-1D model}}
 \label{sect:action_fh_derive}

Here, we present a derivation of the fluctuating hydrodynamic description for the quasi-1D model introduced in the main text, following a similar route that was used earlier in~\cite{2024_Saha_Large,2007_Lefevre_Dynamics} for the symmetric simple exclusion process (SSEP) and related stochastic lattice gases. The dynamics of our model is defined in items (\ref{diff_drift})--(\ref{lane_cross}) of the main text and schematically shown in Fig.~\ref{fig:quasi 1d model}.

The configuration of the system at any time $\tau$ is described by $\boldsymbol{n}(\tau)\equiv\big\{n_{i,j}^+(\tau),n_{i,j}^-(\tau)\big\}$ where the occupation variables take the values
\begin{equation}
\big(n^+_{i,j}(\tau),n^-_{i,j}(\tau)\big)=\big\{(1,0)\,,\,(0,1)\,,\,(0,0)\big\}
\end{equation}
depending on whether the $i$th site of the $j$ lane is occupied by a particle of species $(+)$ or $(-)$, or unoccupied, respectively. The simple exclusion interactions ensure that at any given time, no more than one particle of either species occupies a site. The system configuration evolves in time due to same-lane biased hopping, tumbling and lane crossing. The changes in occupation variables of a site relate to counters associated to the dynamical events in the infinitesimal time interval between $\tau$ and $\tau+d\tau$:
\begin{subequations}\label{eq:micro dynamics}
\begin{align}
n_{i,1}^\pm(\tau+\mathrm{d}\tau)-n_{i,1}^\pm(\tau)&=\mathcal{B}_{i-1,1}^\pm(\tau)-\mathcal{B}_{i,1}^\pm(\tau)\mp\mathcal{T}_{i,1}(\tau)\nonumber\\
&\;-\mathcal{L}_{i}^\pm(\tau),\\
n_{i,2}^\pm(\tau+\mathrm{d}\tau)-n_{i,2}^\pm(\tau)&=\mathcal{B}_{i-1,2}^\pm(\tau)-\mathcal{B}_{i,2}^\pm(\tau)\mp\mathcal{T}_{i,2}(\tau)\nonumber\\
&\;+\mathcal{L}_{i}^\pm(\tau),
\end{align}
\end{subequations}
where $\mathcal{B}_{i,j}^\sigma(\tau)$ is the net number of same-lane hopping of the $\sigma$-species particles between the $i{\text{th}}$ and $(i+1){\text{th}}$ sites of the $j{\text{th}}$ lane; $\mathcal{L}_i^\sigma$ is the net number of lane crossings of the $\sigma$-species particles from the $(i,1){\text{th}}$ site to the $(i,2){\text{th}}$ site; $\mathcal{T}_{i,j}(\tau)$ corresponds the net number of $(+)$-species to $(-)$-species tumblings at the $i{\text{th}}$ site of the $j$ lane. For an infinitesimal $\mathrm{d}\tau$, the various possible values that these variables can take are listed in Table~\ref{possible_current_and_rate}. The rates ensure the exclusion condition and preserve the total number of particles.

Our interest lies in determining the probability that the system starting from an initial configuration $\boldsymbol{n}(0)$ evolves into a final configuration $\boldsymbol{n}(T)$. This transition probability can be written as a path integral over all possible evolution of the trajectories $\{\boldsymbol{n}(\tau),\boldsymbol{\mathcal{B}}(\tau),\boldsymbol{\mathcal{T}}(\tau),\boldsymbol{\mathcal{L}}(\tau)\}$ as
\begin{widetext}
\begin{align}
\mathrm{Pr}\left[\boldsymbol{n}(T)\,|\,\boldsymbol{n}(0)\right]=\int_{\boldsymbol{n}(0)}^{\boldsymbol{n}(T)}\left[\mathcal{D}\boldsymbol{n}\right]\Bigg<\prod_{\tau=0}^{T-\mathrm{d}\tau}\prod_{i=1}^L\Big(&\delta_{n_{i,1}^\pm(\tau+\mathrm{d}\tau)-n_{i,1}^\pm(\tau),\mathcal{B}_{i-1,1}^\pm(\tau)-\mathcal{B}_{i,1}^\pm(\tau)\mp\mathcal{T}_{i,1}(\tau)-\mathcal{L}_{i}^\pm(\tau)}\nonumber\\
&\times\delta_{n_{i,2}^\pm(\tau+\mathrm{d}\tau)-n_{i,2}^\pm(\tau),\mathcal{B}_{i-1,2}^\pm(\tau)-\mathcal{B}_{i,2}^\pm(\tau)\mp\mathcal{T}_{i,2}(\tau)+\mathcal{L}_{i}^\pm(\tau)}\Big)\Bigg>_{\left\{\boldsymbol{\mathcal{B}}(\tau),\boldsymbol{\mathcal{T}}(\tau),\boldsymbol{\mathcal{L}}(\tau)\right\}},
\end{align}
\end{widetext}
where the Kronecker-delta function $\delta_{a,b}$ incorporates the particle conservation in~\eqref{eq:micro dynamics}, and the angular brackets denote the average over histories of $\{\boldsymbol{\mathcal{B}}(\tau),\boldsymbol{\mathcal{T}}(\tau),\boldsymbol{\mathcal{L}}(\tau)\}$. The path integral measure on the occupation variables is defined as
\begin{equation}
\int\left[\mathcal{D}\boldsymbol{n}\right]:=\prod_{\tau=0}^{T}\prod_{i=1}^L\prod_{j=1}^2\prod_{\sigma=(\pm)}\sum_{n_{i,j}^\sigma(\tau)=0}^1.
\end{equation}

We use an integral representation of the Kronecker-delta function $\delta_{a,b}=(2\pi\mathrm{i})^{-1}\int_{-\mathrm{i}\pi}^{\mathrm{i}\pi}\mathrm{d}\widehat{n}\,\mathrm{e}^{-\widehat{n}(a-b)}$ and introduce a response variable $\widehat{n}_{i,j}^\sigma(\tau)$ at every site for both species. Subsequently, performing the averages over the variables $\mathcal{B}_{i,j}^\sigma$, $\mathcal{T}_{i,j}$ and $\mathcal{L}^\sigma$ using Table \ref{possible_current_and_rate}, we write the transition probability in an action formulation,
\begin{equation}
\mathrm{Pr}\left[\boldsymbol{n}(T)\,|\,\boldsymbol{n}(0)\right]=\int_{\boldsymbol{n}(0)}^{\boldsymbol{n}(T)}\left[\mathcal{D}\boldsymbol{n}\right]\left[\mathcal{D}\boldsymbol{\widehat{n}}\right]\mathrm{e}^{\mathcal{K}\left[\boldsymbol{n},\boldsymbol{\widehat{n}}\right]+\mathcal{H}\left[\boldsymbol{n},\boldsymbol{\widehat{n}}\right]} \label{exact_micro_path_int}
\end{equation}
with the ``kinetic'' term
\begin{widetext}
\begin{align}
\mathcal{K}\big[\boldsymbol{n},\boldsymbol{\widehat{n}}\big]=\sum_{i=1}^L\sum_{j=1}^2\sum_{\sigma=(\pm)}\Bigg[n_{i,j}^\sigma(0)\,\widehat{n}_{i,j}^\sigma(0)-n_{i,j}^\sigma(T)\,\widehat{n}_{i,j}^\sigma(T)
\;+\int_0^T\mathrm{d}\tau\,\bigg(n_{i,j}^\sigma(\tau)\,\frac{\mathrm{d}\widehat{n}_{i,j}^\sigma(\tau)}{\mathrm{d}\tau}\bigg)\Bigg] \label{exact_micro_kinetic}
\end{align}
and the ``Hamiltonian'' term
\begin{align}
\mathcal{H}\big[\boldsymbol{n},\boldsymbol{\widehat{n}}\big]=&\,\frac{\ld^2}{\tp}\sum_{i=1}^L\sum_{j=1}^2\sum_{\sigma=(\pm)}\int_0^T\mathrm{d}\tau\,\Big[\big(\mathrm{e}^{\widehat{n}_{i+1,j}^\sigma-\widehat{n}_{i,j}^\sigma}-1\big)\,n_{i,j}^\sigma\,(1-n_{i+1,j})+\big(\mathrm{e}^{\widehat{n}_{i,j}^\sigma-\widehat{n}_{i+1,j}^\sigma}-1\big)\,n_{i+1,j}^\sigma\,(1-n_{i,j})\Big]\nonumber\\
&+\frac{\lp}{2\tp}\sum_{i=1}^L\sum_{j=1}^2\int_0^T\mathrm{d}\tau\,\Big[\big(\mathrm{e}^{\widehat{n}_{i+1,j}^+-\widehat{n}_{i,j}^+}-1\big)\,n_{i,j}^+\,(1-n_{i+1,j})-\big(\mathrm{e}^{\widehat{n}_{i,j}^+-\widehat{n}_{i+1,j}^+}-1\big)\,n_{i+1,j}^+\,(1-n_{i,j})\nonumber\\
&\qquad\qquad\qquad\qquad\qquad-\big(\mathrm{e}^{\widehat{n}_{i+1,j}^--\widehat{n}_{i,j}^-}-1\big)\,n_{i,j}^-\,(1-n_{i+1,j})+\big(\mathrm{e}^{\widehat{n}_{i,j}^--\widehat{n}_{i+1,j}^-}-1\big)\,n_{i+1,j}^-\,(1-n_{i,j})\Big]\nonumber\\
&+\frac{1}{\tp}\sum_{i=1}^L\sum_{j=1}^2\int_0^T\mathrm{d}\tau\,\Big[\big(\mathrm{e}^{\widehat{n}_{i,j}^--\widehat{n}_{i,j}^+}-1\big)\,n_{i,j}^++\big(\mathrm{e}^{\widehat{n}_{i,j}^+-\widehat{n}_{i,j}^-}-1\big)\,n_{i,j}^-\Big]\nonumber\\
&+\frac{1}{\tau_\times}\sum_{i=1}^L\sum_{\sigma=(\pm)}\int_0^T\mathrm{d}\tau\,\Big[\big(\mathrm{e}^{\widehat{n}_{i,2}^\sigma-\widehat{n}_{i,1}^\sigma}-1\big)\,n_{i,1}^\sigma\,(1-n_{i,2})+\big(\mathrm{e}^{\widehat{n}_{i,1}^\sigma-\widehat{n}_{i,2}^\sigma}-1\big)\,n_{i,2}^\sigma\,(1-n_{i,1})\Big], \label{exact_micro_Hamilton}
\end{align}
\end{widetext}
where $n_{i,j}(\tau)=\sum_\sigma n_{i,j}^\sigma(\tau)$ denotes the total occupation number of the $i{\text{th}}$ site of $j$ lane at time $\tau$. The path integral measure on the conjugate response variable is defined as
\begin{equation}
\int\left[\mathcal{D}\boldsymbol{\widehat{n}}\right]:=\prod_{\tau=0}^T\,\prod_{i=1}^L\prod_{j=1}^2\prod_{\sigma=(\pm)}\int_{-\mathrm{i}\pi}^{\mathrm{i}\pi}\frac{\mathrm{d}\widehat{n}_{i,j}^\sigma(\tau)}{2\pi\mathrm{i}}.
\end{equation}

\subsection{The hydrodynamic limit}
Given the diffusive nature of the dynamics for our quasi-1D model, the local statistics of occupation variables in a region of length scale $\ld$ reach a local equilibrium over a period of time scale $\tp=\ld^2/D$ with respect to a slowly varying density profile. Considering that, in these length and time scales, the diffusion dominates over drift or tumbling, this local equilibrium measure is best approximated by the product measure,
\begin{subnumcases}
{\label{local_equil_measure}\big(n_{i,j}^+(\tau),n_{i,j}^-(\tau)\big)=}
(1,0) & with prob. $\rho_i^+(\tau)$\\
(0,1) & with prob. $\rho_i^-(\tau)$\\
(0,0) & with prob. $1-\rho_i(\tau)$
\end{subnumcases}
similar to the SSEP on two lanes. Here, $\rho_i^\pm(\tau)=\rho_\pm(i/\ld,\tau/\tp)$ are slowly varying local average densities with $\rho_i(\tau)=\rho_i^+(\tau)+\rho_i^-(\tau)$.

An important assumption is that $\rho_i^\pm(\tau)$ in the two lanes are the same. This is justified considering that in the time scale $\tp\gg \tau_\times$, particles frequently switch lanes before diffusing on the lane. This results in fast equilibration of adjacent sites on the two lanes. 

To be consistent with this local equilibrium assumption, the response variables are also considered as slowly varying and the same on the two lanes. More precisely,
\begin{equation}
\widehat{n}_{i,j}^\pm(\tau)=\widehat{\rho}_\pm\Big(\frac{i}{\ld},\frac{\tau}{\tp}\Big) \equiv\widehat{\rho}_i^\pm(\tau).
\label{slow_response_field}
\end{equation}
In the hydrodynamic scale, defined by the rescaled coordinates $(x,t)\equiv(i/\ld,\tau/\tp)$, the hydrodynamics description of the model is in terms of the evolution of these slowly varying fields $\rho_\pm(x,t)$ and $\widehat{\rho}_\pm(x,t)$ gives. 

In obtaining the transition probability between the initial and final hydrodynamic densities, $\boldsymbol{\rho}(0)$ and $\boldsymbol{\rho}(T)$, from the action formulation in~(\ref{exact_micro_path_int})--(\ref{exact_micro_Hamilton}), the leading contribution comes from typical microscopic configurations associated to the hydrodynamic fields, such that
\begin{align}
\sum_{i} n_{i,j}^\sigma(0)\,\widehat{n}_{i,j}^\sigma(0)&\simeq\sum_{i}\rho_i^\sigma(0)\,\widehat{\rho}_i^\sigma(0)\text{, and}\nonumber\\
\sum_{i} n_{i,j}^\sigma(T)\,\widehat{n}_{i,j}^\sigma(T)&\simeq\sum_{i}\rho_i^\sigma(T)\,\widehat{\rho}_i^\sigma(T) \label{initial_final_slow_fields}
\end{align}
for large $\ld$ and $\tp$. Taking average of $\mathrm{Pr}\left[\boldsymbol{n}(T)\,|\,\boldsymbol{n}(0)\right]$ in~(\ref{exact_micro_path_int})--(\ref{exact_micro_Hamilton}) over occupation variables using the local equilibrium measure~\eqref{local_equil_measure} give the transition probability in hydrodynamic scale,
\begin{equation}
\mathrm{Pr}\big[\boldsymbol{\rho}(T)|\boldsymbol{\rho}(0)\big]=\int_{\boldsymbol{\rho}(0)}^{\boldsymbol{\rho}(T)}\left[\mathcal{D}\boldsymbol{\rho}\right]\left[\mathcal{D}\boldsymbol{\widehat{\rho}}\right]\Big<\mathrm{e}^{\mathcal{K}\left[\boldsymbol{n},\boldsymbol{\widehat{\rho}}\right]+\mathcal{H}\left[\boldsymbol{n},\boldsymbol{\widehat{\rho}}\right]}\Big>_{\left\{\boldsymbol{n}(\tau)\right\}}
\end{equation}
with $\mathcal{K}$ and $\mathcal{H}$ in~\eqref{exact_micro_kinetic} and~\eqref{exact_micro_Hamilton}, respectively.

In performing the average, in the large $\ld$ and $\tp$ limit, the local equilibrium probability measure at different times are assumed to be independent of one another, except correlations through their slowly varying density fields. This way, the leading-order behavior of the transition probability,
\begin{widetext}
\begin{align}
\mathrm{Pr}\big[\boldsymbol{\rho}(T)\,|\,\boldsymbol{\rho}(0)\big]\simeq\int_{\boldsymbol{\rho}(0)}^{\boldsymbol{\rho}(T)}\left[\mathcal{D}\boldsymbol{\rho}\right]\left[\mathcal{D}\boldsymbol{\widehat{\rho}}\right]\prod_{\tau=0}^T\Big<\mathrm{e}^{\mathrm{d}\tau\,\mathcal{S}(\tau)}\Big>_{\boldsymbol{n}(\tau)}
\,\mathrm{e}^{2\sum_{i=1}^L\sum_{\sigma=(\pm)}\left(\rho_i^\sigma(0)\,\widehat{\rho}_i^\sigma(0)-\rho_i^\sigma(T)\,\widehat{\rho}_i^\sigma(T)\right)} \label{trans_prob_before_avg}
\end{align}
with the action
\begin{subequations}
\begin{align}
\mathcal{S}(\tau)=&\,\sum_{i=1}^L\sum_{j=1}^2\Bigg\{\sum_{\sigma=(\pm)}\bigg(n_{i,j}^\sigma\,\frac{\mathrm{d}\widehat{\rho}_i^\sigma}{\mathrm{d}\tau}\bigg)+\frac{\ld^2}{\tp}\sum_{\sigma=(\pm)}\Big[\big(\mathrm{e}^{\widehat{\rho}_{i+1}^\sigma-\widehat{\rho}_i^\sigma}-1\big)\,n_{i,j}^\sigma\,(1-n_{i+1,j})+\big(\mathrm{e}^{\widehat{\rho}_i^\sigma-\widehat{\rho}_{i+1}^\sigma}-1\big)\,n_{i+1,j}^\sigma\,(1-n_{i,j})\Big]\nonumber\\
&+\frac{\lp}{2\tp}\Big[\big(\mathrm{e}^{\widehat{\rho}_{i+1}^+-\widehat{\rho}_i^+}-1\big)\,n_{i,j}^+\,(1-n_{i+1,j})-\big(\mathrm{e}^{\widehat{\rho}_i^+-\widehat{\rho}_{i+1}^+}-1\big)\,n_{i+1,j}^+\,(1-n_{i,j})-\big(\mathrm{e}^{\widehat{\rho}_{i+1}^--\widehat{\rho}_i^-}-1\big)\,n_{i,j}^-\,(1-n_{i+1,j})\nonumber\\
&+\big(\mathrm{e}^{\widehat{\rho}_i^--\widehat{\rho}_{i+1}^-}-1\big)\,n_{i+1,j}^-\,(1-n_{i,j})\Big]+\frac{1}{\tp}\Big[\big(\mathrm{e}^{\widehat{\rho}_i^--\widehat{\rho}_i^+}-1\big)\,n_{i,j}^++\big(\mathrm{e}^{\widehat{\rho}_i^+-\widehat{\rho}_i^-}-1\big)\,n_{i,j}^-\Big]\Bigg\}, \label{action_before_avg}
\end{align}
\end{subequations}
\end{widetext}
where we used~\eqref{initial_final_slow_fields}. Notably, the contribution in the action from lane-crossing (involving $\tau_\times$) vanishes due to the fast equilibration assumption between the two lanes, equating their local densities and conjugate fields in~\eqref{slow_response_field}.

Further simplification comes from the $\mathrm{d}\tau\to0$ limit, where
\begin{equation}
\prod_\tau\Big<\mathrm{e}^{\mathrm{d}\tau\,\mathcal{S}(\tau)}\Big>_{\boldsymbol{n}(\tau)}\simeq\prod_\tau\mathrm{e}^{\mathrm{d}\tau\langle\mathcal{S}(\tau)\rangle_{\boldsymbol{n}(\tau)}}.
\end{equation}
The averaging of the action essentially reduces to replacing $n_{i,1}^\pm(\tau)$ and $n_{i,2}^\pm(\tau)$ by $\rho_i^\pm(\tau)$ in~\eqref{action_before_avg} This leads to
\begin{align}
\mathrm{Pr}\big[\boldsymbol{\rho}(T)\,|\,\boldsymbol{\rho}(0)\big]=\int_{\boldsymbol{\rho}(0)}^{\boldsymbol{\rho}(T)}\left[\mathcal{D}\boldsymbol{\rho}\right]\left[\mathcal{D}\boldsymbol{\widehat{\rho}}\right]\mathrm{e}^{2\int_0^T\mathrm{d}\tau\sum_{i=1}^L\mathcal{S}_i},
\end{align}
where the action is
\begin{widetext}
\begin{align}
\mathcal{S}_i=&-\widehat{\rho}_i^+\,\frac{\mathrm{d}\rho_i^+}{\mathrm{d}\tau}-\widehat{\rho}_i^-\,\frac{\mathrm{d}\rho_i^-}{\mathrm{d}\tau}+\frac{\ld^2}{\tp}\Big[\big(\mathrm{e}^{\widehat{\rho}_{i+1}^+-\widehat{\rho}_i^+}-1\big)\,\rho_i^+\,(1-\rho_{i+1})+\big(\mathrm{e}^{\widehat{\rho}_{i+1}^--\widehat{\rho}_i^-}-1\big)\,\rho_i^-\,(1-\rho_{i+1})+\big(\mathrm{e}^{\widehat{\rho}_i^+-\widehat{\rho}_{i+1}^+}-1\big)\,\rho_{i+1}^+\,(1-\rho_i)\nonumber\\
&+\big(\mathrm{e}^{\widehat{\rho}_i^--\widehat{\rho}_{i+1}^-}-1\big)\,\rho_{i+1}^-\,(1-\rho_i)\Big]+\frac{\lp}{2\tp}\Big[\big(\mathrm{e}^{\widehat{\rho}_{i+1}^+-\widehat{\rho}_i^+}-1\big)\,\rho_i^+\,(1-\rho_{i+1})-\big(\mathrm{e}^{\widehat{\rho}_i^+-\widehat{\rho}_{i+1}^+}-1\big)\,\rho_{i+1}^+\,(1-\rho_i)\nonumber\\
&-\big(\mathrm{e}^{\widehat{\rho}_{i+1}^--\widehat{\rho}_i^-}-1\big)\,\rho_i^-\,(1-\rho_{i+1})+\big(\mathrm{e}^{\widehat{\rho}_i^--\widehat{\rho}_{i+1}^-}-1\big)\,\rho_{i+1}^-\,(1-\rho_i)\Big]+\frac{1}{\tp}\Big[\big(\mathrm{e}^{\widehat{\rho}_{i,j}^--\widehat{\rho}_{i,j}^+}-1\big)\,\rho_{i,j}^++\big(\mathrm{e}^{\widehat{\rho}_{i,j}^+-\widehat{\rho}_{i,j}^-}-1\big)\,\rho_{i,j}^-\Big]. \label{eq:Si}
\end{align}
\end{widetext}
In writing~\eqref{eq:Si}, we used an integration by parts in the $\tau$ variable to cancel the exponential term outside the averaging in~\eqref{trans_prob_before_avg}.

For writing the path integral in terms of the hydrodynamic fields, we use the definition of $\rho_\pm(x,t)$ and $\widehat{\rho}_\pm(x,t)$ in~(\ref{local_equil_measure}) and (\ref{slow_response_field}) and perform gradient expansions, for large $\ld$ and $\tp$. Keeping the leading-order terms
\begin{widetext}
\begin{equation}\label{eq:trsn hydro r rb}
\mathrm{Pr}\left[\rho_\pm(x,\mathbb{T})\,\big|\,\rho_\pm(x,0)\right]=\int_{\rho_\pm(x,0)}^{\rho_\pm(x,\mathbb{T})}\left[\mathcal{D}\rho_\pm\right]\left[\mathcal{D}\widehat{\rho}_{\pm}\right]\mathrm{e}^{-2\ld\int_0^\mathbb{T}\mathrm{d}t\,\left[\int_0^\mathbb{L}\mathrm{d}x\,\big(\widehat{\rho}_+\,\partial_t\rho_++\widehat{\rho}_-\,\partial_t\rho_-\big)-H\left[\widehat{\rho}_\pm\,,\,\rho_\pm\right]\right]}
\end{equation}
\end{widetext}
with the effective Hamiltonian
\begin{align}
H=&\int_0^\mathbb{L}\mathrm{d}x\,\Big\{\rho_+\,(1-\rho)\,(\partial_x\widehat{\rho}_+)^2+\rho_-\,(1-\rho)\,(\partial_x\widehat{\rho}_-)^2\nonumber\\
&-\big[(1-\rho_-)\,\partial_x\rho_++\rho_+\,\partial_x\rho_--\mathrm{Pe}\,\rho_+\,(1-\rho)\big]\,\partial_x\widehat{\rho}_+\nonumber\\
&-\big[(1-\rho_+)\,\partial_x\rho_-+\rho_-\,\partial_x\rho_++\mathrm{Pe}\,\rho_-\,(1-\rho)\big]\,\partial_x\widehat{\rho}_-\nonumber\\
&+\big(\mathrm{e}^{-\widehat{\rho}_++\widehat{\rho}_-}-1\big)\,\rho_++\big(\mathrm{e}^{\widehat{\rho}_+-\widehat{\rho}_-}-1\big)\,\rho_-\Big\}
\end{align}
where we denote $T/\tp=\mathbb{T}$ and $L/\ld=\mathbb{L}$. Note, that in our choice of rates, the thermal diffusivity, $D=\ld^2/\tp$ and P\'eclet number, $\mathrm{Pe}=\lp/\ld$ are kept finite.

\subsubsection{\bfseries Action in terms of density and polarization fields} 

The action in~\eqref{eq:trsn hydro r rb} describes the stochastic hydrodynamics of the model. More convenient hydrodynamic fields for active matter are the total density ($\rho=\rho_++\rho_-$) and the polarization ($m=\rho_+-\rho_-$) fields. By defining the corresponding response fields $\widehat{\rho}=(\widehat{\rho}_++\widehat{\rho}_-)/2$ and $\widehat{m}=(\widehat{\rho}_+-\widehat{\rho}_-)/2$ in~\eqref{eq:trsn hydro r rb} we get the transition probability
\begin{widetext}
\begin{align}
\mathrm{Pr}\left[\rho(x,\mathbb{T}),m(x,\mathbb{T})\,\big|\,\rho(x,0),m(x,0)\right]\simeq\int\left[\mathcal{D}\rho\right]\left[\mathcal{D}m\right]\left[\mathcal{D}\widehat{\rho}\right]\left[\mathcal{D}\widehat{m}\right]\mathrm{e}^{-2\ld\int_0^\mathbb{T}\mathrm{d}t\,\left[\int_0^\mathbb{L}\mathrm{d}x\,\big(\widehat{\rho}\,\partial_t\rho+\widehat{m}\,\partial_tm\big)-H\left[\widehat{\rho},\widehat{m},\rho,m\right]\right]} \label{trans_prob_tot_dens_polar}
\end{align}
\end{widetext}
with the effective Hamiltonian
\begin{align}\label{eq:Ham non gaus}
H=&\int_0^\mathbb{L}\mathrm{d}x\,\Big\{\rho\,(1-\rho)\,(\partial_x\widehat{\rho})^2+\rho\,(1-\rho)\,(\partial_x\widehat{m})^2\nonumber\\
&+2\,m\,(1-\rho)\,\partial_x\widehat{\rho}\,\partial_x\widehat{m}-\big[\partial_x\rho-\mathrm{Pe}\,m\,(1-\rho)\big]\,\partial_x\widehat{\rho}\nonumber\\
&-\big[(1-\rho)\,\partial_xm+m\,\partial_x\rho-\mathrm{Pe}\,\rho\,(1-\rho)\big]\,\partial_x\widehat{m}\nonumber\\
&+2\,\rho\sinh^2{\widehat{m}}-m\sinh{2\widehat{m}}\Big\}.
\end{align}

\subsubsection{\bfseries Gaussian approximation} 

The effective Hamiltonian in~\eqref{eq:Ham non gaus} is non-Gaussian in $\widehat{m}$. For most states away from transition lines, fluctuations are small and it is reasonable to assume $\widehat{m}$ to be small. A similar quadratic approximation was done in~\cite{2021_Agranov_Exact} for a related dynamics. Keeping up to the quadratic order terms in small $\widehat{m}$, the Hamiltonian is given by
\begin{align}
H&=\int_0^\mathbb{L}\mathrm{d}x\,\bigg[\frac{S_{\rho,\rho}}{2}\,(\partial_x\widehat{\rho})^2+\frac{S_{m,m}}{2}\,(\partial_x\widehat{m})^2+2\,S_{f,f}\,\widehat{m}^2\nonumber\\
&+S_{\rho,m}\,\partial_x\widehat{\rho}\,\partial_x\widehat{m}+\bar{J}_\rho\,\partial_x\widehat{\rho}+\bar{J}_m\,\partial_x\widehat{m}-2\,\bar{J}_f\,\widehat{m}\bigg] \label{hamilton_tot_dens_polar}
\end{align}
with $S_{\rho,\rho}=S_{m,m}=2\,\rho\,(1-\rho)$, $S_{f,f}=\rho$, $S_{\rho,m}=2\,m\,(1-\rho)$ and the average conserved currents due to diffusion drift
\begin{align}
\bar{J}_\rho&=-\partial_x\rho+\mathrm{Pe}\,m\,(1-\rho) \label{avg_tot_den_curr}\\
\bar{J}_m&=-(1-\rho)\,\partial_xm-m\,\partial_x\rho+\mathrm{Pe}\,\rho\,(1-\rho) \label{avg_polar_curr}
\end{align}
and the nonconserved tumbling current
\begin{equation}
\bar{J}_f=m \label{avg_tumble_curr}
\end{equation}

\subsection{\label{app_fl_hy_eqns}The fluctuating hydrodynamics equations}

The hydrodynamic action in~\eqref{trans_prob_tot_dens_polar} with the Gaussian effective Hamiltonian~\eqref{hamilton_tot_dens_polar} is the Martin-Siggia-Rose-Janssen-De Dominicis action~\cite{1973_Martin_Statistical,1976_Janssen_On,1978_Dominicis_Dynamics,1978_Dominicis_Field} of a fluctuating hydrodynamics equation,
\begin{subequations}
\begin{align}
\partial_t\rho&=-\partial_x\bigg(\bar{J_\rho}-\frac{1}{\sqrt{\ld}}\,\eta_\rho\bigg), \label{fl_hy_rho}\\
\partial_tm&=-\partial_x\bigg(\bar{J_m}-\frac{1}{\sqrt{\ld}}\,\eta_m\bigg)-2\,\bigg(\bar{J}_f-\frac{1}{\sqrt{\ld}}\,\eta_f\bigg), \label{fl_hy_m}
\end{align}
\label{fluct_hydro_quasi_1d_rho_m}\end{subequations}
where $\boldsymbol{\eta}\equiv(\eta_\rho\;\eta_m\;\eta_f)^\mathrm{T}$ is a multivariate Gaussian noise vector, whose probability distribution is
\begin{equation}
\mathrm{Pr}(\boldsymbol{\eta})\sim\mathrm{e}^{-2^{-1}\int\mathrm{d}t\,\mathrm{d}t'\,\mathrm{d}x\,\mathrm{d}x'\,\boldsymbol{\eta}(x,t)^\mathrm{T}\,\boldsymbol{\Sigma}^{-1}\,\boldsymbol{\eta}(x',t')}
\end{equation}
with the covariance matrix
\begin{equation}
\boldsymbol{\Sigma}=\delta(x-x')\,\delta(t-t')
\begin{pmatrix}
S_{\rho,\rho} & S_{\rho,m} & 0\\
S_{\rho,m} & S_{m,m} & 0\\
0 & 0 & S_{f,f}
\end{pmatrix}.
\end{equation}

\section{{Generalization to $d$~dimensions}} \label{d_dimensions_general}
Here, we generalize the action formalism and thus, the fluctuating hydrodynamic description for our model on a $d$-dimensional hyper-cubic lattice with $L^d$ sites with periodic boundaries. The lattice sites are indexed by their Cartesian coordinates $(i_1,i_2,\ldots,i_d)$ with $i_k=\{1,2,\ldots,L\}$ for $k=1,2,\ldots,d$. We discuss two possible generalizations of the active dynamics, as described in the main text.

\subsection{Case I: Two-species generalization}
This case corresponds to the system composed of particles of two-species [denoted as $(\pm)$] similar to the quasi-1D model. The $(\pm)$ particles move persistently in the $\pm\hat{x}_1$ direction (which we refer to as \emph{active} direction) with the motion being symmetric along all other directions $\big\{\pm\hat{x}_2,\ldots,\pm\hat{x}_d\big\}$ (which we refer to as \emph{passive} directions). The microscopic dynamics is defined as
\begin{itemize}
\item Biased hopping: A $(+)$ particle at site $(i_1,i_2,\ldots,i_d)$ hops to the site $(i_1\pm1,i_2,\ldots,i_d)$ at rate $(\ld^2\pm\lp/2)/\tp$ and to the sites $(i_1,i_2\pm1,\ldots,i_d),\ldots,(i_1,i_2,\ldots,i_d\pm1)$ at rate $\ld^2/\tp$, while a $(-)$ particle at site $(i_1,i_2,\ldots,i_d)$ hops to the site $(i_1\pm1,i_2,\ldots,i_d)$ at rate $(\ld^2\mp\lp/2)/\tp$ and to the sites $(i_1,i_2\pm1,\ldots,i_d),\ldots,(i_1,i_2,\ldots,i_d\pm1)$ at rate $\ld^2/\tp$, provided that the target site is unoccupied
\item Tumbling: A $(+)$ particle converts to a $(-)$ particle at rate $1/\tp$ and vice versa 
\end{itemize}

Starting from an initial configuration $\rho_\pm(x,t_{\text{ini}})$, the probability that the system eventually reaches a final configuration $\rho_\pm(x,t_{\text{fin}})$ has a hydrodynamic path integral description
\begin{widetext}
\begin{align}
\mathrm{Pr}\left[\rho_\pm(\vec{x},t_{\text{fin}})\,\big|\,\rho_\pm(\vec{x},t_{\text{ini}})\right]=\int\left[\mathcal{D}\rho_\pm\right]\left[\mathcal{D}\widehat{\rho}_{\pm}\right]\mathrm{e}^{-\ld^d\int\mathrm{d}t\,\left[\int\mathrm{d}\vec{x}\,\left(\widehat{\rho}_+\,\partial_t\rho_++\widehat{\rho}_-\,\partial_t\rho_-\right)-H\left[\widehat{\rho}_\pm\,,\,\rho_\pm\right]\right]}
\end{align}
\end{widetext}
with the Hamiltonian given by
\begin{align}
&H=\int\mathrm{d}\vec{x}\,\Big\{\rho_+\,(1-\rho)\,\big|\vec{\nabla}\widehat{\rho}_+\big|^2+\rho_-\,(1-\rho)\,\big|\vec{\nabla}\widehat{\rho}_-\big|^2\nonumber\\
&-\big[(1-\rho_-)\,\vec{\nabla}\rho_++\rho_+\,\vec{\nabla}\rho_--\mathrm{Pe}\,\rho_+\,(1-\rho)\,\hat{x}_1\big]\cdot\vec{\nabla}\widehat{\rho}_+\nonumber\\
&-\big[(1-\rho_+)\,\vec{\nabla}\rho_-+\rho_-\,\vec{\nabla}\rho_++\mathrm{Pe}\,\rho_-\,(1-\rho)\,\hat{x}_1\big]\cdot\vec{\nabla}\widehat{\rho}_-\nonumber\\
&+\big(\mathrm{e}^{-\widehat{\rho}_++\widehat{\rho}_-}-1\big)\,\rho_++\big(\mathrm{e}^{\widehat{\rho}_+-\widehat{\rho}_-}-1\big)\,\rho_-\Big\}
\end{align}
and where we have denoted $\rho=\rho_++\rho_-$. The derivation is following similar analysis as presented in Appendix \ref{sect:action_fh_derive}.

The fluctuating hydrodynamic equation corresponding to this action formalism is given by
\begin{align}
\partial_t\rho_\pm=&\,(1-\rho)\,\nabla^2\rho_\pm+\rho_\pm\,\nabla^2\rho\mp\mathrm{Pe}\,\partial_{x_1}[\rho_\pm\,(1-\rho)]\nonumber\\
&\mp(\rho_+-\rho_-)+\frac{1}{\ld^{d/2}}\,\big(\vec{\nabla}\cdot\vec{\eta}_\pm\pm\eta_f\big)
\end{align}
with the Gaussian noises having zero means and covariances $\big<\big(\vec{\eta}_\pm(\vec{x},t)\cdot\hat{x}_k\big)\,\big(\vec{\eta}_\pm(\vec{x}',t')\cdot\hat{x}_{k'}\big)\big>=2\,\rho_\pm\,(1-\rho)\,\delta_{k,k'}\,\delta(\vec{x}-\vec{x}')\,\delta(t-t')$ and $\big<\eta_f(\vec{x},t)\,\eta_f(\vec{x}',t')\big>=\rho\,\delta(\vec{x}-\vec{x}')\,\delta(t-t')$.

\subsection{Case II: $2d$-species generalization}
In this generalization, the system is composed of particles belonging to $2d$ number of species (denoted as $\sigma=\left\{(\pm)_k\right\}$ with $k=1,2,\ldots,d$), corresponding to the possible internal orientations in the spatial directions $\hat{r}_k=\left\{\pm\hat{x}_k\right\}$ with $k=1,2,\ldots,d$. This corresponds to the isotropic generalization of our quasi-1D model in higher dimensions. The microscopic dynamics is defined as
\begin{itemize}
\item Biased hopping: A particle of $(+)_k$ species at site $(i_1,i_2,\ldots,i_d)$ hops to the site $(i_1,i_2,\ldots,i_k+1,\ldots,i_d)$ at rate $(\ld^2+\lp/2)/\tp$ and to the other adjacent sites at rate $(\ld^2-\lp/2)/\tp$, while a particle of $(-)_k$ species at site $(i_1,i_2,\ldots,i_d)$ hops to the site $(i_1,i_2,\ldots,i_k-1,\ldots,i_d)$ at rate $(\ld^2+\lp/2)/\tp$ and to the other adjacent sites at rate $(\ld^2-\lp/2)/\tp$, provided that the target site is unoccupied.
\item Tumbling: A particle of $\sigma$ species converts to any other species at rate $1/\tp$.
\end{itemize}

Starting from an initial configuration $\rho_\sigma(x,t_{\text{ini}})$, the probability that the system eventually reaches a final configuration $\rho_\sigma(x,t_{\text{fin}})$ has a hydrodynamic path integral description
\begin{widetext}
\begin{equation}
\mathrm{Pr}\left[\rho_\sigma(\vec{x},t_{\text{fin}})\,\big|\,\rho_\sigma(\vec{x},t_{\text{ini}})\right]\simeq\int\left[\mathcal{D}\rho\right]\left[\mathcal{D}\widehat{\rho}\right]\mathrm{e}^{-\ld^d\int\mathrm{d}t\,\left[\int\mathrm{d}\vec{x}\sum_\sigma\left(\widehat{\rho}_\sigma\,\partial_t\rho_\sigma\right)-H\left[\rho_\sigma,\widehat{\rho}_\sigma\right]\right]},
\end{equation}
\end{widetext}
where the Hamiltonian is given by
\begin{align}
&H=\int\mathrm{d}\vec{x}\sum_\sigma\bigg\{\rho_\sigma\,(1-\rho)\,\big|\vec{\nabla}\widehat{\rho}_\sigma\big|^2-\big[(1-\rho)\,\vec{\nabla}\rho_\sigma\nonumber\\
&+\rho_\sigma\,\vec{\nabla}\rho-\mathrm{Pe}\,\rho_\sigma\,(1-\rho)\,\hat{r}_\sigma\big]\cdot\vec{\nabla}\widehat{\rho}_\sigma\nonumber\\
&+\sum_{\sigma'\neq\sigma}\rho_\sigma\,\big(\mathrm{e}^{\widehat{\rho}_{\sigma'}-\widehat{\rho}_\sigma}-1\big)\bigg\}
\end{align}
and we have denoted $\rho=\sum_\sigma\rho_\sigma$.

Corresponding to this action formalism, the fluctuating hydrodynamics for the $\sigma$-species particle is then given by
\begin{align}
&\partial_t\rho_\sigma=(1-\rho)\,\nabla^2\rho_\sigma+\rho_\sigma\,\nabla^2\rho-\mathrm{Pe}\,\vec{\nabla}\big[\rho_\sigma\,(1-\rho)\big]\cdot\hat{r}_\sigma\nonumber\\
&-\sum_{\sigma'\neq\sigma}(\rho_\sigma-\rho_{\sigma'})+\frac{1}{\ld^{d/2}}\,\bigg(\vec{\nabla}\cdot\vec{\eta}_\sigma+\sum_{\sigma'\neq\sigma}\eta_{\sigma\to\sigma'}\bigg)
\end{align}
with the Gaussian noises having zero means and covariances
$\big<\big(\vec{\eta}_\sigma(\vec{x},t)\cdot\hat{x}_k\big)\,\big(\vec{\eta}_{\sigma'}(\vec{x}',t')\cdot\hat{x}_{k'}\big)\big>=2\,\rho_\sigma\,(1-\rho)\,\delta_{\sigma,\sigma'}\,\delta_{k,k'}\,\delta(\vec{x}-\vec{x}')\,\delta(t-t')$ and $\big<\eta_{\sigma\to\sigma'}(\vec{x},t)\,\eta_{\psi\to\psi'}(\vec{x}',t')\big>=(\rho_\sigma+\rho_{\sigma'})\,\delta_{\sigma,\psi}\,\delta_{\sigma',\psi'}\,\delta(\vec{x}-\vec{x}')\,\delta(t-t')$.

\section{{The strictly $1$D model with exchange dynamics}}
Until now, we have strictly imposed the exclusion constraint which forbids two particles in adjacent sites to cross each other on the $1$D lattice. However, in certain physical examples, this single-file constraint may be partially relaxed. For example, the dynamics of active particles inside a channel whose width is of the order of the diameter of the particles~\cite{2020_Miron_Phase}. Motivated by this scenario, we discuss the case where two adjacent particles on the 1D lattice are allowed to exchange their positions at a constant rate, in addition to their usual biased-hopping and tumbling dynamics described in items (\ref{diff_drift}) and (\ref{tumble}) of the main text.

Precisely, we consider the model on a one-dimensional lattice with the following dynamics:
\begin{itemize}
\item Biased hopping: A $(+)$ particle at site $i$ hops to site $i\pm1$ at rate $(\ld^2\pm\lp/2)/\tp$ while a $(-)$ particle at site $i$ hops to site $i\pm1$ at rate $(\ld^2\mp\lp/2)/\tp$, provided that the target site is empty.
\item Tumbling: A particle changes its internal orientation (i.e. $(+)$ to $(-)$ and vice versa) at rate $1/\tp$.
\item Particle-position exchanging: A pair of $(\pm)$ particles at sites $i $ and $ i+1$, exchanges their positions at rate $\ell_e^2/\tp$.
\end{itemize}

Following the same procedure outlined in Appendix \ref{sect:action_fh_derive}, we write the transition probability of the system starting from an initial configuration $\rho_\pm(x,t_{\text{ini}})$ and reaching a final configuration $\rho_\pm(x,t_{\text{fin}})$ in terms of a path integral
\begin{widetext}
\begin{equation}
\mathrm{Pr}\left[\rho(x,t_{\text{fin}}),m(x,t_{\text{fin}})\,\big|\,\rho(x,t_{\text{ini}}),m(x,t_{\text{ini}})\right]\simeq\int\left[\mathcal{D}\rho\right]\left[\mathcal{D}m\right]\left[\mathcal{D}\widehat{\rho}\right]\left[\mathcal{D}\widehat{m}\right]\mathrm{e}^{-\ld\int\mathrm{d}t\,\left[\int\mathrm{d}x\,\left(\widehat{\rho}\,\partial_t\rho+\widehat{m}\partial_tm\right)-H\left[\rho,m,\widehat{\rho},\widehat{m}\right]\right]},
\end{equation}
\end{widetext}
for large $\ld$ and $\tp\left(=\ld^2/D\right)$, keeping $\mathcal{E}=\ell_e^2/\ld^2$ finite, with the effective Hamiltonian as given in~\eqref{hamilton_tot_dens_polar} with the modified terms $S_{m,m}=2\,[\rho\,(1-\rho+\mathcal{E}\,\rho)-\mathcal{E}\,m^2]$ and $\bar{J}_m=-(1-\rho+\mathcal{E}\,\rho)\,\partial_xm-(m-\mathcal{E}\,m)\,\partial_x\rho+\mathrm{Pe}\,\rho\,(1-\rho)$.

The corresponding fluctuating hydrodynamic equation for the density field is identical to the one in~\eqref{fl_hy_rho} with the same parameters, while that for the polarization field is similar to the one in~\eqref{fl_hy_m} but with the modified values of $\bar{J}_m$ and $\big<\eta_m(x,t)\,\eta_m(x',t')\big>$.

The $\mathcal{E}=1$ limit corresponds to the lattice-gas model in~\cite{2018_Houssene_Exact} for which the fluctuating hydrodynamic description was derived~\cite{2021_Agranov_Exact} using an ingenious mapping to known integrable models. The mapping does not straightforwardly extend for the dynamics with arbitrary $\mathcal{E}$.

\textit{Remark}. In the $\mathcal{E}=0$ limit, the hydrodynamic description breaks down, as discussed in the main text and also in the next Appendix.

\section{{Numerical evidence for local equilibrium}}\label{numerical_evidence_loc_eq}

\begin{figure}[t]
\subfloat[Low P\'eclet number: $\mathrm{Pe}=2$\label{low_pe_q1d}]{\includegraphics[width=0.5\linewidth]{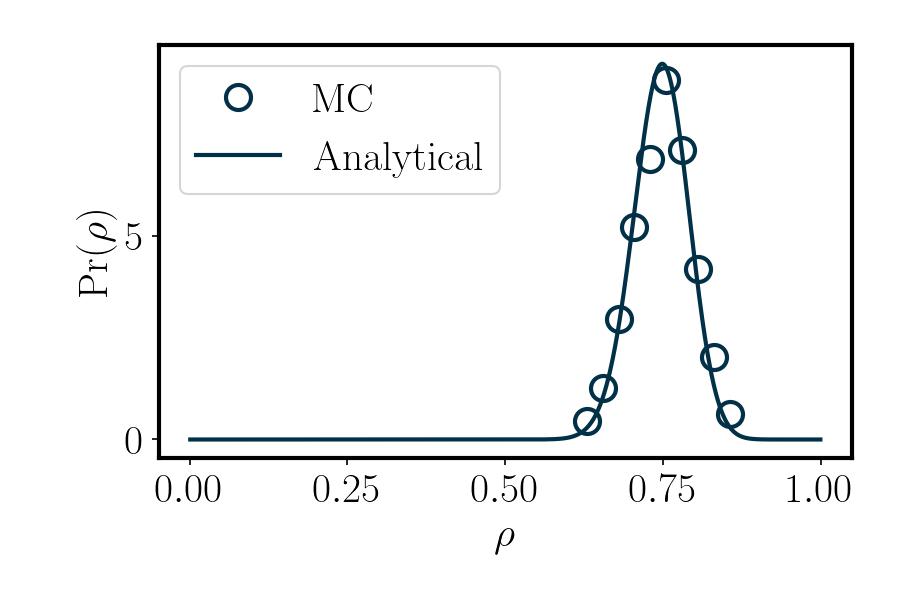}}
\hfill
\subfloat[High P\'eclet number: $\mathrm{Pe}=10$\label{high_pe_q1d}]{\includegraphics[width=0.5\linewidth]{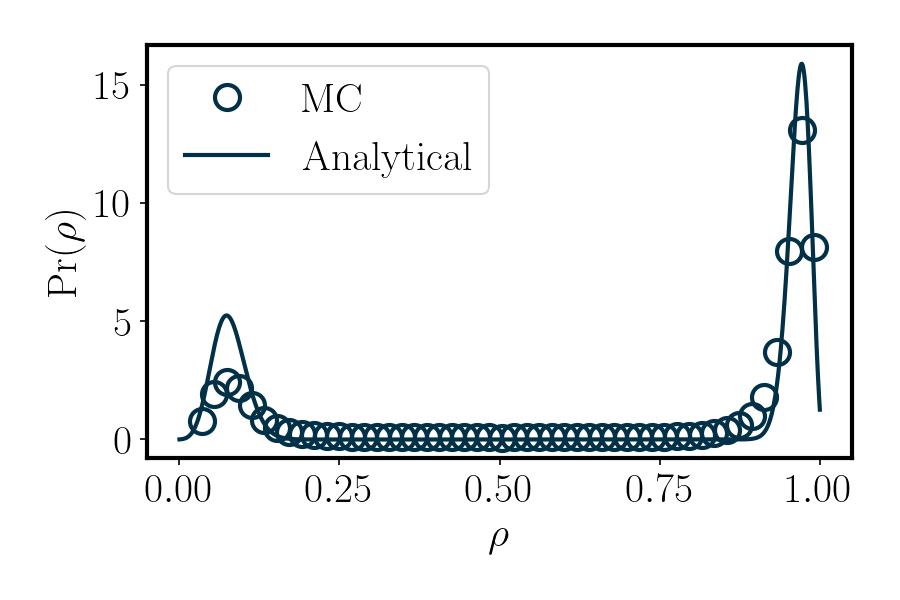}}
\caption{\textbf{For the quasi-1D model}: The number density distribution from Monte Carlo simulation compared with the distributions~\eqref{bernoulli_noMips} and~\eqref{bernoulli_Mips} for appropriate phases in the two-lane ladder model defined in the main text. The bulk-average density is $\rho_0=0.75$. For (a), the parameter values correspond to the homogeneous phase and for (b), they are for the phase-separated state.}
\label{fig:local_eq_q1d}
\end{figure}

Our construction of the fluctuating hydrodynamics crucially relies on an assumption of local equilibrium. For our choice of rates in items (\ref{diff_drift})--(\ref{tumble}) in the main text, diffusion dominates in length scales of order $\ld$, and consequently, the nearby sites within this diffusive length are expected to reach a local equilibrium in times $\tp\sim \ld^2$ around a quasistatic average density defined in~\eqref{local_equil_measure}. A consequence of this particular local equilibrium assumption is that the total number $N$ of particles, irrespective of their polarity, in a region of $l\sim \ld$ sites centered around $i\simeq x\ld$ at time $\tau\simeq t\tp$ is given by the binomial distribution~\cite{2007_Derrida_Non}
\begin{equation}
\mathrm{Pr}_l\big(N\,|\,\rho(x,t)\big)=\binom{l}{N}\,\rho(x,t)^N\,\big(1-\rho(x,t)\big)^{l-N},
\end{equation}
where $\rho(x,t)$ is the slowly varying local average density. 
Taking $N$ and $l$ to be large, while keeping $n=N/l$ finite, a Stirling's approximation gives
\begin{equation}
\mathrm{Pr}_l(n\,|\,\rho)\simeq\exp{\bigg\{-l\bigg[n\ln{\frac{n}{\rho}}+(1-n)\ln{\frac{1-n}{1-\rho}}\bigg]\bigg\}}.
\label{bernoulli}
\end{equation}

In the homogeneous stationary state i.e., outside the binodal region, the most probable local average density profile is uniform $\rho(x)=\rho_0$ everywhere. The probability $\mathrm{Pr}(n)$ of the number-density of particles sampled over \emph{random} boxes of fixed length $l$ in the bulk comes from~\eqref{bernoulli} corresponding to this typical average density $\rho_0$, leading to 
\begin{equation}
\mathrm{Pr}(n)=\mathrm{Pr}_l(n\,|\,\rho_0).
\label{bernoulli_noMips}
\end{equation}

\begin{figure}[t]
\subfloat[Low P\'eclet number: $\mathrm{Pe}=2$\label{low_pe_1d}]{\includegraphics[width=0.5\linewidth]{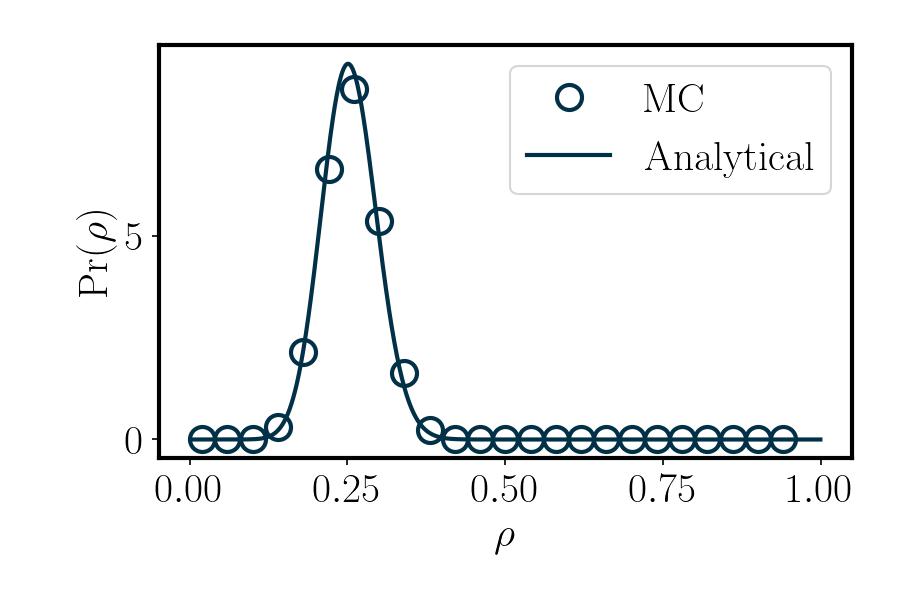}}
\hfill
\subfloat[High P\'eclet number: $\mathrm{Pe}=10$\label{high_pe_1d}]{\includegraphics[width=0.5\linewidth]{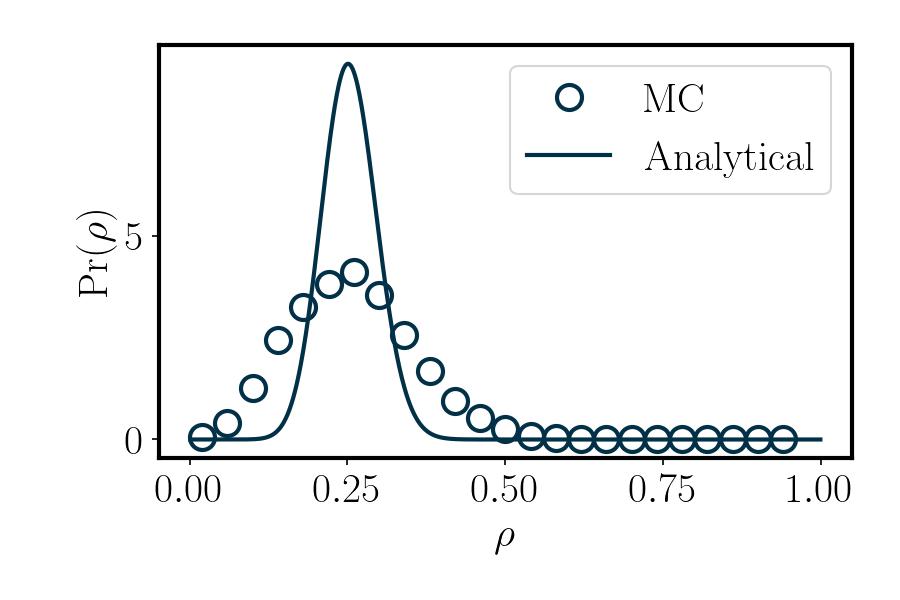}}
\caption{\textbf{For the strictly-1D model}: Comparison of Monte Carlo results of number density distribution and~\eqref{bernoulli_noMips}. Panel (a) show good agreement, supporting local equilibrium for bulk average density $\rho_0 = 0.25$ and low P\'eclet number. However, as shown in panel (b) at higher P\'eclet but at same $\rho_0$, even though the system is in homogeneous phase, the~\eqref{bernoulli_noMips} significantly differs from numerically observed distribution, indicating breakdown of our local equilibrium assumption.}
\label{fig:local_eq_1d}
\end{figure}

\begin{figure*}[t]
\centering
\includegraphics[width=0.99\textwidth]{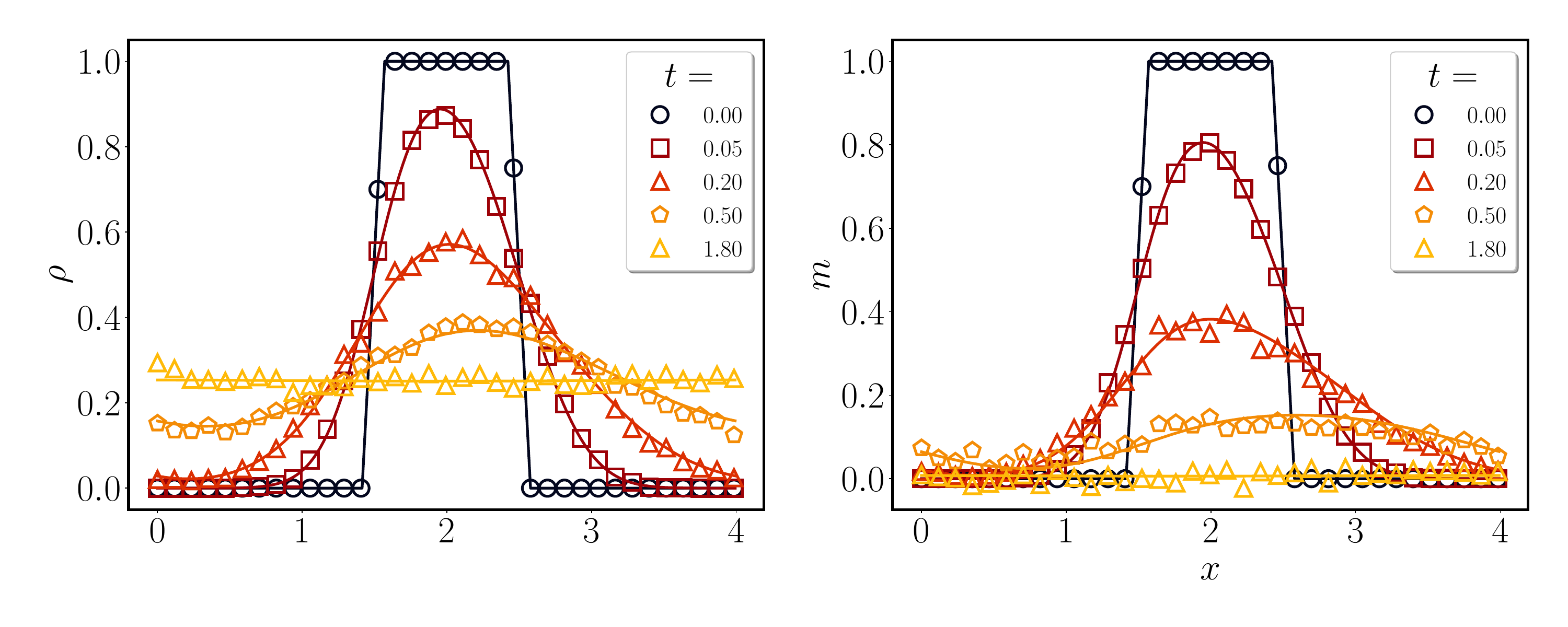}
\caption{\textbf{Validity of noiseless hydrodynamics in the quasi-1D model}: A comparison of the evolution of density $\rho(x,t)$ and polarization $m(x,t)$ from Monte Carlo simulation of the microscopic dynamics and from the noiseless hydrodynamics in~\eqref{fluct_hydro_quasi_1d_rho_m}. The evolution starts from a step initial profile for both $\rho(x,0)$ and $m(x,0)$, centered around the middle of the system. The solid lines represent noiseless hydrodynamics evolution~\eqref{fluct_hydro_quasi_1d_rho_m}, while the markers represent Monte Carlo simulations of the microscopic dynamics. To compare with the noiseless hydrodynamics, the Monte Carlo results are binned over $20$ sites and averaged over $64$ realizations. The hydrodynamic simulations are performed in the domain $[0,L/\ld=4]$ using a pseudospectral method with $2^9$ modes, described later in Appendix \ref{sec:numerical}. The plots are for $\rho_0=0.25$ and the system size $L=4\ld$ with the dynamical parameters $\ld=128$, $\tp=\ld^2$, $\lp=2 \ld$, and $\tau_\times=0.05$, which corresponds to $\mathrm{Pe}=2$. These parameters correspond to a homogeneous stationary state.}
\label{fig:q_1d_nomips}
\end{figure*}

\begin{figure*}[t]
\centering 
\includegraphics[width=0.99\textwidth]{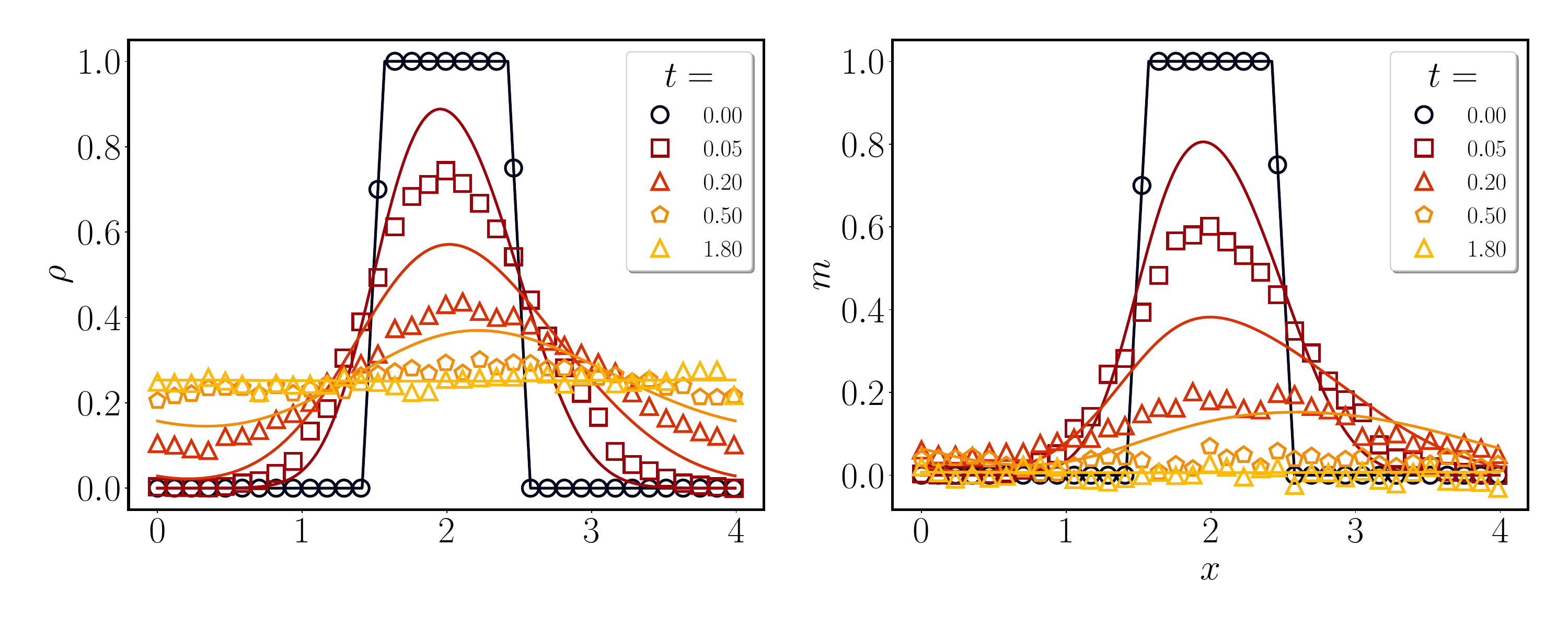}
\caption{\textbf{Breakdown of noiseless hydrodynamics in the strictly-1D model}: A similar comparison between hydrodynamic evolution and Monte Carlo simulations of microscopic dynamics for the single-lane model, for identical parameter values as in Fig.~\ref{fig:q_1d_nomips}. The noticeable deviations between the microscopic and the hydrodynamics indicate unsuitability of the latter for the 1D model.}
\label{fig:1d_nomips}
\end{figure*}

In the phase-separated stationary state, i.e. inside the spinodal region, there are two typical average densities, $\rho_l$ and $\rho_g$, corresponding to the liquid and the gas phases, respectively. In random sampling of the box position, probability for the box to be inside the liquid or the gas region is proportional to their volume fraction. This results in the probability of number density inside randomly sampled boxes,
\begin{equation}
\mathrm{Pr}(n)=v\,\mathrm{Pr}_l(n\,|\,\rho_l)+(1-v)\,\mathrm{Pr}_l(n\,|\,\rho_g),
\label{bernoulli_Mips}
\end{equation}
where $v=(\rho_0-\rho_g)/(\rho_l-\rho_g)$ denotes the volume fraction of the liquid phase.

The number-density distributions in (\ref{bernoulli_noMips}) and (\ref{bernoulli_Mips}) are an indirect test of the local equilibrium assumption. To verify this crucial assumption, we numerically measure $\mathrm{Pr}(n)$ from samples of stationary-state configurations generated using Monte Carlo simulation of the quasi-one-dimensional dynamics. Comparison of the numerical results against the distributions (\ref{bernoulli_noMips}) and (\ref{bernoulli_Mips}) are shown in Fig.~\ref{fig:local_eq_q1d}. The left panel, corresponding to a parameter regime outside the binodal (high density, low activity), thus in the homogeneous phase, shows excellent agreement with~\eqref{bernoulli_noMips}. The right panel depicts a high-density, high-activity regime where MIPS forms, as evidenced by the bimodal distribution. There is good agreement with~\eqref{bernoulli_Mips}, except slight deviations originating from contributions near the liquid-gas interface, which are not included in~\eqref{bernoulli_Mips}. 

A similar test of local equilibrium for the strictly 1D case, shows a different scenario than in the quasi-one-dimensional case. At very low P\'eclet value and low density, the local equilibrium assumption is consistent as shown in the left panel of Fig.~\ref{fig:local_eq_1d}. However, at higher P\'eclet, but still inside the homogeneous phase, the number distribution differs from the theoretical prediction~\eqref{bernoulli_noMips}, suggesting a breakdown of our specific assumption of local equilibrium.

The consequence of local equilibrium for quasi-one-dimension and its breakdown for strict one dimension is reflected in the validity of hydrodynamics in describing the Monte Carlo evolution, shown in Figs.~\ref{fig:q_1d_nomips} and \ref{fig:1d_nomips}.

\section{{The spinodal analysis}} \label{spinodal_calc}
The noiseless hydrodynamics is obtained from~\eqref{fluct_hydro_quasi_1d_rho_m} by setting the noise terms equal to zero
\begin{subequations}
\begin{align}
\partial_t\rho&=-\partial_x\bar{J}_\rho,\\
\partial_tm&=-\partial_x\bar{J}_m-2\bar{J}_f,
\end{align}
\label{noiseless_rho,m}\end{subequations}
where the conservative currents $\bar{J}_\rho$ and $\bar{J}_m$ arising due to biased-diffusion dynamics are given in~\eqref{avg_tot_den_curr} and~\eqref{avg_polar_curr} respectively while the nonconservative current $\bar{J}_f$ arising due to tumbling events is given in~\eqref{avg_tumble_curr}.

A particular solution for the stationary state of~\eqref{noiseless_rho,m} is given by the homogeneous profiles $\rho(x,t)=\rho_0$ and $m(x,t)=0$. A linear stability analysis of the homogeneous solution give the spinodal curves for MIPS. Our analysis is similar to~\cite{2018_Houssene_Exact,2018_Solon_Generalized_PRE} used in a related problem. We perturb the stationary homogeneous profiles as $\rho(x,t)=\rho_0+\delta\rho(x,t)$ and $m(x,t)=\delta m(x,t)$ with $\delta\rho$ and $\delta m$ being small fluctuations in the hydrodynamic fields. This allows us to write the hydrodynamics for the small perturbations up to the leading order as
\begin{subequations}
\begin{align}
\partial_t(\delta\rho)&=\partial_x^2(\delta\rho)-\mathrm{Pe}\,(1-\rho_0)\,\partial_x(\delta m),\\
\partial_t(\delta m)&=(1-\rho_0)\,\partial_x^2(\delta m)-\mathrm{Pe}\,(1-2\rho_0)\,\partial_x(\delta\rho)-2\,\delta m.
\end{align}
\end{subequations}

Defining the continuous Fourier transformation
\begin{equation}
\begin{pmatrix}
\widetilde{\delta\rho}(k,t)\\
\widetilde{\delta m}(k,t)
\end{pmatrix}
=\frac{1}{L/\ld}\int_0^{L/\ld}\mathrm{d}x\,\mathrm{e}^{-\mathrm{i}kx}
\begin{pmatrix}
\delta\rho(x,t)\\
\delta m(x,t)
\end{pmatrix} \label{fourier_k}
\end{equation}
for a large $L/\ld$, we can rewrite the Fourier-transformed noiseless hydrodynamic equations for the small fluctuating fields as
\begin{align}
\partial_t
\begin{pmatrix}
\widetilde{\delta\rho}(k,t)\\
\widetilde{\delta m}(k,t)
\end{pmatrix}
=
&\begin{pmatrix}
-k^2 & -\mathrm{i}\,\mathrm{Pe}\,(1-\rho_0)\,k\\
\mathrm{i}\,\mathrm{Pe}\,(2\rho_0-1)\,k & -(1-\rho_0)\,k^2-2
\end{pmatrix}\nonumber\\
&\times\begin{pmatrix}
\widetilde{\delta\rho}(k,t)\\
\widetilde{\delta m}(k,t)
\end{pmatrix}
,
\end{align}
where $k=2\pi n\ld/L$ with $n=0\,,\,\pm1\,,\,\pm2\,,\,\cdots$. The eigenvalues of the above matrix are given by
\begin{align}
&\lambda_\pm(k)=-1-\Big(1-\frac{\rho_0}{2}\Big)k^2\nonumber\\
&\pm\sqrt{1+\big[\mathrm{Pe}^2(1-\rho_0)(2\rho_0-1)-\rho_0\big]k^2+\frac{\rho_0^{\,2}\,k^4}{4}},
\end{align}
which obey the symmetry $\lambda_\pm(-k)=\lambda_\pm(k)$.

The two eigenvalues $\lambda_+(0)=0$ and $\lambda_-(0)=-2$ corresponding to the $k=0$ mode respectively describe the invariance of the spatially integrated total density $\int\mathrm{d}x\,\rho(x,t)$ and the exponential decay of the spatially integrated polarization $M(t)=\int\mathrm{d}x\,m(x,t)=M(0)\mathrm{e}^{-2t}$, which also directly follows from~\eqref{noiseless_rho,m}.

The homogeneous solutions become linearly unstable at large wavelengths when one of the eigenvalues corresponding to the first mode [i.e., $k=\pm 2\pi/(L/\ld)$] becomes positive, $\lambda_+\,(2\pi/(L/\ld))>0$. This, consequently, puts a condition on the macroscopic size of the system as 
\begin{equation}
L/\ld>\frac{2\pi\sqrt{1-\rho_0}}{\sqrt{\mathrm{Pe}^2\,(1-\rho_0)\,(2\rho_0-1)-2}}
\end{equation}
Note that the term within the square root in the denominator must be positive, i.e., 
\begin{equation}
\mathrm{Pe}^2\,(1-\rho_0)\,(2\rho_0-1)>2.
\end{equation}
Interestingly, this is the same instability condition found earlier~\cite{2018_Houssene_Exact}, where the single-file constraint is not respected and a $\pm$ pair of particles at neighboring sites is allowed to exchange their positions at a rate $\ld^2/\tp$. The above condition is equivalent to 
\begin{equation}
\label{eq:spinodal}
(\rho_0-\rho_l^s)\,(\rho_0-\rho_h^s)<0,
\end{equation}
where 
\begin{equation}
\rho_l^s=\frac{3}{4}-\frac{1}{4}\sqrt{1-\bigg(\frac{4}{\mathrm{Pe}}\bigg)^2}\text{, and }\rho_h^s=\frac{3}{4}+\frac{1}{4}\sqrt{1-\bigg(\frac{4}{\mathrm{Pe}}\bigg)^2}.
\label{rho-lh}
\end{equation}
Therefore, for $\mathrm{Pe}>4$, the system is linearly unstable in the spinodal region given by $\rho_0\in[\rho_l^s\,,\,\rho_h^s]$.

\section{{The binodal analysis}} \label{binodal_calc}
For a system in a phase-separated state, a high-density liquidlike phase and a low-density gaslike phase coexist. The binodal or the coexistence curve gives the two densities, denoted by $\rho_l$ and $\rho_g$ respectively, as a function of a control parameter, which in our case is the P\'eclet number $\mathrm{Pe}$. To compute the binodal, we use the procedure followed in ~\cite{2018_Houssene_Exact,2018_Solon_Generalized_PRE}, which we outline below for the convenience of the readers.

The stationary state is given by $\partial_t\rho=0$, which has a zero flux due to symmetry, which gives us $\bar{J_\rho}=0$. Using~\eqref{avg_tot_den_curr}, we obtain
\begin{equation}
m=\frac{1}{\mathrm{Pe}}\frac{\partial_x\rho}{1-\rho}=-\frac{1}{\mathrm{Pe}}\partial_x\big[\ln{(1-\rho)}\big].
\end{equation}

The stationary state for the polarization field i.e., $\partial_tm=0$ gives us 
$\partial_x\bar{J}_m-2\bar{J}_f=0$, which, using~\eqref{avg_polar_curr},~\eqref{avg_tumble_curr} and the above expression of $m$, can be rewritten as $\partial_xg=0$, where
\begin{align}
g(\rho)&=g_0(\rho)+\Lambda(\rho)\,(\partial_x\rho)^2-\kappa\,\partial_x^2\rho=\text{const, with}\label{g-rho}\\
g_0(\rho)&=\mathrm{Pe}\,\rho(1-\rho)-\frac{2}{\mathrm{Pe}}\ln{(1-\rho)},\label{g0}\\
\Lambda(\rho)&=-\frac{2}{\mathrm{Pe}}\frac{1}{1-\rho}\text{, and }\kappa=\frac{1}{\mathrm{Pe}}.
\end{align}
The spinodal curve is equivalently defined by $g_0'(\rho)=0$, whose two solutions are given by $\rho_l^s$ and $\rho_h^s$ respectively [see~\eqref{rho-lh}].

\subsection{First relation}
In the coexisting phase, density is homogeneous in the liquid ($\rho_l$) and gas ($\rho_g$) phases: $\partial_x\rho=\partial_x^2\rho=0$ in the two phases, giving
\begin{equation}
g_0(\rho_g)=g_0(\rho_l)\equiv\bar{g}(\mathrm{Pe}), \label{binodal_cond_1}
\end{equation}

\begin{figure}
\centering
\includegraphics[width=0.99\linewidth]{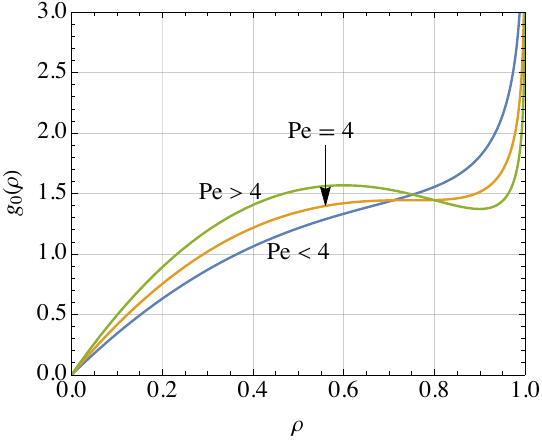}\\
\includegraphics[width=0.99\linewidth]{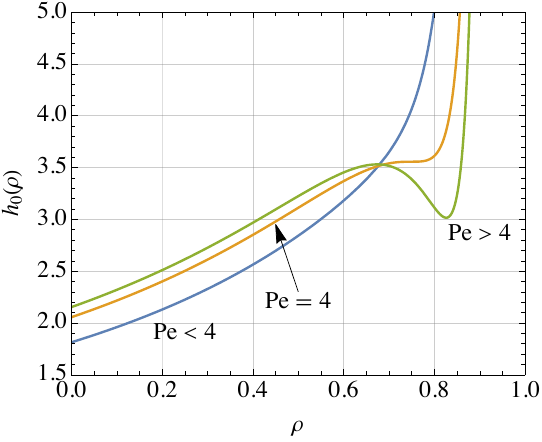}
\caption{The binodal curves are given by the pair of conditions $g_0(\rho_g) = g_0(\rho_l) $ and $h_0(\rho_g) = h_0(\rho_l)$, where the expressions of $g_0(\rho)$ and $h_0(\rho)$ are given by~\eqref{g0} and~\eqref{h0} respectively. Here, $g_0(\rho)$ and $h_0(\rho)$ are plotted as a function of $\rho$ on the left and right figures, respectively. For $\mathrm{Pe} <4$, both are monotonic functions of $\rho$. On the other hand, for $\mathrm{Pe}$, both $g_0(\rho)$ and $h_0(\rho)$ have a minimum and a maximum. For $\mathrm{Pe}=4$, there is an inflection point at $\rho=3/4$, where the first and the second derivatives of both $g_0$ and $h_0$ vanish.}
\label{fig:g0h0}
\end{figure}

For $\mathrm{Pe}<4$,
$g_0(\rho)$ is a monotonic function of $\rho$. On the other hand, it has a maximum and minimum at $\rho_l^s$ and $\rho_h^s$ respectively for $\mathrm{Pe}>4$ [see Fig.~\ref{fig:g0h0}]. For $\mathrm{Pe}=4$, there is an inflection point at $\rho=3/4$, where both $g_0'(\rho)$ and $g_0''(\rho)$ are zero. For $\mathrm{Pe} >4$, any line $g_0(\rho)=\textrm{const}\in\big(g_0(\rho_l^s)\,,\,g_0(\rho_h^s)\big)$, intersects the $g_0(\rho)$ vs. $\rho$ curve at three points, and therefore, there exists an infinite number of solutions for~\eqref{binodal_cond_1}. Hence, we need another relation for fixing the coexistence densities, which we discuss below.

\subsection{Second relation}
Consider the following integral across the liquid-gas interface in a phase-separated state
\begin{equation}
I=\int_{x_g}^{x_l}\mathrm{d}x\,g(\rho)\,\partial_xR(\rho),
\end{equation} 
where $R(\rho)$ is some function of $\rho$, and $x_g$ and $x_l$ are any two points inside the gas and liquid phases respectively. Since $g(\rho)=\bar{g}$ is constant, the above integral yields
\begin{equation}
I=\bar{g}\big[R(\rho_l)-R(\rho_g)\big].
\label{eq:I1}
\end{equation}
On the other hand, substituting the expression of $g(\rho)$ from~\eqref{g-rho} in the integral gives
\begin{align}
&I=\int_{x_g}^{x_l}\mathrm{d}x\,g_0(\rho)\,\partial_xR(\rho)\nonumber\\
&+\int_{x_g}^{x_l}\mathrm{d}x\,\big[\Lambda(\rho)\,R'(\rho)\,(\partial_x\rho)^2-\kappa\,R'(\rho)\,\partial_x^2\rho
\big]\,\partial_x\rho. \label{integral-I2}
\end{align}

\begin{figure*}[t]
\subfloat[No MIPS state forms\label{fig:metastable_no_mips}]{\includegraphics[width=0.495\textwidth]{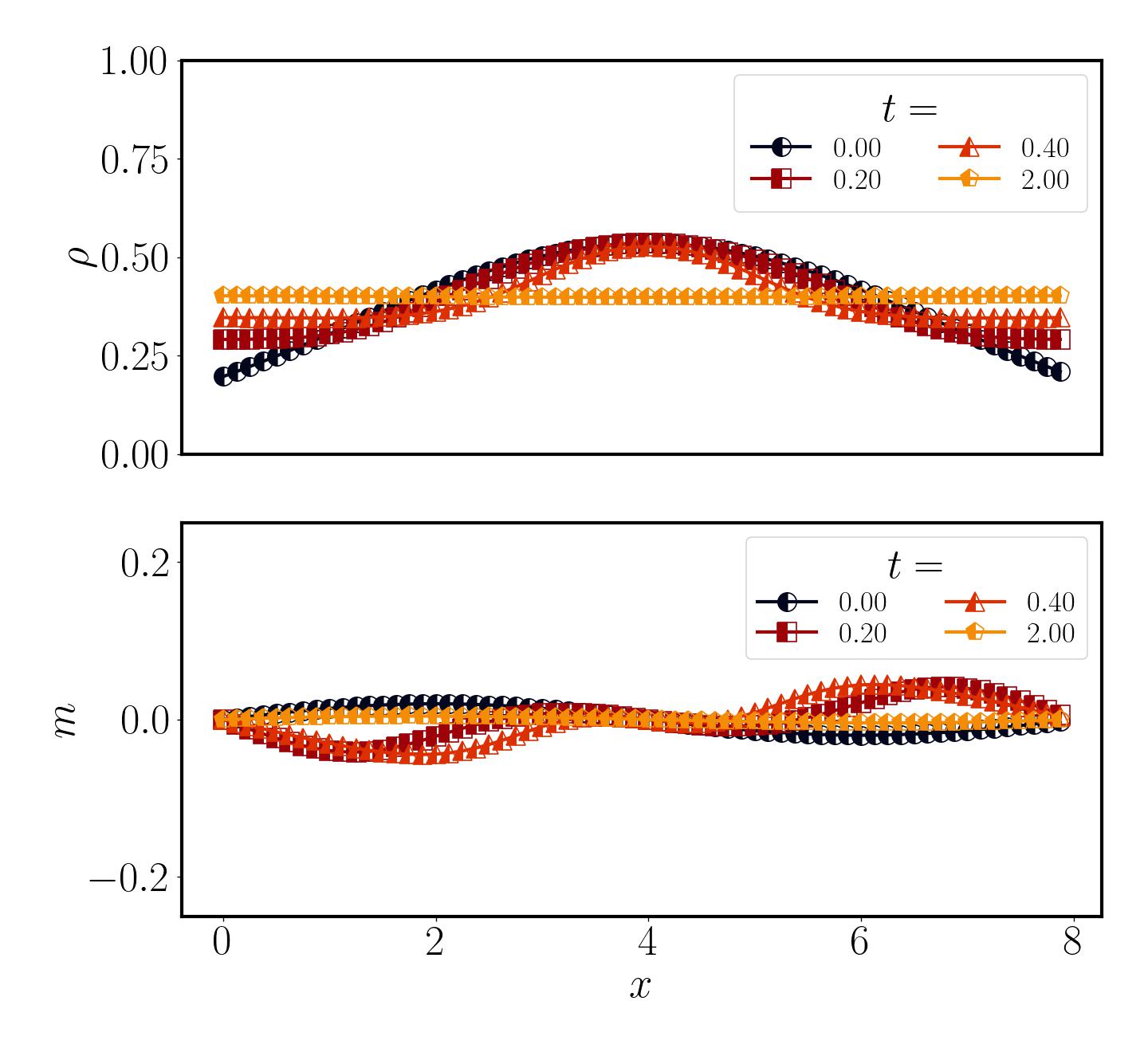}}
\hfill
\subfloat[A MIPS state forms\label{fig:metastable_mips}]{\includegraphics[width=0.495\textwidth]{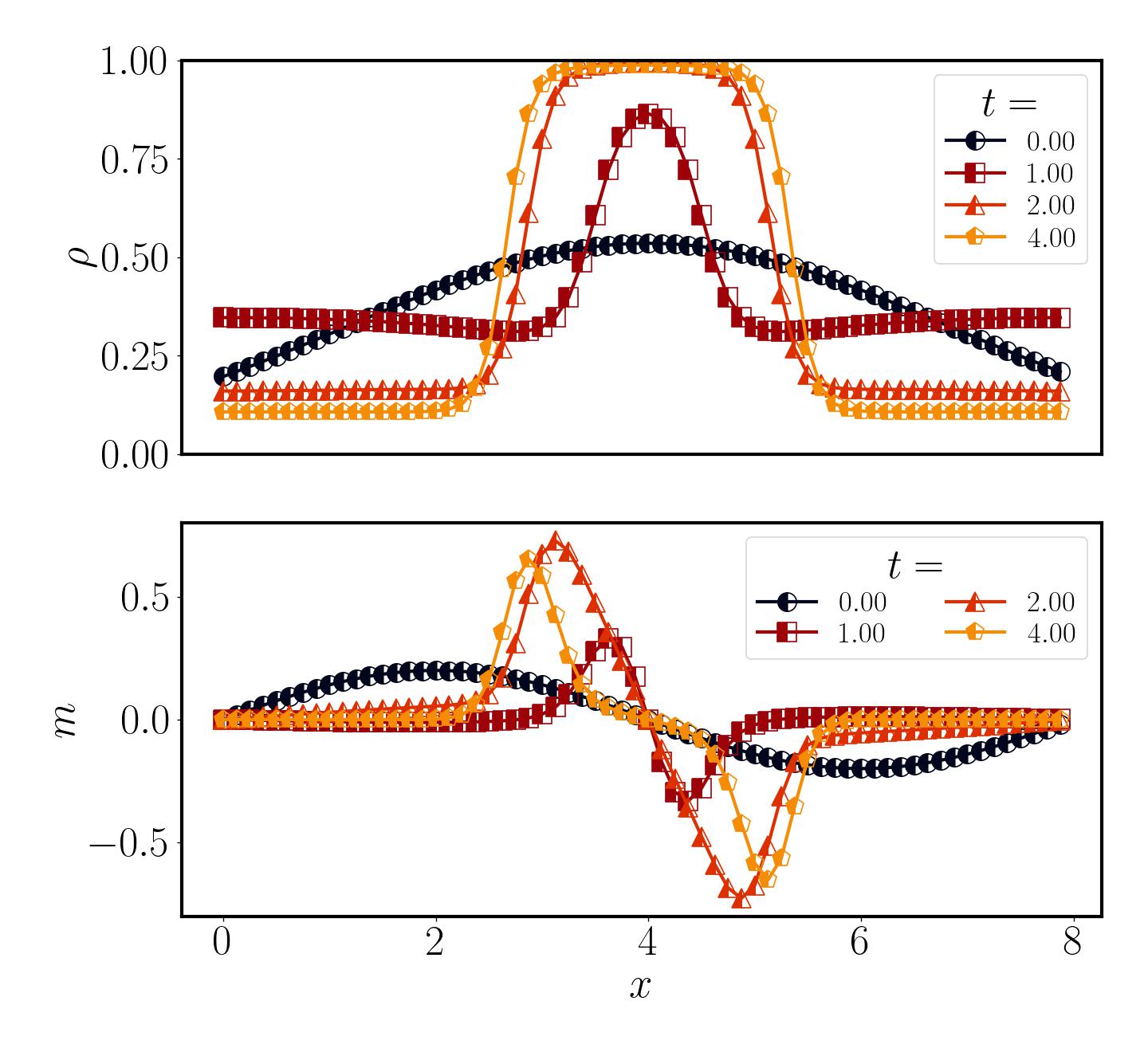}}
\caption{\textbf{Metastability}: Numerical evolution of the noiseless hydrodynamic equation~\eqref{fluct_hydro_quasi_1d_rho_m} for regions outside the spinodal curve but inside the binodal curve demonstrates that a homogeneous solution is linearly stable, but nonlinear instabilities could drive the system to an inhomogeneous MIPS state. The two panels correspond to two initial states with same initial Gaussian total density profile $\rho(x,0)$ but different initial polarization fields $m(x,0)=A \sin{2\pi x/L}$ with amplitude $A=0.02$ for panel (a) and $A=0.2$ for panel (b). The parameters $\mathrm{Pe}=10$ and $\rho_0=0.4$ are for a region between the spinodal and binodal curves. The small inhomogeneity of $m(x,0)$ in panel (a) drives the system towards a uniform final state, whereas larger inhomogeneity in panel (b) leads to an inhomogeneous MIPS state.}
\label{fig:metastability}
\end{figure*}

Now we choose $R(\rho)$ such that the second integral is zero. Consider the relation
\begin{equation}
\partial_x\big[A(\rho)\,(\partial_x\rho)^2\big]=\big[A'(\rho)\,(\partial_x\rho)^2+2A(\rho)\,\partial_x^2\rho\big]\,\partial_x\rho.
\end{equation}
Choosing $2A(\rho)=-\kappa\,R'(\rho)$ and $\kappa\,R''(\rho)=-2\Lambda(\rho)\,R'(\rho)$, the above relation gives
\begin{align}
-\partial_x\bigg[\frac{\kappa}{2}\,R'(\rho)\,(\partial_x\rho)^2\bigg]=\big[\Lambda(\rho)\,R'(\rho)\,(\partial_x\rho)^2\nonumber\\
-\kappa\,R'(\rho)\,\partial_x^2\rho\big]\,\partial_x\rho.
\end{align}
Since $\partial_x\rho=0$ in the liquid and the gas phases, the integral over the above expression across the interface is zero. Hence, defining $g_0(\rho)=\mathrm{d}\phi(R)/\mathrm{d}R$, from~\eqref{integral-I2} we get 
\begin{equation}
I=\phi(R_l)-\phi(R_g)
\label{eq:I2}
\end{equation}
where $R_l=R(\rho_l)$ and $R_g=R(\rho_g)$. Combining the two expressions of $I$,~\eqref{eq:I1} and~\eqref{eq:I2}, we arrive at the relation
\begin{equation}
h_0(R_g)=h_0(R_l)\;\;\mathrm{where}\;\;h_0(R)=R\phi'(R)-\phi(R).
\end{equation}

For the particular form of $\Lambda(\rho)$ given above, the above differential equation of $R(\rho)$ can be solved exactly to give (up to an additive and a multiplicative constant)
\begin{equation}
R(\rho)=\frac{1}{3}\frac{1}{(1-\rho)^3}.
\end{equation}
Inverting the above equation, we can express $\rho(R)=1-(3R)^{-1/3}$, which in turn gives 
\begin{equation}
\phi'(R)=\mathrm{Pe}\big[1-(3R)^{-1/3}\big](3R)^{-1/3}+\frac{2}{\mathrm{3Pe}}\ln{(3R)}.
\end{equation}
Now, integrating with respect to $R$ gives (up to an additive constant, which does not show up in the relation)
\begin{equation}
\phi(R)=\mathrm{Pe}\bigg[\frac{(3R)^{1/3}}{2}-1\bigg](3R)^{1/3}+\frac{2R}{3\mathrm{Pe}}\big[\ln{(3R)}-1\big].
\end{equation}

Therefore, $h_0(R(\rho))$ can be expressed in terms of $\rho$ explicitly as
\begin{equation}
h_0(\rho)=\frac{2}{9\mathrm{Pe}}\frac{1}{(1-\rho)^3}+\frac{\mathrm{Pe}}{6}\frac{3-4\rho}{(1-\rho)^2} \label{h0}
\end{equation} 
and the second coexistence relation can now be written as
\begin{equation}
h_0(\rho_g)=h_0(\rho_l).
\end{equation}

The qualitative behavior of $h_0$ is similar to that of $g_0(\rho)$ [see Fig.~\ref{fig:g0h0}]. For $\mathrm{Pe} <4$, 
$h_0(\rho)$ is a monotonic function of $\rho$, whereas for $\mathrm{Pe}>4$, it has a maximum and a minimum at $\rho_l^s$ and $\rho_h^s$ respectively . For $\mathrm{Pe}=4$, there is an inflection point at $\rho=3/4$, where both $h_0'(\rho)$ and $h_0''(\rho)$ are zero.

Interestingly, the two solutions of $h_0'(\rho)=0$ are same as that of $g_0'(\rho)=0$ and are given by $\rho_l^s$ and $\rho_h^s$ respectively [see~\eqref{rho-lh}]. This follows from using $\phi'(R)=g_0(\rho)$, writing $h_0(\rho)=R(\rho)g_0(\rho)-\phi(R(\rho))$ and taking a derivative with respect to $\rho$
\begin{align}
h_0'(\rho)&=R(\rho)\,g_0'(\rho)+R'(\rho)\,g_0(\rho)-R'(\rho)\,\phi'(R)\nonumber\\
&=R(\rho)\,g_0'(\rho).
\end{align}

The binodal curves are obtained by numerically solving the relations $g_0(\rho_g) = g_0(\rho_l) $ and $h_0(\rho_g) = h_0(\rho_l)$, where the expressions of $g_0(\rho)$ and $h_0(\rho)$ are given by~\eqref{g0} and~\eqref{h0} respectively.

\textit{Remark}. In the intermediate region between the spinodal and binodal curves, the uniform homogeneous solution of the noiseless hydrodynamics is linearly stable, although there are two additional solutions corresponding to the two binodal densities $\rho_l$ and $\rho_g$. In practice, these solutions are selected depending on the initial state, as shown in Fig.~\ref{fig:metastability}.

\section{{The two-point correlations in the stationary-state homogeneous phase}} \label{two_pt_corr}

In this appendix, we present the derivation of the two-point correlations of the hydrodynamic fields in the stationary state of the homogeneous phase of our quasi-1D model. This serves as an indirect check of our fluctuating hydrodynamic description, reported in eq.~(1) of the main text. Our result for the correlations is valid up to the leading order in $1/\ld$ and for this purpose, we use the Gaussian noise approximation, ignoring any subleading terms.

The analysis follows a standard approach~\cite{2009_Bertini_Towards,2016_Sadhu_Correlations,2021_Agranov_Exact} for correlations using fluctuating hydrodynamics. We begin our derivation by adding small fluctuations, $\delta\rho(x,t)$ and $\delta m(x,t)$ which are of at most $O(1/\sqrt{\ld})$ around the stationary-state homogeneous fields, $\rho(x,t)=\rho_0$ and $m(x,t)=0$ respectively. The definition of the two-point correlations of the fluctuating fields then readily follows
\begin{subequations}
\begin{align}
C_{\rho\rho}(x,x')&=\big<\delta\rho(x,t)\,\delta\rho(x',t)\big>,\\
C_{\rho m}(x,x')&=\big<\delta\rho(x,t)\,\delta m(x',t)\big>\text{, and}\\
C_{mm}(x,x')&=\big<\delta m(x,t)\,\delta m(x',t)\big>.
\end{align}
\label{corr_func_defn}\end{subequations}

Putting these perturbed total density and polarization fields in eq.~(1) of the main text and keeping only the terms which are at most linear in $\delta\rho$ and $\delta m$, we obtain
\begin{subequations}
\begin{equation}
\partial_t\begin{pmatrix}
\delta\rho(x,t)\\
\delta m(x,t)
\end{pmatrix}=\boldsymbol{D}_x\begin{pmatrix}
\delta\rho(x,t)\\
\delta m(x,t)
\end{pmatrix}+\boldsymbol{F}(x,t)
\end{equation}
where the deterministic, $\boldsymbol{D}_x$ and fluctuating, $\boldsymbol{F}(x,t)$ components are respectively given as
\begin{align}
\boldsymbol{D}_x&=\begin{pmatrix}
\partial_x^2 & -\mathrm{Pe}\,(1-\rho_0)\,\partial_x\\
-\mathrm{Pe}\,(1-2\rho_0)\,\partial_x & (1-\rho_0)\,\partial_x^2-2
\end{pmatrix},\\
\boldsymbol{F}(x,t)&=\sqrt{\frac{2\rho_0}{\ld}}\begin{pmatrix}
\sqrt{1-\rho_0}\,\partial_x\eta_\rho\\
\sqrt{1-\rho_0}\,\partial_x\eta_m+\sqrt{2}\,\eta_f
\end{pmatrix}
\end{align}
\end{subequations}
where $\eta_\rho(x,t)$, $\eta_m(x,t)$ and $\eta_f(x,t)$ are delta-correlated Gaussian white noises with unit self-covariance and zero cross covariance.

In the large hydrodynamic system size limit $L/\ld\to\infty$, we define the continuous-Fourier-transformed fluctuating hydrodynamic fields as
\begin{equation}
\begin{pmatrix}
\delta\widetilde{\rho}(k,t)\\
\delta\widetilde{m}(k,t)
\end{pmatrix}
=\frac{1}{2\pi}\int_{-\infty}^\infty\mathrm{d}x\,
\begin{pmatrix}
\delta\rho(x,t)\\
\delta m(x,t)
\end{pmatrix}
\,\mathrm{e}^{-\mathrm{i}kx}
\end{equation}
and subsequently rewrite the fluctuating hydrodynamic equations in the Fourier space
\begin{subequations}
\begin{equation}
\partial_t\begin{pmatrix}
\delta\widetilde{\rho}(k,t)\\
\delta\widetilde{m}(k,t)
\end{pmatrix}=\widetilde{\boldsymbol{D}}_k\begin{pmatrix}
\delta\widetilde{\rho}(k,t)\\
\delta\widetilde{m}(k,t)
\end{pmatrix}+\widetilde{\boldsymbol{F}}(k,t)
\label{linear_fluctuate_hydro_fourier}\end{equation}
with the deterministic and fluctuating components in the Fourier space respectively given by
\begin{align}
\widetilde{\boldsymbol{D}}_k&=\begin{pmatrix}
-k^2 & -\mathrm{i}\,\mathrm{Pe}(1-\rho_0)\,k\\
-\mathrm{i}\,\mathrm{Pe}(1-2\rho_0)\,k & -(1-\rho_0)k^2-2
\end{pmatrix},\\
\widetilde{\boldsymbol{F}}(k,t)&=\sqrt{\frac{2\rho_0}{\ld}}
\begin{pmatrix}
\mathrm{i}\,\sqrt{1-\rho_0}\,k\,\widetilde{\eta}_1(k,t)\\
\sqrt{(1-\rho_0)\,k^2+2}\,\widetilde{\eta}_2(k,t)
\end{pmatrix},
\end{align}
where $\widetilde{\eta}_1(k,t)$ and $\widetilde{\eta}_2(k,t)$ are Gaussian noises with mean and covariance
\begin{align}
\big<\eta_\alpha(x,t)\big>&=0\text{ and}\\
\big<\eta_\alpha(x,t)\,\eta_{\alpha'}(x',t')\big>&=\delta_{\alpha,\alpha'}\,\delta(x-x')\,\delta(t-t'),
\end{align}
where $\alpha,\alpha'=\{1,2\}$.
\end{subequations}

The differential equations in~\eqref{linear_fluctuate_hydro_fourier} can now be readily solved which leads to the solutions of the fluctuation fields in Fourier space in terms of $\widetilde{\boldsymbol{D}}_k$ and $\widetilde{\boldsymbol{F}}(k,t)$ as
\begin{equation}
\begin{pmatrix}
\delta\widetilde{\rho}(k,t)\\
\delta\widetilde{m}(k,t)
\end{pmatrix}=\int_{-\infty}^t\mathrm{d}t'\,\mathrm{e}^{\boldsymbol{D}_k(t-t')}\,\boldsymbol{F}(k,t')
\end{equation}

The two-point correlations of the fluctuating hydrodynamic fields in the Fourier space is then obtained using the Fourier-transformed definitions of~\eqref{corr_func_defn} which are given by $C_{\rho\rho}(k,k')=\big<\delta\rho(k,t)\,\delta\rho(k',t)\big>$, $C_{mm}(k,k')=\big<\delta m(k,t)\,\delta m(k',t)\big>$ and $C_{\rho m}(k,k')=\big<\delta\rho(k,t)\,\delta m(k',t)\big>$. For instance, the correlation between the total density fields of two points in Fourier space is given by
\begin{widetext}
\begin{equation}
\widetilde{C}_{\rho\rho}(k,k')=\frac{\rho_0\,(1-\rho_0)\,\Big\{(1-\rho_0)\,(2-\rho_0)\,k^4+\big[4\,(1-\rho_0)+\mathrm{Pe}^2\,(1-\rho_0)^2+\chi^2\big]\,k^2+2\,\big[2+\mathrm{Pe}^2\,(1-\rho_0)\big]\Big\}}{\Big\{(1-\rho_0)\,(2-\rho_0)\,k^4+\big[2\,(1-\rho_0)+(2-\rho_0)\,\chi^2\big]\,k^2+2\,\chi^2\Big\}\,\ld}\,\delta(k+k'), \label{corr_tot_dens_without_zero_k_mode}
\end{equation}
\end{widetext}
where the parameter
\begin{equation}
\chi=\sqrt{2-\mathrm{Pe}^2\,(1-\rho_0)\,(2\rho_0-1)}
\end{equation}
determines the onset of linear instability of the stationary-state homogeneous fields under small perturbations i.e., the spinodal region.

\begin{figure*}[t]
\centering
\includegraphics[width=\linewidth]{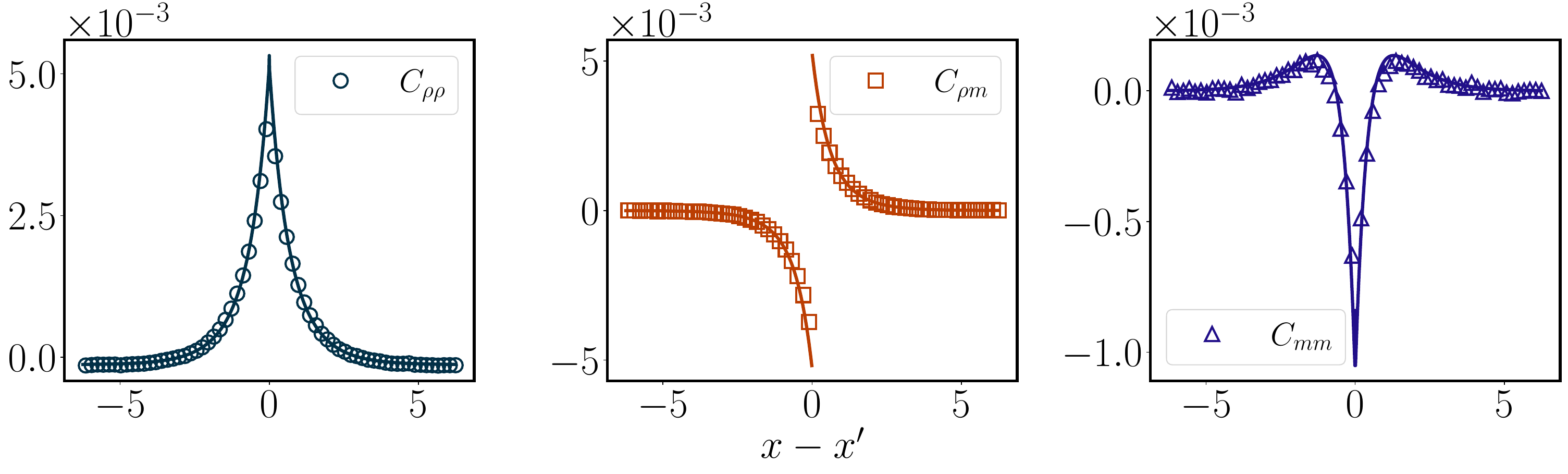}
\caption{\textbf{Correlations in the homogeneous phase in quasi-1D model}: An excellent agreement between the hydrodynamic predictions in~\eqref{eq:correlations}, indicated by solid lines, and their results from a Monte Carlo simulation, indicated by plot markers. The results are for the homogeneous phase with $\rho_0=0.25$, $\mathrm{Pe}=2$. For the simulation, the microscopic correlations $C_{\rho\rho}(i,i')=\langle n_in_{i'}\rangle-\langle n_i\rangle\,\langle n_{i'}\rangle$, $C_{\rho m}(i,i')=\langle n_iM_{i'}\rangle-\langle n_i\rangle\,\langle M_{i'}\rangle$, and $C_{mm}(i,i')=\langle M_iM_{i'}\rangle-\langle M_i\rangle\,\langle M_{i'}\rangle$, where, $i$ and $i'$ denotes the lattice sites, $n_i=\{0,1\}$ and $M_i=\{0,\pm1\}$ denotes the total occupation and polarization of the $i^\mathrm{th}$ site, and $\langle\cdot\rangle$ denotes the averaging over time after reaching the stationary state. Using the translational symmetry, due to periodic boundary conditions, spatial averaging is conducted to further refine the averaging. For comparing with the fluctuating hydrodynamics results, we use the re-scaling as $x-x'=(i-i')/\ld$. Evidently, the $C_{\rho\rho}(x)$ and $C_{mm}(x)$ are symmetric in $x$, while the $C_{\rho m}(x)$ is anti-symmetric in $x$.}
\label{fig:full_correlation-quasi-1d}
\end{figure*}

\begin{figure*}[t!]
\centering
\includegraphics[width=\linewidth]{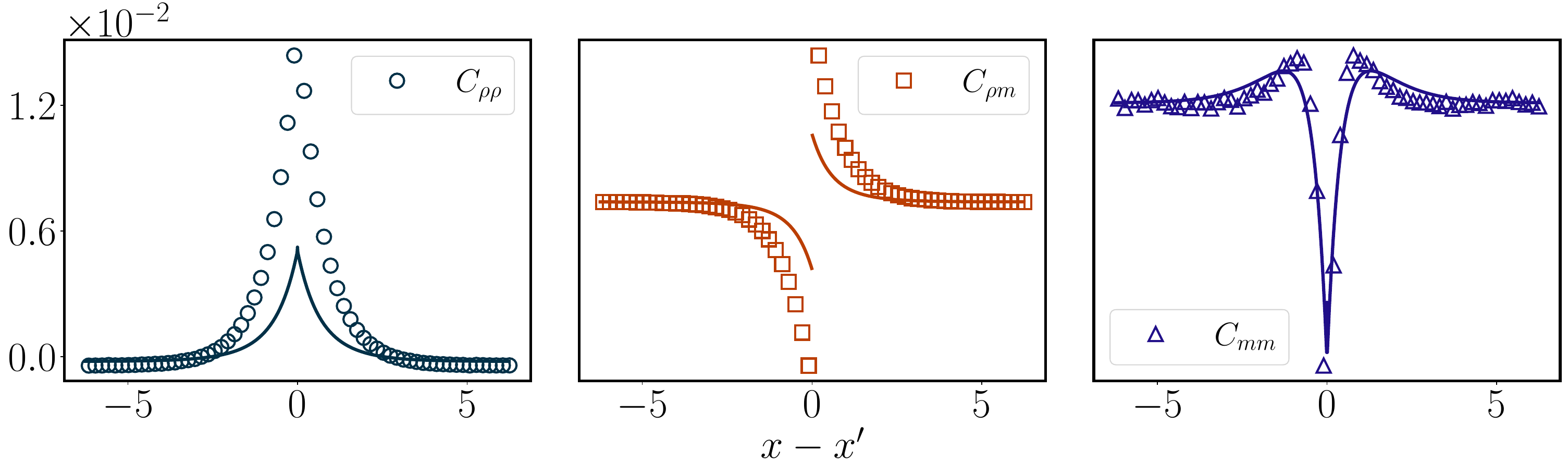}
\caption{\textbf{Correlations in the homogeneous phase in strictly-1D model}: Similar comparison of correlations in the single-lane model for parameters identical to Fig.~\ref{fig:full_correlation-quasi-1d}. The significant deviation between the hydrodynamic and the Monte Carlo results indicate breakdown of the former in the strict 1D geometry.}
\label{fig:full_correlation-1d}
\end{figure*}

The $k=\left(k'=\right)0$ mode relates to the spatial integral of the fluctuations of the total density field, $\int\mathrm{d}x\,\delta\rho(x,t)$. Although the total number of particles in the two-lane system as a whole is conserved, the number of particles in each of the lanes is allowed to fluctuate due to the lane-crossing dynamics. This particle number fluctuation in the two lanes of the system can be assumed to be independent of each other in the long time limit which implies that the $2\times\widetilde{C}_{\rho\rho}(k,k')$ must vanish when $k=\left(k'=\right)0$. We take this into account by including an additional $k=\left(k'=\right)0$ term proportional to the infinitesimal intermode width $\ld/L$, such that the correlation in~\eqref{corr_tot_dens_without_zero_k_mode} is now given as
\begin{widetext}
\begin{align}
\widetilde{C}_{\rho\rho}(k,k')=&\,\Bigg[\frac{\rho_0\,(1-\rho_0)\,\Big\{(1-\rho_0)\,(2-\rho_0)\,k^4+\big[4\,(1-\rho_0)+\mathrm{Pe}^2\,(1-\rho_0)^2+\chi^2\big]\,k^2+2\,\big[2+\mathrm{Pe}^2\,(1-\rho_0)\big]\Big\}}{\Big\{(1-\rho_0)\,(2-\rho_0)\,k^4+\big[2\,(1-\rho_0)+(2-\rho_0)\,\chi^2\big]\,k^2+2\,\chi^2\Big\}\,\ld}\nonumber\\
&-\frac{\rho_0\,(1-\rho_0)\,\big[2+\mathrm{Pe}^2\,(1-\rho_0)\big]}{2\,\chi^2\,L}\,\delta(k)\Bigg]\,\delta(k+k') \label{rho_rho_corr_fourier}
\end{align}

Similarly, the correlations of total density with polarization and polarization with polarization are respectively given as
\begin{align}
\widetilde{C}_{\rho m}(k,k')&=\frac{-2\,\mathrm{i}\,\mathrm{Pe}\,\rho_0^2\,(1-\rho_0)\,\big[(1-\rho_0)\,k^2+2\big]\,k}{\Big\{(1-\rho_0)\,(2-\rho_0)\,k^4+\big[2\,(1-\rho_0)+(2-\rho_0)\,\chi^2\big]\,k^2+2\,\chi^2\Big\}\,\ld}\,\delta(k+k'), \label{rho_m_corr_fourier} \displaybreak\\
\widetilde{C}_{mm}(k,k')&=\frac{\rho_0\,\Big\{(1-\rho_0)\,(2-\rho_0)\,k^4+\big[2\,(2-\rho_0)+\mathrm{Pe}^2\,(1-\rho_0)\,(1-2\rho_0)^2+(1-\rho_0)\,\chi^2\big]\,k^2+2\,\chi^2\Big\}}{\Big\{(1-\rho_0)\,(2-\rho_0)\,k^4+\big[2\,(1-\rho_0)+(2-\rho_0)\,\chi^2\big]\,k^2+2\,\chi^2\Big\}\,\ld}\,\delta(k+k'). \label{m_m_corr_fourier}
\end{align}
\end{widetext}

In the above two expressions of the correlations involving the polarization fields, we do not have any additional $k=\left(k'=\right)0$ term, as the spatially-integrated fluctuating polarization fields $\int\mathrm{d}x\,\delta m(x,t)$ do not vanish.

Finally, we take the inverse Fourier transformation of the expressions in (\ref{rho_rho_corr_fourier}-\ref{rho_m_corr_fourier}), which gives the two-point correlations for the fluctuating hydrodynamic fields in the stationary-state homogeneous phase. These are respectively given by
\begin{widetext}
\begin{subequations}
\begin{align}
C_{\rho\rho}(x,x')&=\frac{\mathrm{Pe}^2\,\rho_0^2\,(1-\rho_0)^2}{[2-\mathrm{Pe}^2\,(1-\rho_0)\,(2-\rho_0)\,(2\rho_0-1)]\,\ld}\,\Bigg[\sqrt{\frac{2}{2-\rho_0}}\,\mathrm{e}^{-\frac{|x-x'|}{\xi_1}}+\sqrt{1-\rho_0}\,\frac{\chi^2-2}{\chi}\,\mathrm{e}^{-\frac{|x-x'|}{\xi_2}}\Bigg]\nonumber\\
&\;+\frac{\rho_0\,(1-\rho_0)}{\ld}\,\delta(x-x')-\frac{\rho_0\,(1-\rho_0)\,\big[2+{\mathrm{Pe}}^2\,(1-\rho_0)\big]}{2\,\chi^2\,L},\\
C_{\rho m}(x,x')&=\mathrm{sgn}(x-x')\,\frac{\mathrm{Pe}\,\rho_0^2\,(1-\rho_0)}{[2-\mathrm{Pe}^2\,(1-\rho_0)\,(2-\rho_0)\,(2\rho_0-1)]\,\ld}\,\bigg[\frac{2}{2-\rho_0}\,\mathrm{e}^{-\frac{|x-x'|}{\xi_1}}+(\chi^2-2)\,\mathrm{e}^{-\frac{|x-x'|}{\xi_2}}\bigg],\\
C_{mm}(x,x')&=\frac{\mathrm{Pe}^2\,\rho_0^2\,(1-\rho_0)\,(1-2\rho_0)}{[2-\mathrm{Pe}^2\,(1-\rho_0)\,(2-\rho_0)\,(2\rho_0-1)]\,\ld}\,\Bigg(\sqrt{\frac{2}{2-\rho_0}}\,\mathrm{e}^{-\frac{|x-x'|}{\xi_1}}-\frac{\chi}{\sqrt{1-\rho_0}}\,\mathrm{e}^{-\frac{|x-x'|}{\xi_2}}\Bigg)+\frac{\rho_0}{\ld}\,\delta(x-x'),
\end{align}
\label{eq:correlations}\end{subequations}
\end{widetext}
where the two ``correlation lengths" corresponding to the length scales of the exponentially decaying long-range correlations are
\begin{equation}
\xi_1=\sqrt{1-\frac{\rho_0}{2}}\quad\text{and}\quad\xi_2=\frac{\sqrt{1-\rho_0}}{\chi}.
\end{equation}

Two important observations are in order at this point. First, all three correlations show a divergence when $2-\mathrm{Pe}^2\,(1-\rho_0)\,(2-\rho_0)\,(2\rho_0-1)\longrightarrow0$. In fact, our results for the correlations are not valid in the region $2-\mathrm{Pe}^2\,(1-\rho_0)\,(2-\rho_0)\,(2\rho_0-1)\le0$. Second, $\xi_1>\xi_2$ in the region $2-\mathrm{Pe}^2\,(1-\rho_0)\,(2-\rho_0)\,(2\rho_0-1)>0$. Consequently, the first exponential term in each of the correlations dominates over the second exponential term when the two points in space are far-separated i.e., $x-x'\gg0$.

There is an excellent agreement, shown in Fig.~\ref{fig:full_correlation-quasi-1d}, between the hydrodynamic results for the correlations in~\eqref{eq:correlations} and corresponding results from a direct Monte Carlo simulation. 

\textit{Remark}. For the strictly 1D model, one would arrive at the same two-point correlations in the stationary-state homogeneous phase owing to the similar fluctuating hydrodynamic equations in the two geometries. Only the additional shift corresponding to the zero momentum mode in the strictly 1D model would be twice that in the quasi-1D model. However, due to a breakdown of hydrodynamics, the exact correlation functions obtained from the Monte Carlo simulations show (see Fig.~\ref{fig:full_correlation-1d}) a clear mismatch with the analytical expressions.

\section{{Details of the numerical simulations}}\label{sec:numerical}

For the microscopic dynamics we use the usual Monte Carlo method. We consider a periodic lattice with two lanes with $L$ sites in each lane. We choose time step $dt=0.01$ to convert the rates to probability for the Monte Carlo simulations. The lane-crossing rate $1/\tau_{\times}$ is taken as 10, 20 or 40 depending on the simulation. 

To solve the hydrodynamic equations we employed the standard Fourier pseudospectral method~\cite{trefethen2000spectral}, with nonlinear terms computed in real space using the $\frac{2}{3}$ dealiasing rule~\cite{trefethen2000spectral}. Time integration was performed in Fourier space utilizing the standard Runge-Kutta fourth order (RK4) method and the Integrating factor Runge-Kutta fourth order (IFRK4) method, with a time step ($\mathrm{d}t$) of $10^{-3}$ or $10^{-4}$, depending on the system's resolution. Some relevant codes for this project can be found in \cite{git_repository}.

\section{{Supplemental videos}}\label{supplement_vid}

\begin{figure}
\subfloat[Video-01]{\includegraphics[width=0.49\linewidth]{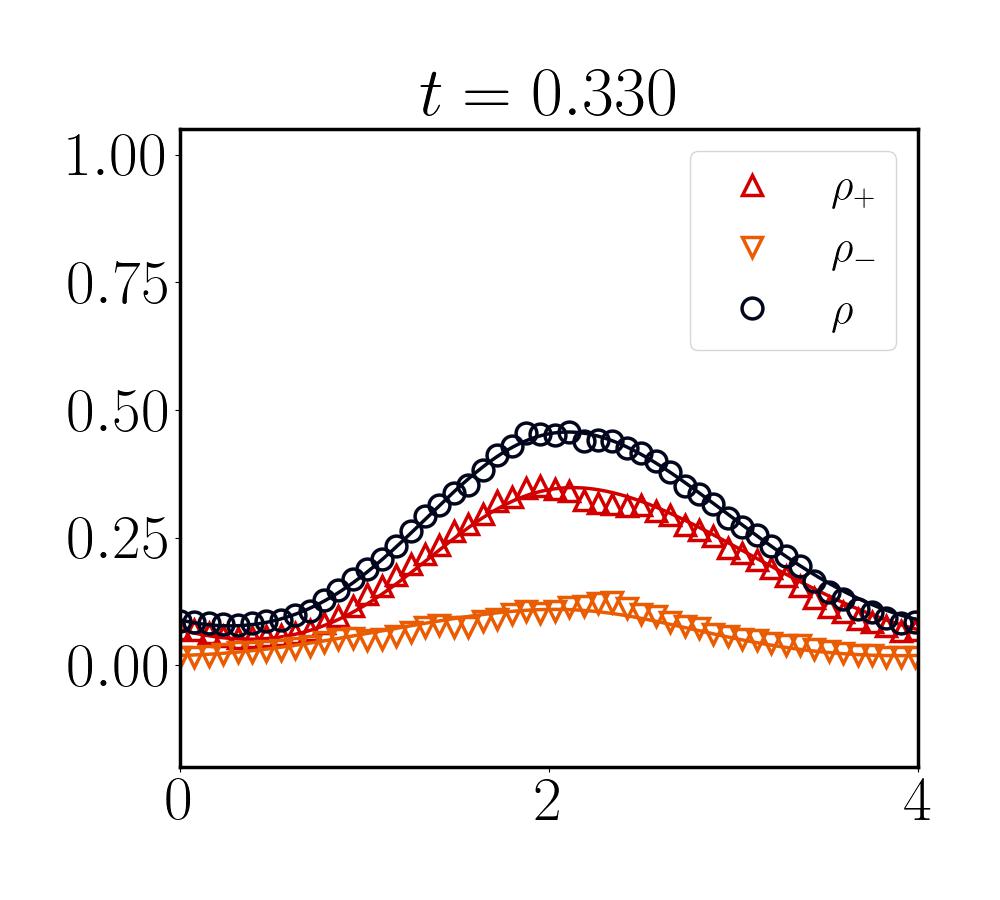}}
\hfill
\subfloat[Video-02]{\includegraphics[width=0.49\linewidth]{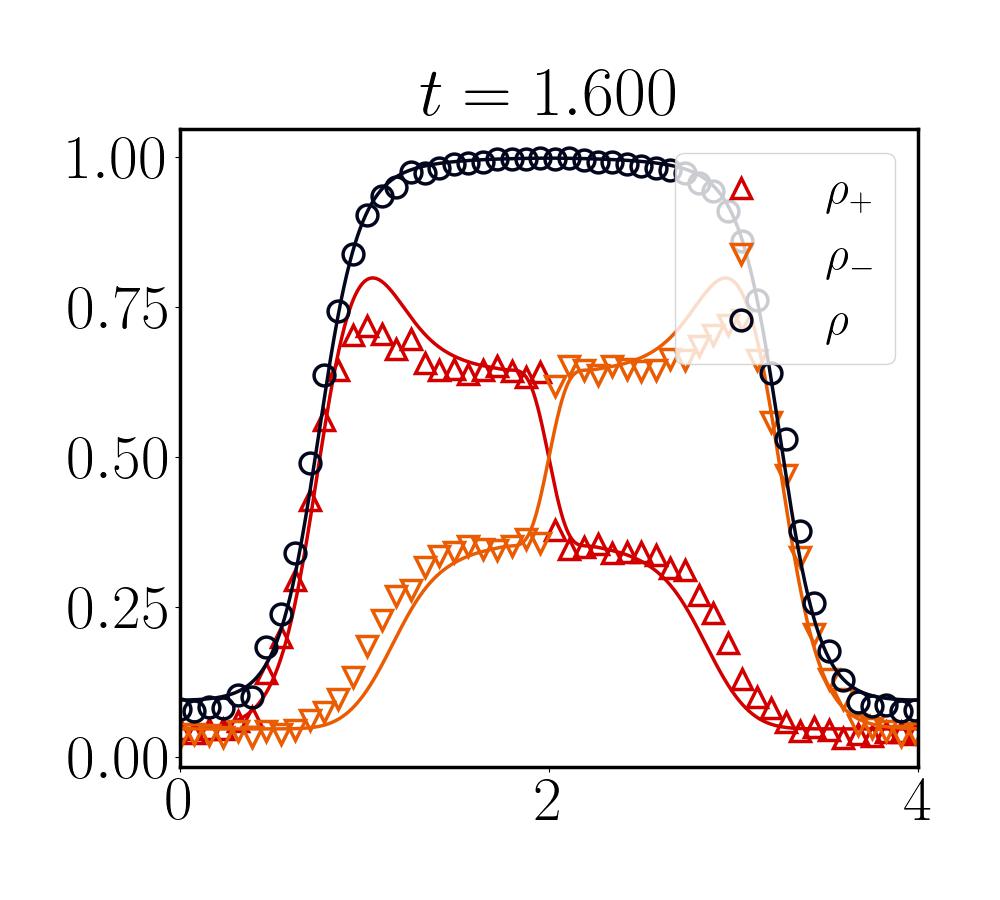}}\\
\subfloat[Video-03]{\includegraphics[width=0.49\linewidth]{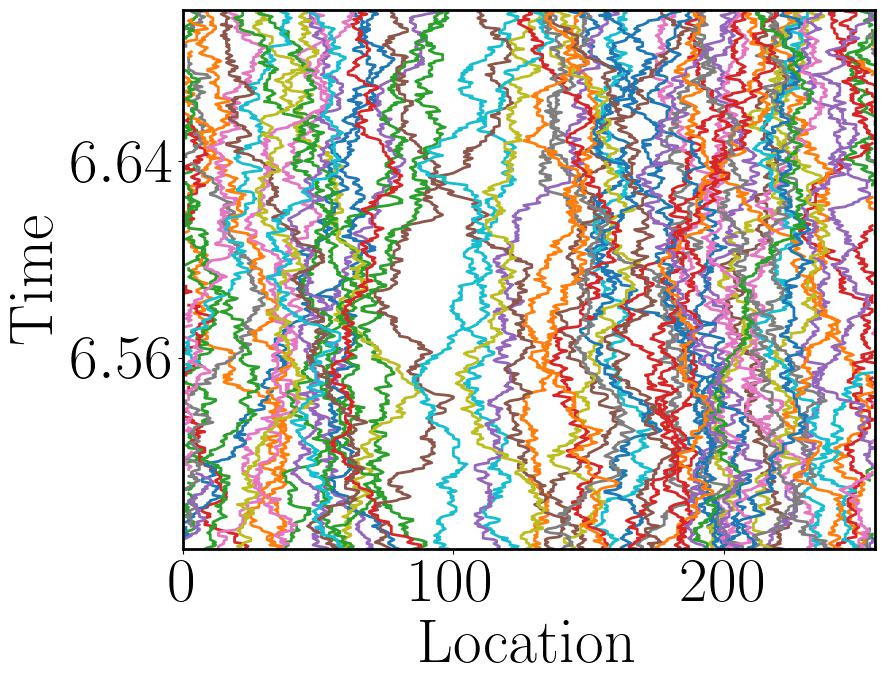}}
\hfill
\subfloat[Video-04]{\includegraphics[width=0.49\linewidth]{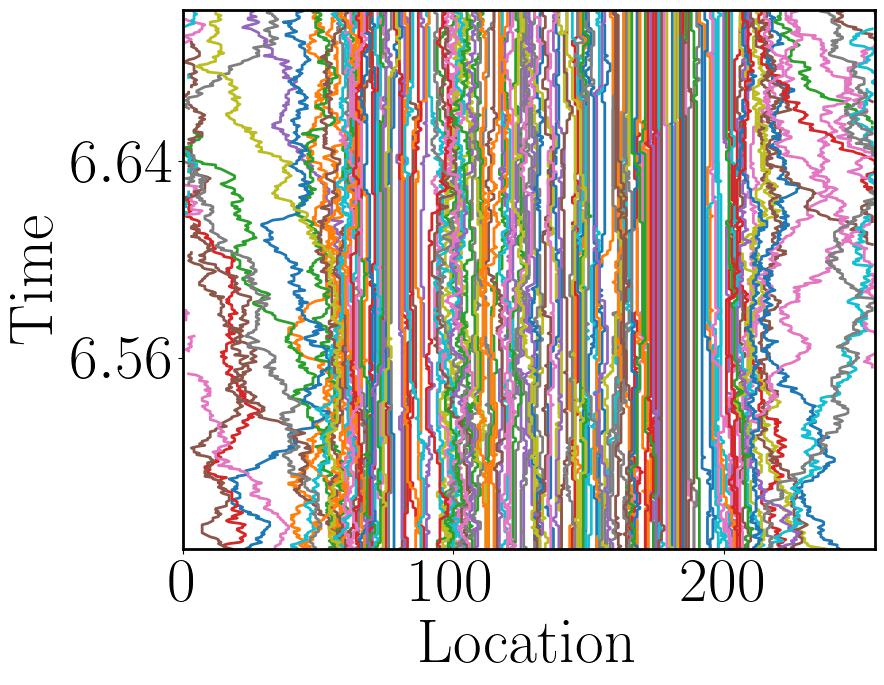}}\\
\subfloat[Video-05]{\includegraphics[width=0.49\linewidth]{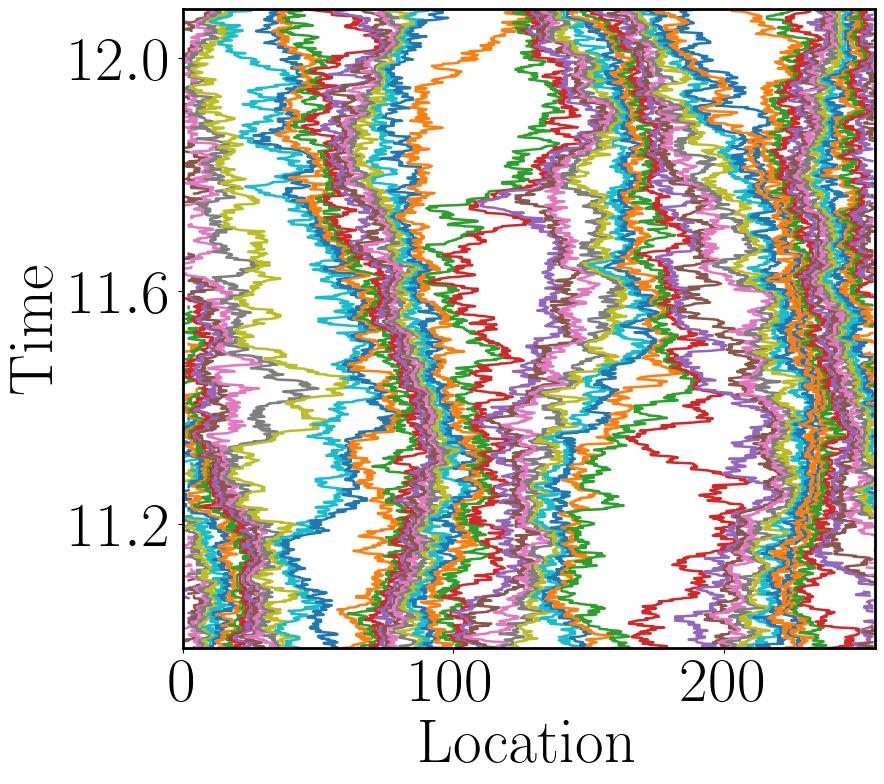}}
\hfill
\subfloat[Video-06]{\includegraphics[width=0.49\linewidth]{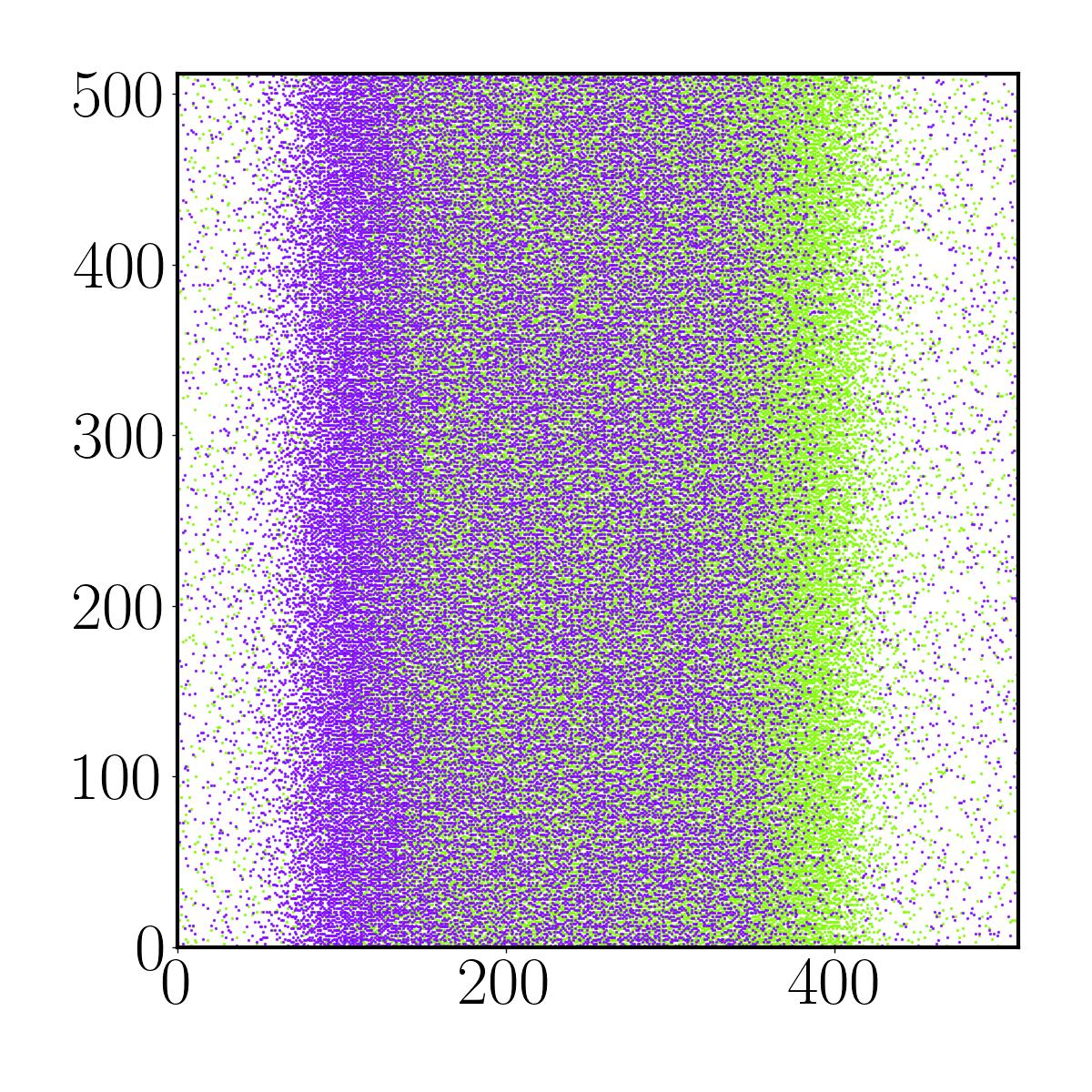}}\\
\subfloat[Video-07]{\includegraphics[width=0.49\linewidth]{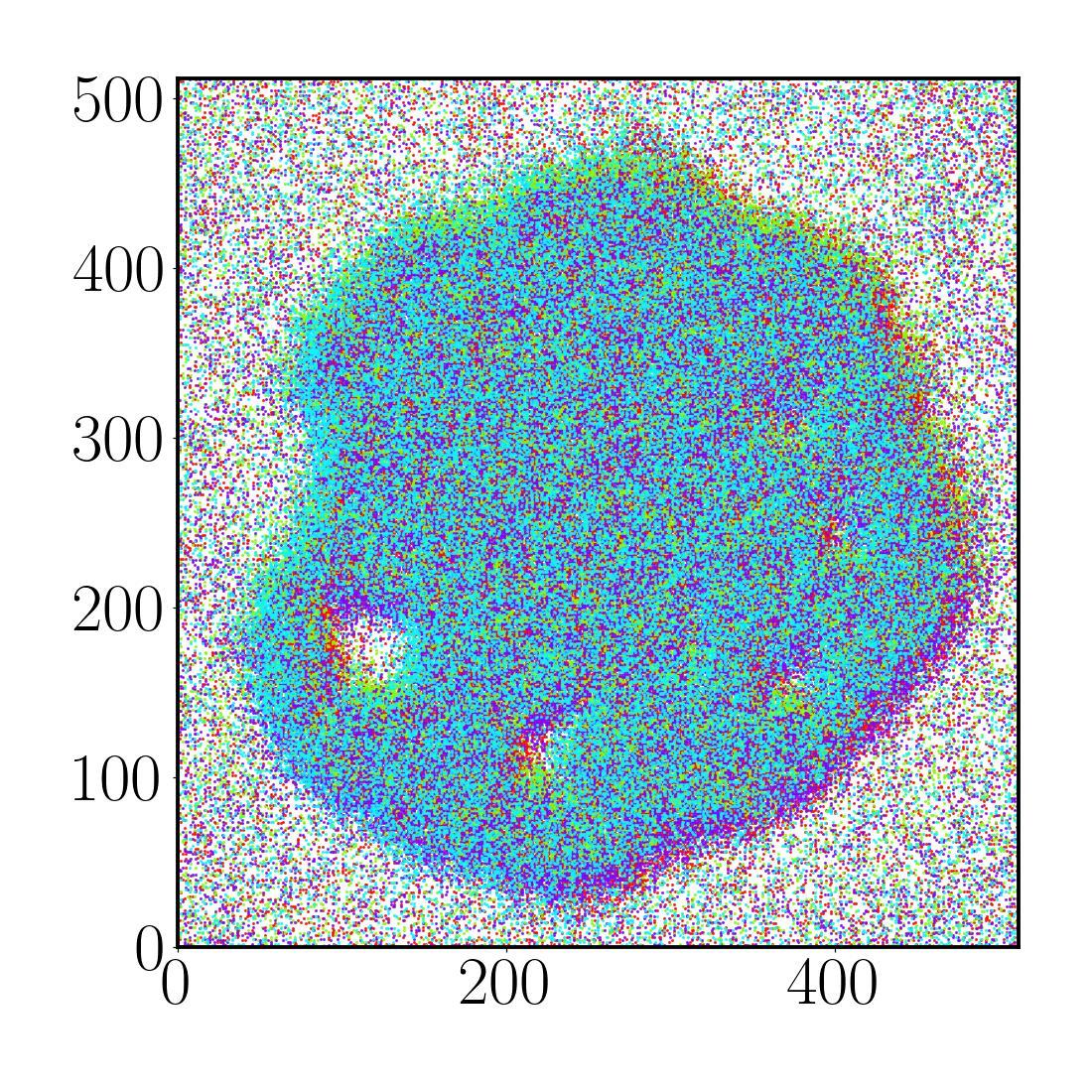}}
\hfill
\subfloat[Video-08]{\includegraphics[width=0.43\linewidth]{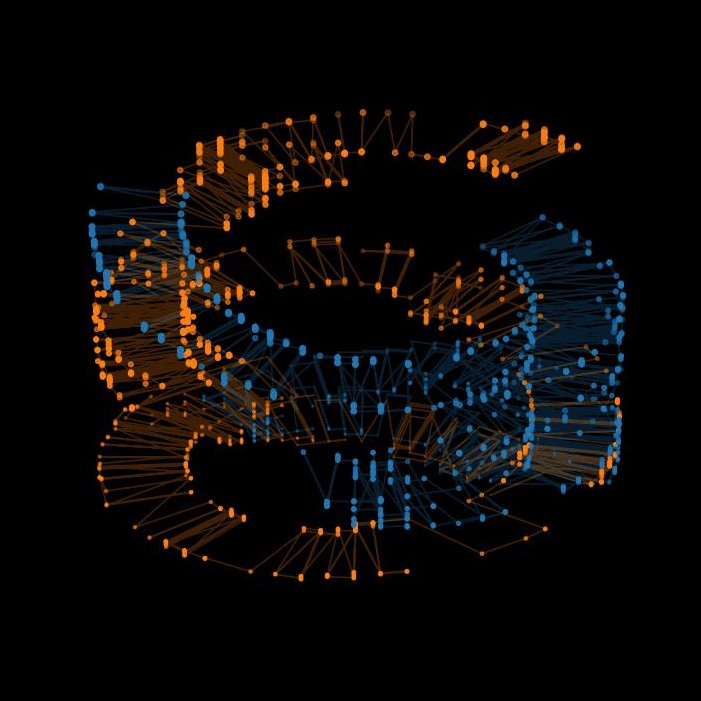}}
\caption{Snapshots of the supplemental videos.}
\label{fig:video}
\end{figure}

Here we provide a brief description along with snapshots of the Supplemental Material videos~\cite{S_M} (Fig.~\ref{fig:video}).

(1) Video-01.mp4: The video shows the time evolution of the coarse-grained density profiles for the quasi-1D two-lane ladder model for parameters outside the binodal. We show the densities only on one lane. Markers represent Monte Carlo simulation and solid lines represent numerical solutions of the hydrodynamics.

(2) Video-02.mp4: The video shows the time evolution for the quasi-1D two-lane ladder model for parameters inside the spinodal. We show the densities only on one lane. Markers represent Monte Carlo simulation and solid lines represent numerical solutions of the hydrodynamics. There is visible evidence for the formation of density inhomogeneity, indicating MIPS.

(3) Video-03.mp4: The video shows the space-time trajectories (kymographs) of particles on the quasi-1D two-lane ladder model for parameters outside the binodal. Particles starting from lane $1$ are tracked over time.

(4) Video-04.mp4: The video shows the space-time trajectories (kymographs) of the particles on a quasi-1D two-lane ladder model for parameters inside the spinodal. Particles starting from lane $1$ are tracked over time. The clustering of trajectories indicates MIPS.

(5) Video-05.mp4: The video shows the space-time trajectories (kymographs) for the strictly 1D single-lane model. The parameters are at low density but moderate Pe, still being outside the binodal. We can clearly see the formation of motile micro-clusters, which merge and break over time, but without coarsening.

(6) Video-06.mp4: The video shows the Monte Carlo evolution of two-species particles on a 2D periodic lattice forming MIPS.

(7) Video-07.mp4: The video shows the Monte Carlo evolution of four-species particles on a 2D periodic lattice forming MIPS.

(8) Video-08.mp4: A visualization of the dynamics of the two-lane ladder model. The video shows the dynamics of two particles represented by two different colors, over time, as they diffuse, switch between lanes, and tumbles. We represent the two lanes as two concentric circles.

%


\begin{thebibliography}{69}%
\makeatletter
\providecommand \@ifxundefined [1]{%
 \@ifx{#1\undefined}
}%
\providecommand \@ifnum [1]{%
 \ifnum #1\expandafter \@firstoftwo
 \else \expandafter \@secondoftwo
 \fi
}%
\providecommand \@ifx [1]{%
 \ifx #1\expandafter \@firstoftwo
 \else \expandafter \@secondoftwo
 \fi
}%
\providecommand \natexlab [1]{#1}%
\providecommand \enquote  [1]{``#1''}%
\providecommand \bibnamefont  [1]{#1}%
\providecommand \bibfnamefont [1]{#1}%
\providecommand \citenamefont [1]{#1}%
\providecommand \href@noop [0]{\@secondoftwo}%
\providecommand \href [0]{\begingroup \@sanitize@url \@href}%
\providecommand \@href[1]{\@@startlink{#1}\@@href}%
\providecommand \@@href[1]{\endgroup#1\@@endlink}%
\providecommand \@sanitize@url [0]{\catcode `\\12\catcode `\$12\catcode
  `\&12\catcode `\#12\catcode `\^12\catcode `\_12\catcode `\%12\relax}%
\providecommand \@@startlink[1]{}%
\providecommand \@@endlink[0]{}%
\providecommand \url  [0]{\begingroup\@sanitize@url \@url }%
\providecommand \@url [1]{\endgroup\@href {#1}{\urlprefix }}%
\providecommand \urlprefix  [0]{URL }%
\providecommand \Eprint [0]{\href }%
\providecommand \doibase [0]{https://doi.org/}%
\providecommand \selectlanguage [0]{\@gobble}%
\providecommand \bibinfo  [0]{\@secondoftwo}%
\providecommand \bibfield  [0]{\@secondoftwo}%
\providecommand \translation [1]{[#1]}%
\providecommand \BibitemOpen [0]{}%
\providecommand \bibitemStop [0]{}%
\providecommand \bibitemNoStop [0]{.\EOS\space}%
\providecommand \EOS [0]{\spacefactor3000\relax}%
\providecommand \BibitemShut  [1]{\csname bibitem#1\endcsname}%
\let\auto@bib@innerbib\@empty
\bibitem [{\citenamefont {Bowick}\ \emph {et~al.}(2022)\citenamefont {Bowick},
  \citenamefont {Fakhri}, \citenamefont {Marchetti},\ and\ \citenamefont
  {Ramaswamy}}]{2022_Bowick_Symmetry}%
  \BibitemOpen
  \bibfield  {author} {\bibinfo {author} {\bibfnamefont {M.~J.}\ \bibnamefont
  {Bowick}}, \bibinfo {author} {\bibfnamefont {N.}~\bibnamefont {Fakhri}},
  \bibinfo {author} {\bibfnamefont {M.~C.}\ \bibnamefont {Marchetti}},\ and\
  \bibinfo {author} {\bibfnamefont {S.}~\bibnamefont {Ramaswamy}},\ }\bibfield
  {title} {\bibinfo {title} {Symmetry, thermodynamics, and topology in active
  matter},\ }\href {https://doi.org/10.1103/PhysRevX.12.010501} {\bibfield
  {journal} {\bibinfo  {journal} {Phys. Rev. X}\ }\textbf {\bibinfo {volume}
  {12}},\ \bibinfo {pages} {010501} (\bibinfo {year} {2022})}\BibitemShut
  {NoStop}%
\bibitem [{\citenamefont {Ramaswamy}(2017)}]{2017_Ramaswamy_Active}%
  \BibitemOpen
  \bibfield  {author} {\bibinfo {author} {\bibfnamefont {S.}~\bibnamefont
  {Ramaswamy}},\ }\bibfield  {title} {\bibinfo {title} {Active matter},\ }\href
  {https://doi.org/10.1088/1742-5468/aa6bc5} {\bibfield  {journal} {\bibinfo
  {journal} {J. Stat. Mech.}\ }\textbf {\bibinfo {volume} {2017}},\ \bibinfo
  {pages} {054002} (\bibinfo {year} {2017})}\BibitemShut {NoStop}%
\bibitem [{\citenamefont {Ramaswamy}(2010)}]{2010_Ramaswamy_The}%
  \BibitemOpen
  \bibfield  {author} {\bibinfo {author} {\bibfnamefont {S.}~\bibnamefont
  {Ramaswamy}},\ }\bibfield  {title} {\bibinfo {title} {The mechanics and
  statistics of active matter},\ }\href
  {https://doi.org/10.1146/annurev-conmatphys-070909-104101} {\bibfield
  {journal} {\bibinfo  {journal} {Annu. Rev. Condens. Matter Phys.}\ }\textbf
  {\bibinfo {volume} {1}},\ \bibinfo {pages} {323} (\bibinfo {year}
  {2010})}\BibitemShut {NoStop}%
\bibitem [{\citenamefont {Marchetti}\ \emph {et~al.}(2013)\citenamefont
  {Marchetti}, \citenamefont {Joanny}, \citenamefont {Ramaswamy}, \citenamefont
  {Liverpool}, \citenamefont {Prost}, \citenamefont {Rao},\ and\ \citenamefont
  {Simha}}]{2013_Marchetti_Hydrodynamics}%
  \BibitemOpen
  \bibfield  {author} {\bibinfo {author} {\bibfnamefont {M.~C.}\ \bibnamefont
  {Marchetti}}, \bibinfo {author} {\bibfnamefont {J.~F.}\ \bibnamefont
  {Joanny}}, \bibinfo {author} {\bibfnamefont {S.}~\bibnamefont {Ramaswamy}},
  \bibinfo {author} {\bibfnamefont {T.~B.}\ \bibnamefont {Liverpool}}, \bibinfo
  {author} {\bibfnamefont {J.}~\bibnamefont {Prost}}, \bibinfo {author}
  {\bibfnamefont {M.}~\bibnamefont {Rao}},\ and\ \bibinfo {author}
  {\bibfnamefont {R.~A.}\ \bibnamefont {Simha}},\ }\bibfield  {title} {\bibinfo
  {title} {Hydrodynamics of soft active matter},\ }\href
  {https://doi.org/10.1103/RevModPhys.85.1143} {\bibfield  {journal} {\bibinfo
  {journal} {Rev. Mod. Phys.}\ }\textbf {\bibinfo {volume} {85}},\ \bibinfo
  {pages} {1143} (\bibinfo {year} {2013})}\BibitemShut {NoStop}%
\bibitem [{\citenamefont {Vicsek}\ \emph {et~al.}(1995)\citenamefont {Vicsek},
  \citenamefont {Czir{\'o}k}, \citenamefont {Ben-Jacob}, \citenamefont
  {Cohen},\ and\ \citenamefont {Shochet}}]{1995_Vicsek_Novel}%
  \BibitemOpen
  \bibfield  {author} {\bibinfo {author} {\bibfnamefont {T.}~\bibnamefont
  {Vicsek}}, \bibinfo {author} {\bibfnamefont {A.}~\bibnamefont {Czir{\'o}k}},
  \bibinfo {author} {\bibfnamefont {E.}~\bibnamefont {Ben-Jacob}}, \bibinfo
  {author} {\bibfnamefont {I.}~\bibnamefont {Cohen}},\ and\ \bibinfo {author}
  {\bibfnamefont {O.}~\bibnamefont {Shochet}},\ }\bibfield  {title} {\bibinfo
  {title} {Novel type of phase transition in a system of self-driven
  particles},\ }\href {https://doi.org/10.1103/PhysRevLett.75.1226} {\bibfield
  {journal} {\bibinfo  {journal} {Phys. Rev. Lett.}\ }\textbf {\bibinfo
  {volume} {75}},\ \bibinfo {pages} {1226} (\bibinfo {year}
  {1995})}\BibitemShut {NoStop}%
\bibitem [{\citenamefont {Deseigne}\ \emph {et~al.}(2012)\citenamefont
  {Deseigne}, \citenamefont {Léonard}, \citenamefont {Dauchot},\ and\
  \citenamefont {Chaté}}]{2012_Deseigne_Vibrated}%
  \BibitemOpen
  \bibfield  {author} {\bibinfo {author} {\bibfnamefont {J.}~\bibnamefont
  {Deseigne}}, \bibinfo {author} {\bibfnamefont {S.}~\bibnamefont {Léonard}},
  \bibinfo {author} {\bibfnamefont {O.}~\bibnamefont {Dauchot}},\ and\ \bibinfo
  {author} {\bibfnamefont {H.}~\bibnamefont {Chaté}},\ }\bibfield  {title}
  {\bibinfo {title} {Vibrated polar disks: Spontaneous motion, binary
  collisions, and collective dynamics},\ }\href
  {https://doi.org/10.1039/C2SM25186H} {\bibfield  {journal} {\bibinfo
  {journal} {Soft Matter}\ }\textbf {\bibinfo {volume} {8}},\ \bibinfo {pages}
  {5629} (\bibinfo {year} {2012})}\BibitemShut {NoStop}%
\bibitem [{\citenamefont {Deseigne}\ \emph {et~al.}(2010)\citenamefont
  {Deseigne}, \citenamefont {Dauchot},\ and\ \citenamefont
  {Chat{\'e}}}]{2010_Deseigne_Collective}%
  \BibitemOpen
  \bibfield  {author} {\bibinfo {author} {\bibfnamefont {J.}~\bibnamefont
  {Deseigne}}, \bibinfo {author} {\bibfnamefont {O.}~\bibnamefont {Dauchot}},\
  and\ \bibinfo {author} {\bibfnamefont {H.}~\bibnamefont {Chat{\'e}}},\
  }\bibfield  {title} {\bibinfo {title} {Collective motion of vibrated polar
  disks},\ }\href {https://doi.org/10.1103/PhysRevLett.105.098001} {\bibfield
  {journal} {\bibinfo  {journal} {Phys. Rev. Lett.}\ }\textbf {\bibinfo
  {volume} {105}},\ \bibinfo {pages} {098001} (\bibinfo {year}
  {2010})}\BibitemShut {NoStop}%
\bibitem [{\citenamefont {Wang}\ \emph {et~al.}(2021)\citenamefont {Wang},
  \citenamefont {Phan}, \citenamefont {Li}, \citenamefont {Wombacher},
  \citenamefont {Qu}, \citenamefont {Peng}, \citenamefont {Chen}, \citenamefont
  {Goldman}, \citenamefont {Levin}, \citenamefont {Austin},\ and\ \citenamefont
  {Liu}}]{2021_Wang_Emergent}%
  \BibitemOpen
  \bibfield  {author} {\bibinfo {author} {\bibfnamefont {G.}~\bibnamefont
  {Wang}}, \bibinfo {author} {\bibfnamefont {T.~V.}\ \bibnamefont {Phan}},
  \bibinfo {author} {\bibfnamefont {S.}~\bibnamefont {Li}}, \bibinfo {author}
  {\bibfnamefont {M.}~\bibnamefont {Wombacher}}, \bibinfo {author}
  {\bibfnamefont {J.}~\bibnamefont {Qu}}, \bibinfo {author} {\bibfnamefont
  {Y.}~\bibnamefont {Peng}}, \bibinfo {author} {\bibfnamefont {G.}~\bibnamefont
  {Chen}}, \bibinfo {author} {\bibfnamefont {D.~I.}\ \bibnamefont {Goldman}},
  \bibinfo {author} {\bibfnamefont {S.~A.}\ \bibnamefont {Levin}}, \bibinfo
  {author} {\bibfnamefont {R.~H.}\ \bibnamefont {Austin}},\ and\ \bibinfo
  {author} {\bibfnamefont {L.}~\bibnamefont {Liu}},\ }\bibfield  {title}
  {\bibinfo {title} {Emergent field-driven robot swarm states},\ }\href
  {https://doi.org/10.1103/PhysRevLett.126.108002} {\bibfield  {journal}
  {\bibinfo  {journal} {Phys. Rev. Lett.}\ }\textbf {\bibinfo {volume} {126}},\
  \bibinfo {pages} {108002} (\bibinfo {year} {2021})}\BibitemShut {NoStop}%
\bibitem [{\citenamefont {Webster-Wood}\ \emph {et~al.}(2022)\citenamefont
  {Webster-Wood}, \citenamefont {Guix}, \citenamefont {Xu}, \citenamefont
  {Behkam}, \citenamefont {Sato}, \citenamefont {Sarkar}, \citenamefont
  {Sanchez}, \citenamefont {Shimizu},\ and\ \citenamefont
  {Parker}}]{2022_Wood_Biohybrid}%
  \BibitemOpen
  \bibfield  {author} {\bibinfo {author} {\bibfnamefont {V.~A.}\ \bibnamefont
  {Webster-Wood}}, \bibinfo {author} {\bibfnamefont {M.}~\bibnamefont {Guix}},
  \bibinfo {author} {\bibfnamefont {N.~W.}\ \bibnamefont {Xu}}, \bibinfo
  {author} {\bibfnamefont {B.}~\bibnamefont {Behkam}}, \bibinfo {author}
  {\bibfnamefont {H.}~\bibnamefont {Sato}}, \bibinfo {author} {\bibfnamefont
  {D.}~\bibnamefont {Sarkar}}, \bibinfo {author} {\bibfnamefont
  {S.}~\bibnamefont {Sanchez}}, \bibinfo {author} {\bibfnamefont
  {M.}~\bibnamefont {Shimizu}},\ and\ \bibinfo {author} {\bibfnamefont {K.~K.}\
  \bibnamefont {Parker}},\ }\bibfield  {title} {\bibinfo {title} {Biohybrid
  robots: Recent progress, challenges, and perspectives},\ }\href
  {https://doi.org/10.1088/1748-3190/ac9c3b} {\bibfield  {journal} {\bibinfo
  {journal} {Bioinspir. Biomim.}\ }\textbf {\bibinfo {volume} {18}},\ \bibinfo
  {pages} {015001} (\bibinfo {year} {2022})}\BibitemShut {NoStop}%
\bibitem [{\citenamefont {Khasseh}\ \emph {et~al.}(2023)\citenamefont
  {Khasseh}, \citenamefont {Wald}, \citenamefont {Moessner}, \citenamefont
  {Weber},\ and\ \citenamefont {Heyl}}]{2023_Khasseh_Active}%
  \BibitemOpen
  \bibfield  {author} {\bibinfo {author} {\bibfnamefont {R.}~\bibnamefont
  {Khasseh}}, \bibinfo {author} {\bibfnamefont {S.}~\bibnamefont {Wald}},
  \bibinfo {author} {\bibfnamefont {R.}~\bibnamefont {Moessner}}, \bibinfo
  {author} {\bibfnamefont {C.~A.}\ \bibnamefont {Weber}},\ and\ \bibinfo
  {author} {\bibfnamefont {M.}~\bibnamefont {Heyl}},\ }\href@noop {} {\bibinfo
  {title} {Active quantum flocks}} (\bibinfo {year} {2023}),\ \Eprint
  {https://arxiv.org/abs/2308.01603v2} {arXiv:2308.01603v2 [quant-ph]}
  \BibitemShut {NoStop}%
\bibitem [{\citenamefont {Frangipane}\ \emph {et~al.}(2018)\citenamefont
  {Frangipane}, \citenamefont {Dell'Arciprete}, \citenamefont {Petracchini},
  \citenamefont {Maggi}, \citenamefont {Saglimbeni}, \citenamefont {Bianchi},
  \citenamefont {Vizsnyiczai}, \citenamefont {Bernardini},\ and\ \citenamefont
  {{Di Leonardo}}}]{2018_Frangipane_Dynamic}%
  \BibitemOpen
  \bibfield  {author} {\bibinfo {author} {\bibfnamefont {G.}~\bibnamefont
  {Frangipane}}, \bibinfo {author} {\bibfnamefont {D.}~\bibnamefont
  {Dell'Arciprete}}, \bibinfo {author} {\bibfnamefont {S.}~\bibnamefont
  {Petracchini}}, \bibinfo {author} {\bibfnamefont {C.}~\bibnamefont {Maggi}},
  \bibinfo {author} {\bibfnamefont {F.}~\bibnamefont {Saglimbeni}}, \bibinfo
  {author} {\bibfnamefont {S.}~\bibnamefont {Bianchi}}, \bibinfo {author}
  {\bibfnamefont {G.}~\bibnamefont {Vizsnyiczai}}, \bibinfo {author}
  {\bibfnamefont {M.~L.}\ \bibnamefont {Bernardini}},\ and\ \bibinfo {author}
  {\bibfnamefont {R.}~\bibnamefont {{Di Leonardo}}},\ }\bibfield  {title}
  {\bibinfo {title} {Dynamic density shaping of photokinetic \textit{E.
  coli}},\ }\href {https://doi.org/10.7554/eLife.36608} {\bibfield  {journal}
  {\bibinfo  {journal} {eLife}\ }\textbf {\bibinfo {volume} {7}},\ \bibinfo
  {pages} {e36608} (\bibinfo {year} {2018})}\BibitemShut {NoStop}%
\bibitem [{\citenamefont {Fodor}\ and\ \citenamefont
  {Cates}(2021)}]{2021_Fodor_Active}%
  \BibitemOpen
  \bibfield  {author} {\bibinfo {author} {\bibfnamefont {{\'E}.}~\bibnamefont
  {Fodor}}\ and\ \bibinfo {author} {\bibfnamefont {M.~E.}\ \bibnamefont
  {Cates}},\ }\bibfield  {title} {\bibinfo {title} {Active engines
  thermodynamics moves forward},\ }\href
  {https://doi.org/10.1209/0295-5075/134/10003} {\bibfield  {journal} {\bibinfo
   {journal} {EPL}\ }\textbf {\bibinfo {volume} {134}},\ \bibinfo {pages}
  {10003} (\bibinfo {year} {2021})}\BibitemShut {NoStop}%
\bibitem [{\citenamefont {Cates}(2022)}]{2022_Cates_Active}%
  \BibitemOpen
  \bibfield  {author} {\bibinfo {author} {\bibfnamefont {M.~E.}\ \bibnamefont
  {Cates}},\ }\bibfield  {title} {\bibinfo {title} {Active field theories},\
  }in\ \href {https://doi.org/10.1093/oso/9780192858313.003.0006} {\emph
  {\bibinfo {booktitle} {{Active Matter and Nonequilibrium Statistical Physics:
  Lecture Notes of the Les Houches Summer School: Volume 112, September
  2018}}}}\ (\bibinfo  {publisher} {Oxford University Press},\ \bibinfo {year}
  {2022})\ Chap.~\bibinfo {chapter} {6}, p.\ \bibinfo {pages} {180}\BibitemShut
  {NoStop}%
\bibitem [{\citenamefont {Shaebani}\ \emph {et~al.}(2020)\citenamefont
  {Shaebani}, \citenamefont {Wysocki}, \citenamefont {Winkler}, \citenamefont
  {Gompper},\ and\ \citenamefont {Rieger}}]{2020_Shaebani_Computational}%
  \BibitemOpen
  \bibfield  {author} {\bibinfo {author} {\bibfnamefont {M.~R.}\ \bibnamefont
  {Shaebani}}, \bibinfo {author} {\bibfnamefont {A.}~\bibnamefont {Wysocki}},
  \bibinfo {author} {\bibfnamefont {R.~G.}\ \bibnamefont {Winkler}}, \bibinfo
  {author} {\bibfnamefont {G.}~\bibnamefont {Gompper}},\ and\ \bibinfo {author}
  {\bibfnamefont {H.}~\bibnamefont {Rieger}},\ }\bibfield  {title} {\bibinfo
  {title} {Computational models for active matter},\ }\href
  {https://doi.org/10.1038/s42254-020-0152-1} {\bibfield  {journal} {\bibinfo
  {journal} {Nat. Rev. Phys.}\ }\textbf {\bibinfo {volume} {2}},\ \bibinfo
  {pages} {181} (\bibinfo {year} {2020})}\BibitemShut {NoStop}%
\bibitem [{\citenamefont {Klamser}\ \emph {et~al.}(2021)\citenamefont
  {Klamser}, \citenamefont {Dauchot},\ and\ \citenamefont
  {Tailleur}}]{2021_Klamser_Kinetic}%
  \BibitemOpen
  \bibfield  {author} {\bibinfo {author} {\bibfnamefont {J.~U.}\ \bibnamefont
  {Klamser}}, \bibinfo {author} {\bibfnamefont {O.}~\bibnamefont {Dauchot}},\
  and\ \bibinfo {author} {\bibfnamefont {J.}~\bibnamefont {Tailleur}},\
  }\bibfield  {title} {\bibinfo {title} {Kinetic {Monte Carlo} algorithms for
  active matter systems},\ }\href
  {https://doi.org/10.1103/PhysRevLett.127.150602} {\bibfield  {journal}
  {\bibinfo  {journal} {Phys. Rev. Lett.}\ }\textbf {\bibinfo {volume} {127}},\
  \bibinfo {pages} {150602} (\bibinfo {year} {2021})}\BibitemShut {NoStop}%
\bibitem [{\citenamefont {Sabass}\ \emph {et~al.}(2023)\citenamefont {Sabass},
  \citenamefont {Winkler}, \citenamefont {Auth}, \citenamefont {Elgeti},
  \citenamefont {Fedosov}, \citenamefont {Ripoll}, \citenamefont
  {Vliegenthart},\ and\ \citenamefont {Gompper}}]{2023_Sabass_Computational}%
  \BibitemOpen
  \bibfield  {author} {\bibinfo {author} {\bibfnamefont {B.}~\bibnamefont
  {Sabass}}, \bibinfo {author} {\bibfnamefont {R.~G.}\ \bibnamefont {Winkler}},
  \bibinfo {author} {\bibfnamefont {T.}~\bibnamefont {Auth}}, \bibinfo {author}
  {\bibfnamefont {J.}~\bibnamefont {Elgeti}}, \bibinfo {author} {\bibfnamefont
  {D.~A.}\ \bibnamefont {Fedosov}}, \bibinfo {author} {\bibfnamefont
  {M.}~\bibnamefont {Ripoll}}, \bibinfo {author} {\bibfnamefont {G.~A.}\
  \bibnamefont {Vliegenthart}},\ and\ \bibinfo {author} {\bibfnamefont
  {G.}~\bibnamefont {Gompper}},\ }\bibfield  {title} {\bibinfo {title}
  {Computational physics of active matter},\ }in\ \href
  {https://doi.org/10.1039/9781839169465-00354} {\emph {\bibinfo {booktitle}
  {Out-of-equilibrium Soft Matter}}}\ (\bibinfo  {publisher} {The Royal Society
  of Chemistry},\ \bibinfo {year} {2023})\ Chap.~\bibinfo {chapter} {10}, p.\
  \bibinfo {pages} {354}\BibitemShut {NoStop}%
\bibitem [{\citenamefont {Cates}\ and\ \citenamefont
  {Tailleur}(2015)}]{2015_Cates_Motility}%
  \BibitemOpen
  \bibfield  {author} {\bibinfo {author} {\bibfnamefont {M.~E.}\ \bibnamefont
  {Cates}}\ and\ \bibinfo {author} {\bibfnamefont {J.}~\bibnamefont
  {Tailleur}},\ }\bibfield  {title} {\bibinfo {title} {Motility-induced phase
  separation},\ }\href
  {https://doi.org/10.1146/annurev-conmatphys-031214-014710} {\bibfield
  {journal} {\bibinfo  {journal} {Annu. Rev. Condens. Matter Phys.}\ }\textbf
  {\bibinfo {volume} {6}},\ \bibinfo {pages} {219} (\bibinfo {year}
  {2015})}\BibitemShut {NoStop}%
\bibitem [{\citenamefont {O'Byrne}\ \emph {et~al.}(2023)\citenamefont
  {O'Byrne}, \citenamefont {Solon}, \citenamefont {Tailleur},\ and\
  \citenamefont {Zhao}}]{2023_Byrne_An}%
  \BibitemOpen
  \bibfield  {author} {\bibinfo {author} {\bibfnamefont {J.}~\bibnamefont
  {O'Byrne}}, \bibinfo {author} {\bibfnamefont {A.}~\bibnamefont {Solon}},
  \bibinfo {author} {\bibfnamefont {J.}~\bibnamefont {Tailleur}},\ and\
  \bibinfo {author} {\bibfnamefont {Y.}~\bibnamefont {Zhao}},\ }\bibinfo
  {title} {An introduction to motility-induced phase separation},\ in\ \href
  {https://doi.org/10.1039/9781839169465-00107} {\emph {\bibinfo {booktitle}
  {Out-of-equilibrium Soft Matter}}}\ (\bibinfo  {publisher} {The Royal Society
  of Chemistry},\ \bibinfo {year} {2023})\ Chap.~\bibinfo {chapter} {4}, p.\
  \bibinfo {pages} {107–150}\BibitemShut {NoStop}%
\bibitem [{\citenamefont {Bertini}\ \emph {et~al.}(2015)\citenamefont
  {Bertini}, \citenamefont {{De Sole}}, \citenamefont {Gabrielli},
  \citenamefont {Jona-Lasinio},\ and\ \citenamefont
  {Landim}}]{2015_Bertini_Macroscopic}%
  \BibitemOpen
  \bibfield  {author} {\bibinfo {author} {\bibfnamefont {L.}~\bibnamefont
  {Bertini}}, \bibinfo {author} {\bibfnamefont {A.}~\bibnamefont {{De Sole}}},
  \bibinfo {author} {\bibfnamefont {D.}~\bibnamefont {Gabrielli}}, \bibinfo
  {author} {\bibfnamefont {G.}~\bibnamefont {Jona-Lasinio}},\ and\ \bibinfo
  {author} {\bibfnamefont {C.}~\bibnamefont {Landim}},\ }\bibfield  {title}
  {\bibinfo {title} {Macroscopic fluctuation theory},\ }\href
  {https://doi.org/10.1103/RevModPhys.87.593} {\bibfield  {journal} {\bibinfo
  {journal} {Rev. Mod. Phys.}\ }\textbf {\bibinfo {volume} {87}},\ \bibinfo
  {pages} {593} (\bibinfo {year} {2015})}\BibitemShut {NoStop}%
\bibitem [{\citenamefont {Doyon}(2018)}]{2018_Doyon_Exact}%
  \BibitemOpen
  \bibfield  {author} {\bibinfo {author} {\bibfnamefont {B.}~\bibnamefont
  {Doyon}},\ }\bibfield  {title} {\bibinfo {title} {Exact large-scale
  correlations in integrable systems out of equilibrium},\ }\href
  {https://doi.org/10.21468/SciPostPhys.5.5.054} {\bibfield  {journal}
  {\bibinfo  {journal} {SciPost Phys.}\ }\textbf {\bibinfo {volume} {5}},\
  \bibinfo {pages} {054} (\bibinfo {year} {2018})}\BibitemShut {NoStop}%
\bibitem [{\citenamefont {Spohn}(2014)}]{2014_Spohn_Nonlinear}%
  \BibitemOpen
  \bibfield  {author} {\bibinfo {author} {\bibfnamefont {H.}~\bibnamefont
  {Spohn}},\ }\bibfield  {title} {\bibinfo {title} {Nonlinear fluctuating
  hydrodynamics for anharmonic chains},\ }\href
  {https://doi.org/10.1007/s10955-014-0933-y} {\bibfield  {journal} {\bibinfo
  {journal} {J. Stat. Phys.}\ }\textbf {\bibinfo {volume} {154}},\ \bibinfo
  {pages} {1191} (\bibinfo {year} {2014})}\BibitemShut {NoStop}%
\bibitem [{\citenamefont {Bertini}\ \emph {et~al.}(2016)\citenamefont
  {Bertini}, \citenamefont {Collura}, \citenamefont {{De Nardis}},\ and\
  \citenamefont {Fagotti}}]{2016_Bertini_Transport}%
  \BibitemOpen
  \bibfield  {author} {\bibinfo {author} {\bibfnamefont {B.}~\bibnamefont
  {Bertini}}, \bibinfo {author} {\bibfnamefont {M.}~\bibnamefont {Collura}},
  \bibinfo {author} {\bibfnamefont {J.}~\bibnamefont {{De Nardis}}},\ and\
  \bibinfo {author} {\bibfnamefont {M.}~\bibnamefont {Fagotti}},\ }\bibfield
  {title} {\bibinfo {title} {Transport in out-of-equilibrium {XXZ} chains:
  Exact profiles of charges and currents},\ }\href
  {https://doi.org/10.1103/PhysRevLett.117.207201} {\bibfield  {journal}
  {\bibinfo  {journal} {Phys. Rev. Lett.}\ }\textbf {\bibinfo {volume} {117}},\
  \bibinfo {pages} {207201} (\bibinfo {year} {2016})}\BibitemShut {NoStop}%
\bibitem [{\citenamefont {{Castro-Alvaredo}}\ \emph {et~al.}(2016)\citenamefont
  {{Castro-Alvaredo}}, \citenamefont {Doyon},\ and\ \citenamefont
  {Yoshimura}}]{2016_Castro_Emergent}%
  \BibitemOpen
  \bibfield  {author} {\bibinfo {author} {\bibfnamefont {O.~A.}\ \bibnamefont
  {{Castro-Alvaredo}}}, \bibinfo {author} {\bibfnamefont {B.}~\bibnamefont
  {Doyon}},\ and\ \bibinfo {author} {\bibfnamefont {T.}~\bibnamefont
  {Yoshimura}},\ }\bibfield  {title} {\bibinfo {title} {Emergent hydrodynamics
  in integrable quantum systems out of equilibrium},\ }\href
  {https://doi.org/10.1103/PhysRevX.6.041065} {\bibfield  {journal} {\bibinfo
  {journal} {Phys. Rev. X}\ }\textbf {\bibinfo {volume} {6}},\ \bibinfo {pages}
  {041065} (\bibinfo {year} {2016})}\BibitemShut {NoStop}%
\bibitem [{\citenamefont {{De Nardis}}\ \emph {et~al.}(2018)\citenamefont {{De
  Nardis}}, \citenamefont {Bernard},\ and\ \citenamefont
  {Doyon}}]{2018_Nardis_Hydrodynamic}%
  \BibitemOpen
  \bibfield  {author} {\bibinfo {author} {\bibfnamefont {J.}~\bibnamefont {{De
  Nardis}}}, \bibinfo {author} {\bibfnamefont {D.}~\bibnamefont {Bernard}},\
  and\ \bibinfo {author} {\bibfnamefont {B.}~\bibnamefont {Doyon}},\ }\bibfield
   {title} {\bibinfo {title} {Hydrodynamic diffusion in integrable systems},\
  }\href {https://doi.org/10.1103/PhysRevLett.121.160603} {\bibfield  {journal}
  {\bibinfo  {journal} {Phys. Rev. Lett.}\ }\textbf {\bibinfo {volume} {121}},\
  \bibinfo {pages} {160603} (\bibinfo {year} {2018})}\BibitemShut {NoStop}%
\bibitem [{\citenamefont {Simha}\ and\ \citenamefont
  {Ramaswamy}(2002)}]{2002_Simha_Hydrodynamic}%
  \BibitemOpen
  \bibfield  {author} {\bibinfo {author} {\bibfnamefont {R.~A.}\ \bibnamefont
  {Simha}}\ and\ \bibinfo {author} {\bibfnamefont {S.}~\bibnamefont
  {Ramaswamy}},\ }\bibfield  {title} {\bibinfo {title} {Hydrodynamic
  fluctuations and instabilities in ordered suspensions of self-propelled
  particles},\ }\href {https://doi.org/10.1103/PhysRevLett.89.058101}
  {\bibfield  {journal} {\bibinfo  {journal} {Phys. Rev. Lett.}\ }\textbf
  {\bibinfo {volume} {89}},\ \bibinfo {pages} {058101} (\bibinfo {year}
  {2002})}\BibitemShut {NoStop}%
\bibitem [{\citenamefont {Toner}\ \emph {et~al.}(2005)\citenamefont {Toner},
  \citenamefont {Tu},\ and\ \citenamefont
  {Ramaswamy}}]{2005_Toner_Hydrodynamics}%
  \BibitemOpen
  \bibfield  {author} {\bibinfo {author} {\bibfnamefont {J.}~\bibnamefont
  {Toner}}, \bibinfo {author} {\bibfnamefont {Y.}~\bibnamefont {Tu}},\ and\
  \bibinfo {author} {\bibfnamefont {S.}~\bibnamefont {Ramaswamy}},\ }\bibfield
  {title} {\bibinfo {title} {Hydrodynamics and phases of flocks},\ }\href
  {https://doi.org/10.1016/j.aop.2005.04.011} {\bibfield  {journal} {\bibinfo
  {journal} {Ann. Phys.}\ }\textbf {\bibinfo {volume} {318}},\ \bibinfo {pages}
  {170} (\bibinfo {year} {2005})}\BibitemShut {NoStop}%
\bibitem [{\citenamefont {J\"ulicher}\ \emph {et~al.}(2018)\citenamefont
  {J\"ulicher}, \citenamefont {Grill},\ and\ \citenamefont
  {Salbreux}}]{2018_Julicher_Hydrodynamic}%
  \BibitemOpen
  \bibfield  {author} {\bibinfo {author} {\bibfnamefont {F.}~\bibnamefont
  {J\"ulicher}}, \bibinfo {author} {\bibfnamefont {S.~W.}\ \bibnamefont
  {Grill}},\ and\ \bibinfo {author} {\bibfnamefont {G.}~\bibnamefont
  {Salbreux}},\ }\bibfield  {title} {\bibinfo {title} {Hydrodynamic theory of
  active matter},\ }\href {https://doi.org/10.1088/1361-6633/aab6bb} {\bibfield
   {journal} {\bibinfo  {journal} {Rep. Prog. Phys.}\ }\textbf {\bibinfo
  {volume} {81}},\ \bibinfo {pages} {076601} (\bibinfo {year}
  {2018})}\BibitemShut {NoStop}%
\bibitem [{\citenamefont {Partridge}\ and\ \citenamefont
  {Lee}(2019)}]{2019_Partridge_Critical}%
  \BibitemOpen
  \bibfield  {author} {\bibinfo {author} {\bibfnamefont {B.}~\bibnamefont
  {Partridge}}\ and\ \bibinfo {author} {\bibfnamefont {C.~F.}\ \bibnamefont
  {Lee}},\ }\bibfield  {title} {\bibinfo {title} {Critical motility-induced
  phase separation belongs to the {Ising} universality class},\ }\href
  {https://doi.org/10.1103/PhysRevLett.123.068002} {\bibfield  {journal}
  {\bibinfo  {journal} {Phys. Rev. Lett.}\ }\textbf {\bibinfo {volume} {123}},\
  \bibinfo {pages} {068002} (\bibinfo {year} {2019})}\BibitemShut {NoStop}%
\bibitem [{\citenamefont {Langford}\ and\ \citenamefont
  {Omar}(2024)}]{Langford2023}%
  \BibitemOpen
  \bibfield  {author} {\bibinfo {author} {\bibfnamefont {L.}~\bibnamefont
  {Langford}}\ and\ \bibinfo {author} {\bibfnamefont {A.~K.}\ \bibnamefont
  {Omar}},\ }\bibfield  {title} {\bibinfo {title} {Theory of capillary tension
  and interfacial dynamics of motility-induced phases},\ }\href
  {https://doi.org/10.1103/PhysRevE.110.054604} {\bibfield  {journal} {\bibinfo
   {journal} {Phys. Rev. E}\ }\textbf {\bibinfo {volume} {110}},\ \bibinfo
  {pages} {054604} (\bibinfo {year} {2024})}\BibitemShut {NoStop}%
\bibitem [{\citenamefont {Kourbane-Houssene}\ \emph {et~al.}(2018)\citenamefont
  {Kourbane-Houssene}, \citenamefont {Erignoux}, \citenamefont {Bodineau},\
  and\ \citenamefont {Tailleur}}]{2018_Houssene_Exact}%
  \BibitemOpen
  \bibfield  {author} {\bibinfo {author} {\bibfnamefont {M.}~\bibnamefont
  {Kourbane-Houssene}}, \bibinfo {author} {\bibfnamefont {C.}~\bibnamefont
  {Erignoux}}, \bibinfo {author} {\bibfnamefont {T.}~\bibnamefont {Bodineau}},\
  and\ \bibinfo {author} {\bibfnamefont {J.}~\bibnamefont {Tailleur}},\
  }\bibfield  {title} {\bibinfo {title} {Exact hydrodynamic description of
  active lattice gases},\ }\href
  {https://doi.org/10.1103/PhysRevLett.120.268003} {\bibfield  {journal}
  {\bibinfo  {journal} {Phys. Rev. Lett.}\ }\textbf {\bibinfo {volume} {120}},\
  \bibinfo {pages} {268003} (\bibinfo {year} {2018})}\BibitemShut {NoStop}%
\bibitem [{\citenamefont {Erignoux}(2024)}]{2024_Erignoux}%
  \BibitemOpen
  \bibfield  {author} {\bibinfo {author} {\bibfnamefont {C.}~\bibnamefont
  {Erignoux}},\ }\bibfield  {title} {\bibinfo {title} {On the hydrodynamics of
  active matter models on a lattice},\ }\href
  {https://doi.org/10.61102/1024-2953-mprf.2024.30.1.002} {\bibfield  {journal}
  {\bibinfo  {journal} {Markov Process. Related Fields}\ }\textbf {\bibinfo
  {volume} {30}},\ \bibinfo {pages} {57} (\bibinfo {year} {2024})}\BibitemShut
  {NoStop}%
\bibitem [{\citenamefont {Agranov}\ \emph {et~al.}(2021)\citenamefont
  {Agranov}, \citenamefont {Ro}, \citenamefont {Kafri},\ and\ \citenamefont
  {Lecomte}}]{2021_Agranov_Exact}%
  \BibitemOpen
  \bibfield  {author} {\bibinfo {author} {\bibfnamefont {T.}~\bibnamefont
  {Agranov}}, \bibinfo {author} {\bibfnamefont {S.}~\bibnamefont {Ro}},
  \bibinfo {author} {\bibfnamefont {Y.}~\bibnamefont {Kafri}},\ and\ \bibinfo
  {author} {\bibfnamefont {V.}~\bibnamefont {Lecomte}},\ }\bibfield  {title}
  {\bibinfo {title} {Exact fluctuating hydrodynamics of active lattice gases
  — typical fluctuations},\ }\href {https://doi.org/10.1088/1742-5468/ac1406}
  {\bibfield  {journal} {\bibinfo  {journal} {J. Stat. Mech.}\ }\textbf
  {\bibinfo {volume} {2021}},\ \bibinfo {pages} {083208} (\bibinfo {year}
  {2021})}\BibitemShut {NoStop}%
\bibitem [{\citenamefont {Agranov}\ \emph {et~al.}(2023)\citenamefont
  {Agranov}, \citenamefont {Ro}, \citenamefont {Kafri},\ and\ \citenamefont
  {Lecomte}}]{2023_Agranov_Macroscopic}%
  \BibitemOpen
  \bibfield  {author} {\bibinfo {author} {\bibfnamefont {T.}~\bibnamefont
  {Agranov}}, \bibinfo {author} {\bibfnamefont {S.}~\bibnamefont {Ro}},
  \bibinfo {author} {\bibfnamefont {Y.}~\bibnamefont {Kafri}},\ and\ \bibinfo
  {author} {\bibfnamefont {V.}~\bibnamefont {Lecomte}},\ }\bibfield  {title}
  {\bibinfo {title} {Macroscopic fluctuation theory and current fluctuations in
  active lattice gases},\ }\href
  {https://doi.org/10.21468/SciPostPhys.14.3.045} {\bibfield  {journal}
  {\bibinfo  {journal} {SciPost Phys.}\ }\textbf {\bibinfo {volume} {14}},\
  \bibinfo {pages} {045} (\bibinfo {year} {2023})}\BibitemShut {NoStop}%
\bibitem [{\citenamefont {Digregorio}\ \emph {et~al.}(2018)\citenamefont
  {Digregorio}, \citenamefont {Levis}, \citenamefont {Suma}, \citenamefont
  {Cugliandolo}, \citenamefont {Gonnella},\ and\ \citenamefont
  {Pagonabarraga}}]{2018_Digregorio_Full}%
  \BibitemOpen
  \bibfield  {author} {\bibinfo {author} {\bibfnamefont {P.}~\bibnamefont
  {Digregorio}}, \bibinfo {author} {\bibfnamefont {D.}~\bibnamefont {Levis}},
  \bibinfo {author} {\bibfnamefont {A.}~\bibnamefont {Suma}}, \bibinfo {author}
  {\bibfnamefont {L.~F.}\ \bibnamefont {Cugliandolo}}, \bibinfo {author}
  {\bibfnamefont {G.}~\bibnamefont {Gonnella}},\ and\ \bibinfo {author}
  {\bibfnamefont {I.}~\bibnamefont {Pagonabarraga}},\ }\bibfield  {title}
  {\bibinfo {title} {Full phase diagram of active {Brownian} disks: From
  melting to motility-induced phase separation},\ }\href
  {https://doi.org/10.1103/PhysRevLett.121.098003} {\bibfield  {journal}
  {\bibinfo  {journal} {Phys. Rev. Lett.}\ }\textbf {\bibinfo {volume} {121}},\
  \bibinfo {pages} {098003} (\bibinfo {year} {2018})}\BibitemShut {NoStop}%
\bibitem [{\citenamefont {Klamser}\ \emph {et~al.}(2018)\citenamefont
  {Klamser}, \citenamefont {Kapfer},\ and\ \citenamefont
  {W.~Krauth}}]{2018_Klamser_Thermodynamic}%
  \BibitemOpen
  \bibfield  {author} {\bibinfo {author} {\bibfnamefont {J.~U.}\ \bibnamefont
  {Klamser}}, \bibinfo {author} {\bibfnamefont {S.~C.}\ \bibnamefont
  {Kapfer}},\ and\ \bibinfo {author} {\bibfnamefont {W.}~\bibnamefont
  {W.~Krauth}},\ }\bibfield  {title} {\bibinfo {title} {Thermodynamic phases in
  two-dimensional active matter},\ }\href
  {https://doi.org/10.1038/s41467-018-07491-5} {\bibfield  {journal} {\bibinfo
  {journal} {Nat. Commun.}\ }\textbf {\bibinfo {volume} {9}},\ \bibinfo {pages}
  {5045} (\bibinfo {year} {2018})}\BibitemShut {NoStop}%
\bibitem [{\citenamefont {Caballero}\ \emph {et~al.}(2018)\citenamefont
  {Caballero}, \citenamefont {Nardini},\ and\ \citenamefont
  {Cates}}]{2018_Caballero_From}%
  \BibitemOpen
  \bibfield  {author} {\bibinfo {author} {\bibfnamefont {F.}~\bibnamefont
  {Caballero}}, \bibinfo {author} {\bibfnamefont {C.}~\bibnamefont {Nardini}},\
  and\ \bibinfo {author} {\bibfnamefont {M.~E.}\ \bibnamefont {Cates}},\
  }\bibfield  {title} {\bibinfo {title} {From bulk to microphase separation in
  scalar active matter: a perturbative renormalization group analysis},\ }\href
  {https://doi.org/10.1088/1742-5468/aaf321} {\bibfield  {journal} {\bibinfo
  {journal} {J. Stat. Mech.}\ }\textbf {\bibinfo {volume} {2018}},\ \bibinfo
  {pages} {123208} (\bibinfo {year} {2018})}\BibitemShut {NoStop}%
\bibitem [{\citenamefont {Szamel}\ and\ \citenamefont
  {Flenner}(2021)}]{2021_Szamel_Long}%
  \BibitemOpen
  \bibfield  {author} {\bibinfo {author} {\bibfnamefont {G.}~\bibnamefont
  {Szamel}}\ and\ \bibinfo {author} {\bibfnamefont {E.}~\bibnamefont
  {Flenner}},\ }\bibfield  {title} {\bibinfo {title} {Long-ranged velocity
  correlations in dense systems of self-propelled particles},\ }\href
  {https://doi.org/10.1209/0295-5075/133/60002} {\bibfield  {journal} {\bibinfo
   {journal} {EPL}\ }\textbf {\bibinfo {volume} {133}},\ \bibinfo {pages}
  {60002} (\bibinfo {year} {2021})}\BibitemShut {NoStop}%
\bibitem [{\citenamefont {Ferro}\ \emph {et~al.}(2019)\citenamefont {Ferro},
  \citenamefont {Can}, \citenamefont {Turner}, \citenamefont {ElShenawy},\ and\
  \citenamefont {Yildiz}}]{2019_Ferro_Kinesin}%
  \BibitemOpen
  \bibfield  {author} {\bibinfo {author} {\bibfnamefont {L.~S.}\ \bibnamefont
  {Ferro}}, \bibinfo {author} {\bibfnamefont {S.}~\bibnamefont {Can}}, \bibinfo
  {author} {\bibfnamefont {M.~A.}\ \bibnamefont {Turner}}, \bibinfo {author}
  {\bibfnamefont {M.~M.}\ \bibnamefont {ElShenawy}},\ and\ \bibinfo {author}
  {\bibfnamefont {A.}~\bibnamefont {Yildiz}},\ }\bibfield  {title} {\bibinfo
  {title} {Kinesin and dynein use distinct mechanisms to bypass obstacles},\
  }\href {https://doi.org/10.7554/eLife.48629} {\bibfield  {journal} {\bibinfo
  {journal} {eLife}\ }\textbf {\bibinfo {volume} {1}},\ \bibinfo {pages}
  {031080} (\bibinfo {year} {2019})}\BibitemShut {NoStop}%
\bibitem [{\citenamefont {Wienand}\ \emph {et~al.}(2024)\citenamefont
  {Wienand}, \citenamefont {Karch}, \citenamefont {Impertro}, \citenamefont
  {Schweizer}, \citenamefont {McCulloch}, \citenamefont {Vasseur},
  \citenamefont {Gopalakrishnan}, \citenamefont {Aidelsburger},\ and\
  \citenamefont {Bloch}}]{2024_Wienand_Emergence}%
  \BibitemOpen
  \bibfield  {author} {\bibinfo {author} {\bibfnamefont {J.~F.}\ \bibnamefont
  {Wienand}}, \bibinfo {author} {\bibfnamefont {S.}~\bibnamefont {Karch}},
  \bibinfo {author} {\bibfnamefont {A.}~\bibnamefont {Impertro}}, \bibinfo
  {author} {\bibfnamefont {C.}~\bibnamefont {Schweizer}}, \bibinfo {author}
  {\bibfnamefont {E.}~\bibnamefont {McCulloch}}, \bibinfo {author}
  {\bibfnamefont {R.}~\bibnamefont {Vasseur}}, \bibinfo {author} {\bibfnamefont
  {S.}~\bibnamefont {Gopalakrishnan}}, \bibinfo {author} {\bibfnamefont
  {M.}~\bibnamefont {Aidelsburger}},\ and\ \bibinfo {author} {\bibfnamefont
  {I.}~\bibnamefont {Bloch}},\ }\bibfield  {title} {\bibinfo {title} {Emergence
  of fluctuating hydrodynamics in chaotic quantum systems},\ }\href
  {https://doi.org/10.1038/s41567-024-02611-z} {\bibfield  {journal} {\bibinfo
  {journal} {Nat. Phys.}\ }\textbf {\bibinfo {volume} {20}},\ \bibinfo {pages}
  {1732} (\bibinfo {year} {2024})}\BibitemShut {NoStop}%
\bibitem [{\citenamefont {Siems}\ \emph {et~al.}(2012)\citenamefont {Siems},
  \citenamefont {Kreuter}, \citenamefont {Erbe}, \citenamefont {Schwierz},
  \citenamefont {Sengupta}, \citenamefont {Leiderer},\ and\ \citenamefont
  {Nielaba}}]{2012_Siems_Non}%
  \BibitemOpen
  \bibfield  {author} {\bibinfo {author} {\bibfnamefont {U.}~\bibnamefont
  {Siems}}, \bibinfo {author} {\bibfnamefont {C.}~\bibnamefont {Kreuter}},
  \bibinfo {author} {\bibfnamefont {A.}~\bibnamefont {Erbe}}, \bibinfo {author}
  {\bibfnamefont {N.}~\bibnamefont {Schwierz}}, \bibinfo {author}
  {\bibfnamefont {S.}~\bibnamefont {Sengupta}}, \bibinfo {author}
  {\bibfnamefont {P.}~\bibnamefont {Leiderer}},\ and\ \bibinfo {author}
  {\bibfnamefont {P.}~\bibnamefont {Nielaba}},\ }\bibfield  {title} {\bibinfo
  {title} {Non-monotonic crossover from single-file to regular diffusion in
  micro-channels},\ }\href {https://doi.org/10.1038/srep01015} {\bibfield
  {journal} {\bibinfo  {journal} {Sci. Rep.}\ }\textbf {\bibinfo {volume}
  {2}},\ \bibinfo {pages} {1015} (\bibinfo {year} {2012})}\BibitemShut
  {NoStop}%
\bibitem [{\citenamefont {Miron}\ \emph {et~al.}(2020)\citenamefont {Miron},
  \citenamefont {Mukamel},\ and\ \citenamefont {Posch}}]{2020_Miron_Phase}%
  \BibitemOpen
  \bibfield  {author} {\bibinfo {author} {\bibfnamefont {A.}~\bibnamefont
  {Miron}}, \bibinfo {author} {\bibfnamefont {D.}~\bibnamefont {Mukamel}},\
  and\ \bibinfo {author} {\bibfnamefont {H.~A.}\ \bibnamefont {Posch}},\
  }\bibfield  {title} {\bibinfo {title} {Phase transition in a {1D} driven
  tracer model},\ }\href {https://doi.org/10.1088/1742-5468/ab8c35} {\bibfield
  {journal} {\bibinfo  {journal} {J. Stat. Mech.}\ }\textbf {\bibinfo {volume}
  {2020}},\ \bibinfo {pages} {063216} (\bibinfo {year} {2020})}\BibitemShut
  {NoStop}%
\bibitem [{\citenamefont {Wilke}\ \emph {et~al.}(2018)\citenamefont {Wilke},
  \citenamefont {Reithmann},\ and\ \citenamefont {Frey}}]{2018_Wilke_Two}%
  \BibitemOpen
  \bibfield  {author} {\bibinfo {author} {\bibfnamefont {P.}~\bibnamefont
  {Wilke}}, \bibinfo {author} {\bibfnamefont {E.}~\bibnamefont {Reithmann}},\
  and\ \bibinfo {author} {\bibfnamefont {E.}~\bibnamefont {Frey}},\ }\bibfield
  {title} {\bibinfo {title} {Two-species active transport along cylindrical
  biofilaments is limited by emergent topological hindrance},\ }\href
  {https://doi.org/10.1103/PhysRevX.8.031063} {\bibfield  {journal} {\bibinfo
  {journal} {Phys Rev X}\ }\textbf {\bibinfo {volume} {8}},\ \bibinfo {pages}
  {031063} (\bibinfo {year} {2018})}\BibitemShut {NoStop}%
\bibitem [{\citenamefont {Nandi}\ \emph {et~al.}(2021)\citenamefont {Nandi},
  \citenamefont {Täuber},\ and\ \citenamefont {Priyanka}}]{2021_Nandi_Dynein}%
  \BibitemOpen
  \bibfield  {author} {\bibinfo {author} {\bibfnamefont {R.}~\bibnamefont
  {Nandi}}, \bibinfo {author} {\bibfnamefont {U.~C.}\ \bibnamefont {Täuber}},\
  and\ \bibinfo {author} {\bibnamefont {Priyanka}},\ }\bibfield  {title}
  {\bibinfo {title} {Dynein-inspired multilane exclusion process with open
  boundary conditions},\ }\href {https://doi.org/10.3390/e23101343} {\bibfield
  {journal} {\bibinfo  {journal} {Entropy}\ }\textbf {\bibinfo {volume} {23}},\
  \bibinfo {pages} {1343} (\bibinfo {year} {2021})}\BibitemShut {NoStop}%
\bibitem [{\citenamefont {Martin}\ \emph {et~al.}(1973)\citenamefont {Martin},
  \citenamefont {Siggia},\ and\ \citenamefont
  {Rose}}]{1973_Martin_Statistical}%
  \BibitemOpen
  \bibfield  {author} {\bibinfo {author} {\bibfnamefont {P.~C.}\ \bibnamefont
  {Martin}}, \bibinfo {author} {\bibfnamefont {E.~D.}\ \bibnamefont {Siggia}},\
  and\ \bibinfo {author} {\bibfnamefont {H.~A.}\ \bibnamefont {Rose}},\
  }\bibfield  {title} {\bibinfo {title} {Statistical dynamics of classical
  systems},\ }\href {https://doi.org/10.1103/PhysRevA.8.423} {\bibfield
  {journal} {\bibinfo  {journal} {Phys. Rev. A}\ }\textbf {\bibinfo {volume}
  {8}},\ \bibinfo {pages} {423} (\bibinfo {year} {1973})}\BibitemShut {NoStop}%
\bibitem [{\citenamefont {Janssen}(1976)}]{1976_Janssen_On}%
  \BibitemOpen
  \bibfield  {author} {\bibinfo {author} {\bibfnamefont {H.~K.}\ \bibnamefont
  {Janssen}},\ }\bibfield  {title} {\bibinfo {title} {On a lagrangean for
  classical field dynamics and renormalization group calculations of dynamical
  critical properties},\ }\href {https://doi.org/10.1007/BF01316547} {\bibfield
   {journal} {\bibinfo  {journal} {Z. Physik B}\ }\textbf {\bibinfo {volume}
  {23}},\ \bibinfo {pages} {377} (\bibinfo {year} {1976})}\BibitemShut
  {NoStop}%
\bibitem [{\citenamefont {{De Dominicis}}(1978)}]{1978_Dominicis_Dynamics}%
  \BibitemOpen
  \bibfield  {author} {\bibinfo {author} {\bibfnamefont {C.}~\bibnamefont {{De
  Dominicis}}},\ }\bibfield  {title} {\bibinfo {title} {Dynamics as a
  substitute for replicas in systems with quenched random impurities},\ }\href
  {https://doi.org/10.1103/PhysRevB.18.4913} {\bibfield  {journal} {\bibinfo
  {journal} {Phys. Rev. B}\ }\textbf {\bibinfo {volume} {18}},\ \bibinfo
  {pages} {4913} (\bibinfo {year} {1978})}\BibitemShut {NoStop}%
\bibitem [{\citenamefont {{De Dominicis}}\ and\ \citenamefont
  {Peliti}(1978)}]{1978_Dominicis_Field}%
  \BibitemOpen
  \bibfield  {author} {\bibinfo {author} {\bibfnamefont {C.}~\bibnamefont {{De
  Dominicis}}}\ and\ \bibinfo {author} {\bibfnamefont {L.}~\bibnamefont
  {Peliti}},\ }\bibfield  {title} {\bibinfo {title} {Field-theory
  renormalization and critical dynamics above {$T_c$}: Helium,
  antiferromagnets, and liquid-gas systems},\ }\href
  {https://doi.org/10.1103/PhysRevB.18.353} {\bibfield  {journal} {\bibinfo
  {journal} {Phys. Rev. B}\ }\textbf {\bibinfo {volume} {18}},\ \bibinfo
  {pages} {353} (\bibinfo {year} {1978})}\BibitemShut {NoStop}%
\bibitem [{\citenamefont {Saha}\ and\ \citenamefont
  {Sadhu}(2024)}]{2024_Saha_Large}%
  \BibitemOpen
  \bibfield  {author} {\bibinfo {author} {\bibfnamefont {S.}~\bibnamefont
  {Saha}}\ and\ \bibinfo {author} {\bibfnamefont {T.}~\bibnamefont {Sadhu}},\
  }\bibfield  {title} {\bibinfo {title} {Large deviations in the symmetric
  simple exclusion process with slow boundaries: A hydrodynamic perspective},\
  }\href {https://doi.org/10.21468/SciPostPhys.17.2.033} {\bibfield  {journal}
  {\bibinfo  {journal} {SciPost Phys.}\ }\textbf {\bibinfo {volume} {17}},\
  \bibinfo {pages} {033} (\bibinfo {year} {2024})}\BibitemShut {NoStop}%
\bibitem [{\citenamefont {Trefethen}(2000)}]{trefethen2000spectral}%
  \BibitemOpen
  \bibfield  {author} {\bibinfo {author} {\bibfnamefont {L.~N.}\ \bibnamefont
  {Trefethen}},\ }\href {https://doi.org/10.1137/1.9780898719598} {\emph
  {\bibinfo {title} {Spectral methods in MATLAB}}}\ (\bibinfo  {publisher}
  {SIAM Philadelphia},\ \bibinfo {year} {2000})\BibitemShut {NoStop}%
\bibitem [{\citenamefont {Solon}\ \emph
  {et~al.}(2018{\natexlab{a}})\citenamefont {Solon}, \citenamefont
  {Stenhammar}, \citenamefont {Cates}, \citenamefont {Kafri},\ and\
  \citenamefont {Tailleur}}]{2018_Solon_Generalized_PRE}%
  \BibitemOpen
  \bibfield  {author} {\bibinfo {author} {\bibfnamefont {A.~P.}\ \bibnamefont
  {Solon}}, \bibinfo {author} {\bibfnamefont {J.}~\bibnamefont {Stenhammar}},
  \bibinfo {author} {\bibfnamefont {M.~E.}\ \bibnamefont {Cates}}, \bibinfo
  {author} {\bibfnamefont {Y.}~\bibnamefont {Kafri}},\ and\ \bibinfo {author}
  {\bibfnamefont {J.}~\bibnamefont {Tailleur}},\ }\bibfield  {title} {\bibinfo
  {title} {Generalized thermodynamics of phase equilibria in scalar active
  matter},\ }\href {https://doi.org/10.1103/PhysRevE.97.020602} {\bibfield
  {journal} {\bibinfo  {journal} {Phys. Rev. E}\ }\textbf {\bibinfo {volume}
  {97}},\ \bibinfo {pages} {020602(R)} (\bibinfo {year}
  {2018}{\natexlab{a}})}\BibitemShut {NoStop}%
\bibitem [{\citenamefont {Solon}\ \emph
  {et~al.}(2018{\natexlab{b}})\citenamefont {Solon}, \citenamefont
  {Stenhammar}, \citenamefont {Cates}, \citenamefont {Kafri},\ and\
  \citenamefont {Tailleur}}]{2018_Solon_Generalized_NJP}%
  \BibitemOpen
  \bibfield  {author} {\bibinfo {author} {\bibfnamefont {A.~P.}\ \bibnamefont
  {Solon}}, \bibinfo {author} {\bibfnamefont {J.}~\bibnamefont {Stenhammar}},
  \bibinfo {author} {\bibfnamefont {M.~E.}\ \bibnamefont {Cates}}, \bibinfo
  {author} {\bibfnamefont {Y.}~\bibnamefont {Kafri}},\ and\ \bibinfo {author}
  {\bibfnamefont {J.}~\bibnamefont {Tailleur}},\ }\bibfield  {title} {\bibinfo
  {title} {Generalized thermodynamics of motility-induced phase separation:
  phase equilibria, laplace pressure, and change of ensembles},\ }\href
  {https://doi.org/10.1088/1367-2630/aaccdd} {\bibfield  {journal} {\bibinfo
  {journal} {New J. Phys.}\ }\textbf {\bibinfo {volume} {20}},\ \bibinfo
  {pages} {075001} (\bibinfo {year} {2018}{\natexlab{b}})}\BibitemShut
  {NoStop}%
\bibitem [{\citenamefont {Tjhung}\ \emph {et~al.}(2018)\citenamefont {Tjhung},
  \citenamefont {Nardini},\ and\ \citenamefont {Cates}}]{2018_Tjhung_Cluster}%
  \BibitemOpen
  \bibfield  {author} {\bibinfo {author} {\bibfnamefont {E.}~\bibnamefont
  {Tjhung}}, \bibinfo {author} {\bibfnamefont {C.}~\bibnamefont {Nardini}},\
  and\ \bibinfo {author} {\bibfnamefont {M.~E.}\ \bibnamefont {Cates}},\
  }\bibfield  {title} {\bibinfo {title} {Cluster phases and bubbly phase
  separation in active fluids: Reversal of the {Ostwald} process},\ }\href
  {https://doi.org/10.1103/PhysRevX.8.031080} {\bibfield  {journal} {\bibinfo
  {journal} {Phys. Rev. X}\ }\textbf {\bibinfo {volume} {8}},\ \bibinfo {pages}
  {031080} (\bibinfo {year} {2018})}\BibitemShut {NoStop}%
\bibitem [{\citenamefont {Nejad}\ and\ \citenamefont
  {Yeomans}(2023)}]{2023_Nejad_Spontaneous}%
  \BibitemOpen
  \bibfield  {author} {\bibinfo {author} {\bibfnamefont {M.~R.}\ \bibnamefont
  {Nejad}}\ and\ \bibinfo {author} {\bibfnamefont {J.~M.}\ \bibnamefont
  {Yeomans}},\ }\bibfield  {title} {\bibinfo {title} {Spontaneous rotation of
  active droplets in two and three dimensions},\ }\href
  {https://doi.org/10.1103/PRXLife.1.023008} {\bibfield  {journal} {\bibinfo
  {journal} {PRX Life}\ }\textbf {\bibinfo {volume} {1}},\ \bibinfo {pages}
  {023008} (\bibinfo {year} {2023})}\BibitemShut {NoStop}%
\bibitem [{S_M()}]{S_M}%
  \BibitemOpen
  \href@noop {} {}\bibinfo {note} {See Supplemental Material at \url{http://link.aps.org/supplemental/10.1103/PhysRevE.111.024128} for videos.}\BibitemShut {Stop}%
\bibitem [{\citenamefont {Mukherjee}\ \emph {et~al.}(2023)\citenamefont
  {Mukherjee}, \citenamefont {Raghu},\ and\ \citenamefont
  {Mohanty}}]{2023_Mukherjee_Nonexistence}%
  \BibitemOpen
  \bibfield  {author} {\bibinfo {author} {\bibfnamefont {I.}~\bibnamefont
  {Mukherjee}}, \bibinfo {author} {\bibfnamefont {A.}~\bibnamefont {Raghu}},\
  and\ \bibinfo {author} {\bibfnamefont {P.~K.}\ \bibnamefont {Mohanty}},\
  }\bibfield  {title} {\bibinfo {title} {Nonexistence of motility induced phase
  separation transition in one dimension},\ }\href
  {https://doi.org/10.21468/SciPostPhys.14.6.165} {\bibfield  {journal}
  {\bibinfo  {journal} {SciPost Phys.}\ }\textbf {\bibinfo {volume} {14}},\
  \bibinfo {pages} {165} (\bibinfo {year} {2023})}\BibitemShut {NoStop}%
\bibitem [{\citenamefont {Blondel}\ \emph {et~al.}(2021)\citenamefont
  {Blondel}, \citenamefont {Erignoux},\ and\ \citenamefont
  {Simon}}]{2021_Blondel_Stefan}%
  \BibitemOpen
  \bibfield  {author} {\bibinfo {author} {\bibfnamefont {O.}~\bibnamefont
  {Blondel}}, \bibinfo {author} {\bibfnamefont {C.}~\bibnamefont {Erignoux}},\
  and\ \bibinfo {author} {\bibfnamefont {M.}~\bibnamefont {Simon}},\ }\bibfield
   {title} {\bibinfo {title} {Stefan problem for a nonergodic facilitated
  exclusion process},\ }\href {https://doi.org/10.2140/pmp.2021.2.127}
  {\bibfield  {journal} {\bibinfo  {journal} {Probab. Math. Phys.}\ }\textbf
  {\bibinfo {volume} {2}},\ \bibinfo {pages} {127} (\bibinfo {year}
  {2021})}\BibitemShut {NoStop}%
\bibitem [{\citenamefont {Touzo}\ \emph {et~al.}(2023)\citenamefont {Touzo},
  \citenamefont {{Le Doussal}},\ and\ \citenamefont
  {Schehr}}]{2023_Touzo_Interacting}%
  \BibitemOpen
  \bibfield  {author} {\bibinfo {author} {\bibfnamefont {L.}~\bibnamefont
  {Touzo}}, \bibinfo {author} {\bibfnamefont {P.}~\bibnamefont {{Le
  Doussal}}},\ and\ \bibinfo {author} {\bibfnamefont {G.}~\bibnamefont
  {Schehr}},\ }\bibfield  {title} {\bibinfo {title} {Interacting, running and
  tumbling: The active {Dyson Brownian} motion},\ }\href
  {https://doi.org/10.1209/0295-5075/acdabb} {\bibfield  {journal} {\bibinfo
  {journal} {EPL}\ }\textbf {\bibinfo {volume} {142}},\ \bibinfo {pages}
  {61004} (\bibinfo {year} {2023})}\BibitemShut {NoStop}%
\bibitem [{\citenamefont {Das}\ \emph {et~al.}(2024)\citenamefont {Das},
  \citenamefont {Ghosh}, \citenamefont {Sadhu},\ and\ \citenamefont
  {Klamser}}]{2024_Das}%
  \BibitemOpen
  \bibfield  {author} {\bibinfo {author} {\bibfnamefont {S.}~\bibnamefont
  {Das}}, \bibinfo {author} {\bibfnamefont {S.}~\bibnamefont {Ghosh}}, \bibinfo
  {author} {\bibfnamefont {T.}~\bibnamefont {Sadhu}},\ and\ \bibinfo {author}
  {\bibfnamefont {J.~U.}\ \bibnamefont {Klamser}},\ }\href@noop {} {\bibinfo
  {title} {Role of kinematic constraints in the time reversal symmetry breaking
  of a model active matter}} (\bibinfo {year} {2024}),\ \Eprint
  {https://arxiv.org/abs/2409.10425} {arXiv:2409.10425 [cond-mat.soft]}
  \BibitemShut {NoStop}%
\bibitem [{\citenamefont {{Y-E Keta}}\ \emph {et~al.}(2022)\citenamefont {{Y-E
  Keta}}, \citenamefont {Jack},\ and\ \citenamefont
  {Berthier}}]{2022_Berthier_Disordered}%
  \BibitemOpen
  \bibfield  {author} {\bibinfo {author} {\bibnamefont {{Y-E Keta}}}, \bibinfo
  {author} {\bibfnamefont {R.~L.}\ \bibnamefont {Jack}},\ and\ \bibinfo
  {author} {\bibfnamefont {L.}~\bibnamefont {Berthier}},\ }\bibfield  {title}
  {\bibinfo {title} {Disordered collective motion in dense assemblies of
  persistent particles},\ }\href
  {https://doi.org/10.1103/PhysRevLett.129.048002} {\bibfield  {journal}
  {\bibinfo  {journal} {Phys. Rev. Lett.}\ }\textbf {\bibinfo {volume} {129}},\
  \bibinfo {pages} {048002} (\bibinfo {year} {2022})}\BibitemShut {NoStop}%
\bibitem [{\citenamefont {Mason}\ \emph {et~al.}(2023)\citenamefont {Mason},
  \citenamefont {Erignoux}, \citenamefont {Jack},\ and\ \citenamefont
  {Bruna}}]{2023_Mason_Exact}%
  \BibitemOpen
  \bibfield  {author} {\bibinfo {author} {\bibfnamefont {J.}~\bibnamefont
  {Mason}}, \bibinfo {author} {\bibfnamefont {C.}~\bibnamefont {Erignoux}},
  \bibinfo {author} {\bibfnamefont {R.~L.}\ \bibnamefont {Jack}},\ and\
  \bibinfo {author} {\bibfnamefont {M.}~\bibnamefont {Bruna}},\ }\bibfield
  {title} {\bibinfo {title} {Exact hydrodynamics and onset of phase separation
  for an active exclusion process},\ }\href
  {https://doi.org/10.1098/rspa.2023.0524} {\bibfield  {journal} {\bibinfo
  {journal} {Proc. R. Soc. A}\ }\textbf {\bibinfo {volume} {479}},\ \bibinfo
  {pages} {20230524} (\bibinfo {year} {2023})}\BibitemShut {NoStop}%
\bibitem [{\citenamefont {Bodineau}\ and\ \citenamefont
  {Derrida}(2005)}]{2005_Bodineau_Distribution}%
  \BibitemOpen
  \bibfield  {author} {\bibinfo {author} {\bibfnamefont {T.}~\bibnamefont
  {Bodineau}}\ and\ \bibinfo {author} {\bibfnamefont {B.}~\bibnamefont
  {Derrida}},\ }\bibfield  {title} {\bibinfo {title} {Distribution of current
  in nonequilibrium diffusive systems and phase transitions},\ }\href
  {https://doi.org/10.1103/PhysRevE.72.066110} {\bibfield  {journal} {\bibinfo
  {journal} {Phys. Rev. E}\ }\textbf {\bibinfo {volume} {72}},\ \bibinfo
  {pages} {066110} (\bibinfo {year} {2005})}\BibitemShut {NoStop}%
\bibitem [{\citenamefont {Bunin}\ \emph {et~al.}(2012)\citenamefont {Bunin},
  \citenamefont {Kafri},\ and\ \citenamefont {Podolsky}}]{2012_Bunin_Non}%
  \BibitemOpen
  \bibfield  {author} {\bibinfo {author} {\bibfnamefont {G.}~\bibnamefont
  {Bunin}}, \bibinfo {author} {\bibfnamefont {Y.}~\bibnamefont {Kafri}},\ and\
  \bibinfo {author} {\bibfnamefont {D.}~\bibnamefont {Podolsky}},\ }\bibfield
  {title} {\bibinfo {title} {Non-differentiable large-deviation functionals in
  boundary-driven diffusive systems},\ }\href
  {https://doi.org/10.1088/1742-5468/2012/10/L10001} {\bibfield  {journal}
  {\bibinfo  {journal} {J. Stat. Mech.}\ }\textbf {\bibinfo {volume} {2012}},\
  \bibinfo {pages} {L10001} (\bibinfo {year} {2012})}\BibitemShut {NoStop}%
\bibitem [{\citenamefont {{Tsobgni Nyawo}}\ and\ \citenamefont
  {Touchette}(2016)}]{2016_Nyawo_Large}%
  \BibitemOpen
  \bibfield  {author} {\bibinfo {author} {\bibfnamefont {P.}~\bibnamefont
  {{Tsobgni Nyawo}}}\ and\ \bibinfo {author} {\bibfnamefont {H.}~\bibnamefont
  {Touchette}},\ }\bibfield  {title} {\bibinfo {title} {Large deviations of the
  current for driven periodic diffusions},\ }\href
  {https://doi.org/10.1103/PhysRevE.94.032101} {\bibfield  {journal} {\bibinfo
  {journal} {Phys. Rev. E}\ }\textbf {\bibinfo {volume} {94}},\ \bibinfo
  {pages} {032101} (\bibinfo {year} {2016})}\BibitemShut {NoStop}%
\bibitem [{\citenamefont {Baek}\ \emph {et~al.}(2018)\citenamefont {Baek},
  \citenamefont {Kafri},\ and\ \citenamefont {Lecomte}}]{2018_Baek_Dynamical}%
  \BibitemOpen
  \bibfield  {author} {\bibinfo {author} {\bibfnamefont {Y.}~\bibnamefont
  {Baek}}, \bibinfo {author} {\bibfnamefont {Y.}~\bibnamefont {Kafri}},\ and\
  \bibinfo {author} {\bibfnamefont {V.}~\bibnamefont {Lecomte}},\ }\bibfield
  {title} {\bibinfo {title} {Dynamical phase transitions in the current
  distribution of driven diffusive channels},\ }\href
  {https://doi.org/10.1088/1751-8121/aaa8f9} {\bibfield  {journal} {\bibinfo
  {journal} {J. Phys. A: Math. Theor.}\ }\textbf {\bibinfo {volume} {51}},\
  \bibinfo {pages} {105001} (\bibinfo {year} {2018})}\BibitemShut {NoStop}%
\bibitem [{\citenamefont {Lefèvre}\ and\ \citenamefont
  {Biroli}(2007)}]{2007_Lefevre_Dynamics}%
  \BibitemOpen
  \bibfield  {author} {\bibinfo {author} {\bibfnamefont {A.}~\bibnamefont
  {Lefèvre}}\ and\ \bibinfo {author} {\bibfnamefont {G.}~\bibnamefont
  {Biroli}},\ }\bibfield  {title} {\bibinfo {title} {Dynamics of interacting
  particle systems: Stochastic process and field theory},\ }\href
  {https://doi.org/10.1088/1742-5468/2007/07/P07024} {\bibfield  {journal}
  {\bibinfo  {journal} {J. Stat. Mech.}\ }\textbf {\bibinfo {volume} {2007}},\
  \bibinfo {pages} {P07024} (\bibinfo {year} {2007})}\BibitemShut {NoStop}%
\bibitem [{\citenamefont {Derrida}(2007)}]{2007_Derrida_Non}%
  \BibitemOpen
  \bibfield  {author} {\bibinfo {author} {\bibfnamefont {B.}~\bibnamefont
  {Derrida}},\ }\bibfield  {title} {\bibinfo {title} {Non-equilibrium steady
  states: Fluctuations and large deviations of the density and of the
  current},\ }\href {https://doi.org/10.1088/1742-5468/2007/07/P07023}
  {\bibfield  {journal} {\bibinfo  {journal} {J. Stat. Mech.}\ }\textbf
  {\bibinfo {volume} {2007}},\ \bibinfo {pages} {P07023} (\bibinfo {year}
  {2007})}\BibitemShut {NoStop}%
\bibitem [{\citenamefont {Bertini}\ \emph {et~al.}(2009)\citenamefont
  {Bertini}, \citenamefont {{De Sole}}, \citenamefont {Gabrielli},
  \citenamefont {Jona-Lasinio},\ and\ \citenamefont
  {Landim}}]{2009_Bertini_Towards}%
  \BibitemOpen
  \bibfield  {author} {\bibinfo {author} {\bibfnamefont {L.}~\bibnamefont
  {Bertini}}, \bibinfo {author} {\bibfnamefont {A.}~\bibnamefont {{De Sole}}},
  \bibinfo {author} {\bibfnamefont {D.}~\bibnamefont {Gabrielli}}, \bibinfo
  {author} {\bibfnamefont {G.}~\bibnamefont {Jona-Lasinio}},\ and\ \bibinfo
  {author} {\bibfnamefont {C.}~\bibnamefont {Landim}},\ }\bibfield  {title}
  {\bibinfo {title} {Towards a nonequilibrium thermodynamics: A self-contained
  macroscopic description of driven diffusive systems},\ }\href
  {https://doi.org/10.1088/1742-5468/2010/11/L11001} {\bibfield  {journal}
  {\bibinfo  {journal} {J. Stat. Phys.}\ }\textbf {\bibinfo {volume} {135}},\
  \bibinfo {pages} {857} (\bibinfo {year} {2009})}\BibitemShut {NoStop}%
\bibitem [{\citenamefont {Sadhu}\ and\ \citenamefont
  {Derrida}(2016)}]{2016_Sadhu_Correlations}%
  \BibitemOpen
  \bibfield  {author} {\bibinfo {author} {\bibfnamefont {T.}~\bibnamefont
  {Sadhu}}\ and\ \bibinfo {author} {\bibfnamefont {B.}~\bibnamefont
  {Derrida}},\ }\bibfield  {title} {\bibinfo {title} {Correlations of the
  density and of the current in non-equilibrium diffusive systems},\ }\href
  {https://doi.org/10.1088/1742-5468/2016/11/113202} {\bibfield  {journal}
  {\bibinfo  {journal} {J. Stat. Mech.}\ }\textbf {\bibinfo {volume} {2016}},\
  \bibinfo {pages} {113202} (\bibinfo {year} {2016})}\BibitemShut {NoStop}%
\bibitem [{\citenamefont {Mukherjee}(2024)}]{git_repository}%
  \BibitemOpen
  \bibfield  {author} {\bibinfo {author} {\bibfnamefont {R.}~\bibnamefont
  {Mukherjee}},\ }\href@noop {} {\bibinfo {title} {Active lattice gas}},\
  \bibinfo {howpublished}
  {\url{https://github.com/rikmukherjee/Active-Lattice-Gas.git}} (\bibinfo
  {year} {2024})\BibitemShut {NoStop}%
\end{thebibliography}

\end{document}